\documentclass[onecolumn,superscriptaddress,amsmath,amssymb,floatfix,aps,notitlepage,prx,longbibliography]{revtex4-1}
\usepackage{mathrsfs, hyperref}
\usepackage{amssymb, amsbsy, amsmath, latexsym, dsfont, array, layout, graphics,mathrsfs,braket,amsfonts,amsthm,
  amssymb,graphicx,subfigure, youngtab,color,bm,mathtools,braket,verbatim,url,cleveref,natbib,hypernat,extarrows,inputenc,multirow}
\usepackage[usernames,dvipsnames]{xcolor}
\usepackage[]{threeparttable}

\makeatletter
\newcommand*{\inlineequation}[2][]{%
  \begingroup
    \refstepcounter{equation}%
    \ifx\\#1\\%
    \else
      \label{#1}%
    \fi
    \relpenalty=10000 %
    \binoppenalty=10000 %
    \ensuremath{%
      #2%
    }%
    ~\@eqnnum
  \endgroup
}
\makeatother
\makeatletter
\def\old@comma{,}
\catcode`\,=13
\def,{%
  \ifmmode%
    \old@comma\discretionary{}{}{}%
  \else%
    \old@comma%
  \fi%
}
\makeatother

\newcolumntype{M}[1]{>{\centering\arraybackslash}m{#1}}
\newcolumntype{N}{@{}m{0pt}@{}}

\allowdisplaybreaks

\newcommand{\splitatcommas}[1]{%
  \begingroup
  \begingroup\lccode`~=`, \lowercase{\endgroup
    \edef~{\mathchar\the\mathcode`, \penalty0 \noexpand\hspace{0pt plus 1em}}%
  }\mathcode`,="8000 #1%
  \endgroup
}

\newcommand{\bs}[1]{\boldsymbol{#1}}
\newcommand{\mbf}[1]{\mathbf{#1}}

\newcommand{\externalStress}[1]{{\color{red}#1}}
\newcommand{\inclusionI}[1]{{\color{blue}#1}}
\newcommand{\inclusionJ}[1]{{\color{OliveGreen}#1}}
\newcommand{\imageInclusion}[1]{{\color{Plum}#1}}

\begin{document}
\title{Method of image charges for describing linear deformation of\\
bounded 2D solid structures with circular holes and inclusions}
\author{Siddhartha Sarkar}
\affiliation{Department of Electrical Engineering, Princeton University, Princeton, NJ 08544, USA}
\author{Matja\v{z} \v{C}ebron}
\affiliation{Faculty of Mechanical Engineering, University of Ljubljana, SI-1000 Ljubljana, Slovenia}
\author{Miha Brojan}
\email{miha.brojan@fs.uni-lj.si}
\affiliation{Faculty of Mechanical Engineering, University of Ljubljana, SI-1000 Ljubljana, Slovenia}
\author{Andrej Ko\v{s}mrlj}
\email{andrej@princeton.edu}
\affiliation{Department of Mechanical and Aerospace Engineering, Princeton University, Princeton, NJ 08544, USA}
\affiliation{Princeton Institute for the Science and Technology of Materials, Princeton University, Princeton, NJ 08544, USA}

\begin{abstract}
We present a method for predicting the linear response deformation of finite and semi-infinite 2D solid structures with circular holes and inclusions by employing the analogies with image charges and induction in electrostatics. Charges in electrostatics induce image charges near conductive boundaries and an external electric field induces polarization (dipoles, quadrupoles, and other multipoles) of conductive and dielectric objects. Similarly,  charges in elasticity  induce image charges near boundaries and  external stress induces polarization (quadrupoles and other multipoles) inside holes and inclusions. Stresses generated by these induced elastic multipoles as well as stresses generated by their images near boundaries then lead to interactions between holes and inclusions and with their images, which induce additional polarization and thus additional deformation of holes and inclusions. We present a method that expands induced polarization in a series of elastic multipoles, which systematically takes into account the interactions of inclusions and holes with the external field, between them, and with their images. The results of our method for linear deformation of circular holes and inclusions near straight and curved  boundaries show good agreement with both linear finite element simulations and experiments.
\end{abstract}
\maketitle

\section{Introduction}

Elastic materials with holes and inclusions have been studied extensively in materials science, engineering, and biology. On one hand, the goal is to homogenize the microscale distribution of holes and inclusions to obtain effective material properties on the macroscale~\cite{Eshelby,Hashin,Castaneda,Torquato}, where the detailed micropattern of deformations and stresses is ignored. On the other hand, in mechanical metamaterials~\cite{bertoldi2017flexible}, the geometry, topology and contrasting elastic properties of different materials, are exploited to obtain extraordinary functionalities, such as shape morphing~\cite{ShapeMorph,ShapeMorph2}, mechanical cloaking~\cite{Cloaking,Cloaking2,Cloaking3}, negative Poisson's ratio~\cite{almgren1985isotropic,lakes1987foam,Auxetic2,Auxetic,babaee20133d}, negative thermal expansion~\cite{NegativeThermalExpansion,NegativeThermalExpansion2}, effective negative swelling~\cite{NegativeSwelling,NegativeSwelling2,NegativeSwelling3}, and tunable phononic band gaps~\cite{BandGap3,BandGap,BandGap2}. Furthermore, in biology, the microscale interactions between rigid proteins embedded in biological membranes  structures have important functional consequences~\cite{ProteinAssembly2,ProteinAssembly,Haselwandter,Protein,goulian1993long,park1996interactions,golestanian1996fluctuationPRE,golestanian1996fluctuationEPL,weikl1998interaction,yolcu2012membrane,Deserno,Purohit}.

Linear deformations of infinite thin plates with circular holes under external loading have been studied extensively over the years~\cite{Green,Haddon,Ukadgaonker,Ting,Hoang08}. Green  demonstrated how to construct a solution for infinite thin plates with an arbitrary number of holes~\cite{Green} by expanding the Airy stress function around each hole in terms of the Michell solution for biharmonic functions~\cite{Michell}. However, it remained unclear how  this procedure could be generalized to finite structures with boundaries.

In the two companion papers, we generalized Green's method~\cite{Green} by employing analogies with electrostatics to describe the stress distribution and displacements of cylindrical holes and inclusions embedded either in a thin elastic plate (plane stress) or in an infinitely thick elastic matrix (plane strain) that are under small external loads, which can be treated as a 2D problem with circular holes and inclusions. We considered infinite, semi-infinite as well as finite-size 2D structures.

\begin{figure}[!t]
  \centering
  \includegraphics[scale=.95]{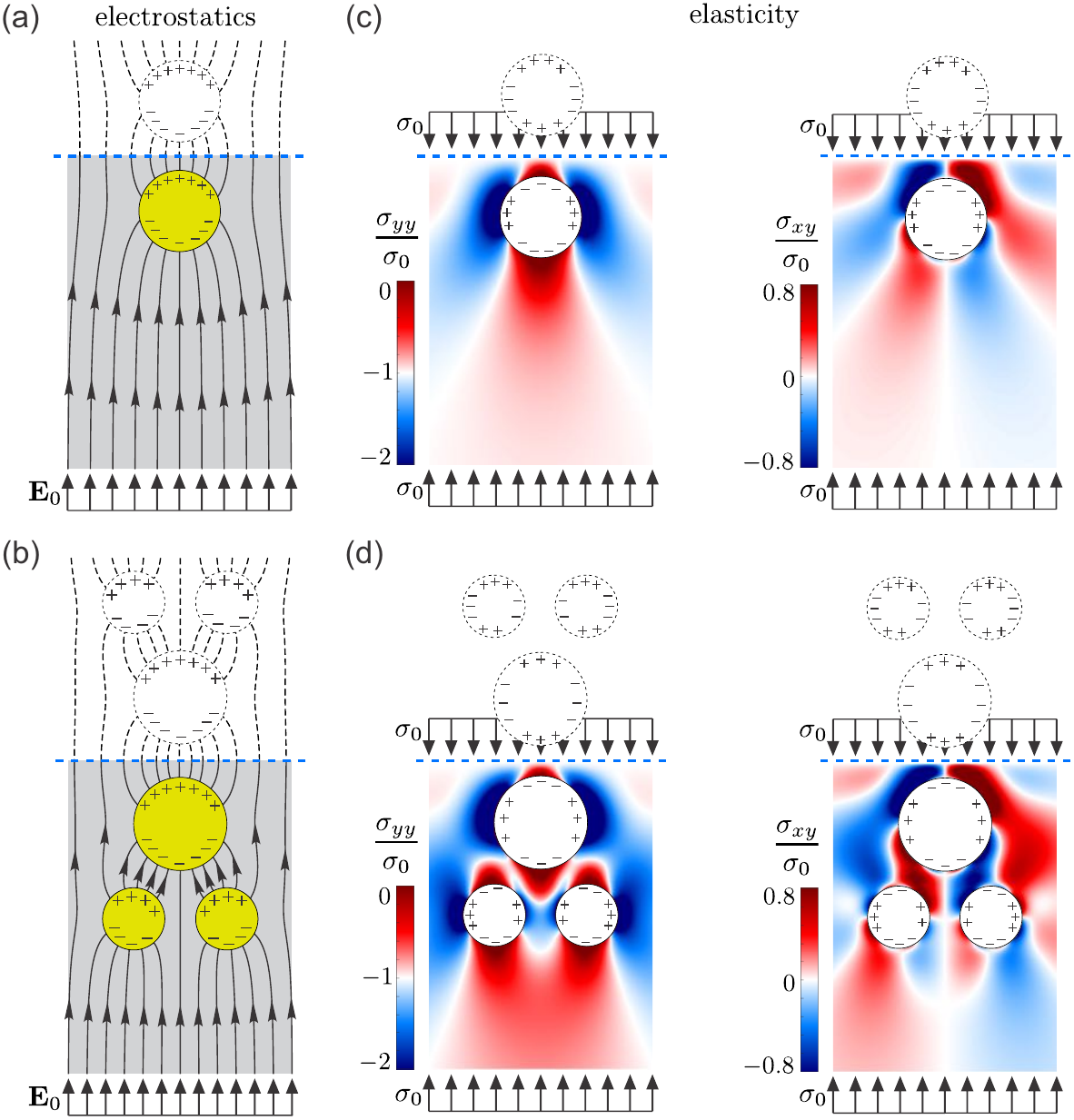}
  \caption{Induction and image charges in electrostatics and elasticity. (a,b)~Induced polarization due to external electric field $\mathbf{E}_0$ of (a)~a single and (b)~multiple conducting spheres (yellow disks) near a conductive wall (dashed blue line), which induces image charges (dashed circles). The induced charge distribution on conductive spheres is influenced by interactions between the spheres as well as by interactions with their images. The resultant electric field lines are shown in grey color.
(c,d)~Induced quadrupoles due to external uniaxial compressive stress $\sigma_0$ in (c)~a single and (d)~multiple circular holes (white disks) embedded in an elastic matrix near an edge with prescribed tractions (dashed blue line), which induces image charges (dashed circles).  The induced charge distribution inside holes is affected by interactions between holes as well as by interactions with their images. Heat maps show the resultant stress fields $\sigma_{yy}$ and $\sigma_{xy}$.} 
  \label{fig:induction}
\end{figure}

Charges in electrostatics induce image charges near conductive boundaries and an external electric field induces polarization (dipoles, quadrupoles, and other multipoles) of conductive and dielectric objects. When such a polarized object is near a conductive plate, it is also influenced by the electric field generated by its image, which leads to further charge redistribution on the surface of this object (see Fig.~\ref{fig:induction}a). Moreover, when multiple conductive objects in an external electric field are placed near conductive plates, they interact with each other as well as with their images via the electric fields generated by induced polarizations and induced image charges (see Fig.~\ref{fig:induction}b).

Similarly, charges in elasticity induce image charges near boundaries and  external stress induces polarization (quadrupoles and other multipoles) inside deformed holes and inclusions. When such a deformed inclusion is  near a boundary it is also influenced by the stress fields produced by its image, which leads to further deformation of the inclusion (see Fig.~\ref{fig:induction}c). When multiple deformed holes and inclusions are located near boundaries, they interact with each other as well as with their images via the stress fields generated by induced elastic multipoles and induced image charges (see Fig.~\ref{fig:induction}d). 

In a companion paper~\cite{sarkar2019elastic}, we presented a method to describe linear deformations of circular holes and inclusions embedded in an \textit{infinite} 2D elastic matrix under small external loads, by systematically expanding induced polarization of each hole/inclusion in terms of elastic multipoles, which are related to terms in the Michell solution for biharmonic functions~\cite{Michell}. Here, we generalize this method to describe linear deformation of \textit{finite} and \textit{semi-infinite} 2D structures with circular holes and inclusions by taking into account also the interaction with induced images near boundaries. The results of this method are compared  with linear finite element simulations and experiments. We show that the error decreases exponentially as the maximum degree of elastic multipoles is increased. 

The remaining part of the paper is organized as follows. In Section~\ref{sec:images}, we review the concept of image charges in electrostatics and  demonstrate how to construct image charges and image multipoles near a straight traction-free edge in 2D linear elasticity. In Sections~\ref{sec:flat_edge_traction} and \ref{sec:strip}, we use image charges and image multipoles to develop a method for evaluating the deformation of semi-infinite structures (Sec.~\ref{sec:flat_edge_traction}) and infinite strips (Sec.~\ref{sec:strip}) with circular holes and inclusions under external stress load. In Section~\ref{sec:flat_edge_displacements}, we show how to adapt this method, when such semi-infinite structures are under displacement-controlled loads. In Section~\ref{sec:disk}, we demonstrate how to use image charges to evaluate the deformation of circular disks with circular holes and inclusion under 3 different loading conditions: hydrostatic stress, no-slip (displacement control), and slip. In all Sections, the results of our method are compared with linear finite element simulations and experiments. In Section~\ref{sec:conclusion}, we give concluding remarks and we discuss potential extensions of this method to structures with other geometries.

\section{Image charges in electrostatics and elasticity}
\label{sec:images}

The method of image charges is commonly used in electrostatics to satisfy various boundary conditions~\cite{Jackson}. Similarly, image charges can be introduced in 2D elasticity to satisfy boundary conditions. Below we first discuss image charges in electrostatics for charges near a straight conductive edge and then  demonstrate how image charges can be constructed in elasticity for charges near a straight traction-free edge. In Sections~\ref{sec:flat_edge_displacements} and~\ref{sec:disk} we also show how to construct image charges near a straight rigid edge (Sec.~\ref{sec:flat_edge_displacements}) and near a curved edge (Sec.~\ref{sec:disk}) with three different types of boundary conditions.

\begin{figure}[!t]
  \centering
  \includegraphics[scale=1]{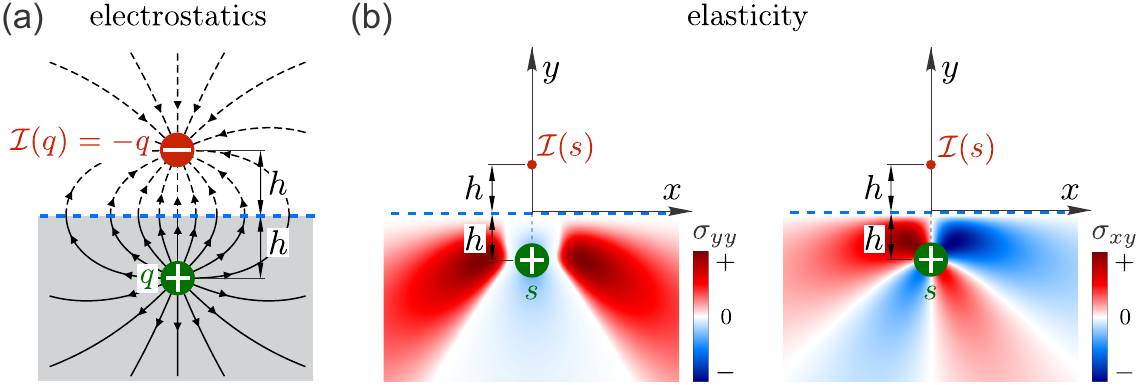}
  \caption{Image charges in electrostatics and elasticity. (a)~A point electric charge $q$ (green) near an infinite conducting plate (dashed blue line) induces an opposite image charge $\mathcal{I}(q)=-q$ (red) to satisfy the boundary condition that the electric field lines (black lines) are orthogonal to the conductive plate. (b)~A point disclination with charge $s$ (green) embedded in a semi-infinite elastic material near a traction-free edge (dashed blue line) induces an image charge $\mathcal{I}(s)$ (red) to satisfy boundary conditions at the traction-free edge ($\sigma_{xy}=\sigma_{yy}=0$). Heat maps show the resultant stress fields $\sigma_{yy}$ and $\sigma_{xy}$.} 
  \label{Fig:ImageIllus}
\end{figure}

Let us consider a 2D semi-infinite dielectric medium with permittivity $\epsilon_e$ filling half-space $y<y_b$ and with a monopole with charge $q$  at position $\mbf{x}_0=(x_0,y_0)=(x_0,y_b-h)$, which is at the distance $h$ from an infinite perfectly conducting straight edge at $y=y_b$, as shown in Fig.~\ref{Fig:ImageIllus}a. The electric potential $U(\mbf{x})$ for this case can be found by solving the governing equation $\Delta U =-\frac{q}{\epsilon_e}\delta(\mbf{x}-\mbf{x}_0)$ subject to the boundary condition imposed by the conducting edge, i.e. $E_x(x,y_b)=0$, where the electric field $\mbf{E}(\mbf{x})=(E_x(\mbf{x}),E_y(\mbf{x}))$ is related to the electric potential $U(\mbf{x})$ as $\mbf{E}=-\bs{\nabla} U$.~\cite{Jackson} The solution for the governing equation in an infinite 2D dielectric medium is $U_m(\mbf{x}-\mbf{x}_0|q) = -\frac{q}{2\pi \epsilon_e}\ln |\mbf{x}-\mbf{x}_0|$, but the corresponding electric field does not satisfy boundary conditions at the conducting edge. Boundary conditions can be satisfied by placing an image monopole charge $-q$ on the opposite side of the conducting edge at position $\mbf{x}^*_0=(x_0^*,y_0^*)=(x_0,y_b+h)$ (see Fig.~\ref{Fig:ImageIllus}a) with the corresponding electric potential 
$\mathcal{I}[U_m(\mbf{x}-\mbf{x}_0|q)]=+\frac{q}{2\pi \epsilon_e}\ln |\mbf{x}-\mbf{x}^*_0|$. The total electric potential is thus $U^\text{tot}(\mbf{x})=U_m(\mbf{x}-\mbf{x}_0|q)+\mathcal{I}[U_m(\mbf{x}-\mbf{x}_0|q)]=-\frac{q}{2\pi \epsilon_e}\ln |\mbf{x}-\mbf{x}_0|+\frac{q}{2\pi \epsilon_e}\ln |\mbf{x}-\mbf{x}^*_0|$.~\cite{Jackson} It is straightforward to check that the corresponding electric field $\mbf{E}^\text{tot}=-\bs{\nabla} U^\text{tot}$ satisfies the boundary condition at the conducting edge (see Fig.~\ref{Fig:ImageIllus}a). Note that the image of the image charge is the original charge itself, i.e. $\mathcal{I}\big[\mathcal{I}[U_m(\mbf{x}-\mbf{x}_0|q)]\big]=U_m(\mbf{x}-\mbf{x}_0|q)$.~\cite{Jackson}

Let us  consider a related problem in 2D elasticity for a semi-infinite elastic medium with the Young's modulus $E$ filling  half-space $y<y_b$  with a topological monopole (disclination) with charge $s$   at position $\mbf{x}_0=(x_0,y_0)=(x_0,y_b-h)$, which is at the distance $h$ from the traction-free edge at $y=y_b$ (see Fig.~\ref{Fig:ImageIllus}b). The Airy stress function $\chi(\mbf{x})$ for this case can be found by solving the governing equation $\Delta \Delta \chi = E s \delta (\mbf{x}-\mbf{x}_0)$  subject to the boundary condition imposed by the traction-free edge, i.e. $\sigma_{xy}(x,y_b)=\sigma_{yy}(x,y_b)=0$, where stress fields $\sigma_{ij}(\mbf{x})$ are related to the Airy stress function $\chi(\mbf{x})$ as $\sigma_{xx}=\frac{\partial^2 \chi}{\partial y^2}$, $\sigma_{yy}=\frac{\partial^2 \chi}{\partial x^2}$, and $\sigma_{xy}=-\frac{\partial^2 \chi}{\partial x \partial y}$. The solution for the governing equation in an infinite 2D elastic medium is $\chi_m^s(\mbf{x}-\mbf{x}_0|s)  = \frac{E s}{8\pi}|\mbf{x}-\mbf{x}_0|^2 \big(\ln|\mbf{x}-\mbf{x}_0| -1/2\big)$,~\cite{Chaikin,Moshe2} but the corresponding stress fields do not satisfy boundary conditions at the traction-free edge. Inspired by electrostatic we try placing an image disclination with charge $-s$ on the opposite side of the traction-free edge at position $\mbf{x}_0^* = (x_0^*,y_0^*)=(x_0,y_b+h)$. The trial Airy stress function is thus $\chi^\text{tr}(\mbf{x})=\chi_m^s(\mbf{x}-\mbf{x}_0|s)+\chi_m^s(\mbf{x}-\mbf{x}^*_0|-s)=\frac{E s}{8\pi}|\mbf{x}-\mbf{x}_0|^2 \big(\ln|\mbf{x}-\mbf{x}_0| -1/2\big)-\frac{E s}{8\pi}|\mbf{x}-\mbf{x}^*_0|^2 \big(\ln|\mbf{x}-\mbf{x}^*_0| -1/2\big)$. This trial Airy stress function $\chi^\text{tr}(\mbf{x})$ satisfies one of the boundary conditions, $\sigma^\textrm{tr}_{yy}(x,y_b)=0$, at the traction-free edge, but the other boundary condition is violated because $\sigma^\textrm{tr}_{xy}(x,y_b)=-\frac{E s (x-x_0) h}{2\pi \left((x-x_0)^2+h^2\right)}\ne 0$. Thus in elasticity additional image  multipoles have to be included  to satisfy both boundary conditions. This difference between electrostatics and elasticity stems from the fact that the electric potential $U(\mbf{x})$ is a harmonic function~\cite{Jackson}, while the Airy stress function $\chi(\mbf{x})$ is a biharmonic function~\cite{Barber}. The additional image multipoles can be found by calculating the Airy stress function $\chi^a(\mbf{x})$ that corresponds to distributed tractions $f_x(x)=-\sigma^\textrm{tr}_{xy}(x,y_b)=\frac{E s (x-x_0) h}{2\pi ((x-x_0)^2+h^2)}$ and $f_y(x)=-\sigma^\textrm{tr}_{yy}(x,y_b)=0$ along the edge $(x,y_b)$. This can be done with the help of the solution of the Boussinesq problem in 2D, where the response to a concentrated force $\mbf{F}=(1,0)$ in the $x$-direction at point $(x',y_b)$ on the edge is described with the Airy stress function $\chi_B(\mbf{x}|x',y_b)=-(1/\pi) (y-y_b) \arctan[(y-y_b)/(x-x')]$, where $\mbf{x}=(x,y)$.~\cite{Landau} The Airy stress function of the additional elastic image multipoles can then be obtained as $\chi^a(\mbf{x}) = \int_{-\infty}^\infty dx' f_x(x') \chi_B(\mbf{x}|x',y_b)=-(E s h/2\pi) (y-y_0^*+h) \ln|\mbf{x}-\mbf{x}^*_0|$. Thus the Airy stress functions $\chi_m^s(\mbf{x}-\mbf{x}_0|s)$ for the monopole disclination and for its image $\mathcal{I}[\chi_m^s(\mbf{x}-\mbf{x}_0|s)]$ can be expressed as
\begin{subequations}
\begin{align}
    \chi_m^s(\mbf{x}-\mbf{x}_0|s) & = +\frac{E s}{8\pi}|\mbf{x}-\mbf{x}_0|^2 \big(\ln|\mbf{x}-\mbf{x}_0| -1/2\big) = +\frac{Es}{8 \pi} r^2 \big(\ln r - 1/2 \big),\\
    \mathcal{I}[\chi_m^s(\mbf{x}-\mbf{x}_0|s)] & = - \frac{E s}{8\pi}|\mbf{x}-\mbf{x}_0^*|^2 \big(\ln|\mbf{x}-\mbf{x}_0^*| -1/2\big)-\frac{E s h}{2\pi} (y-y_0^*+h) \ln|\mbf{x}-\mbf{x}^*_0|, \nonumber\\
    \mathcal{I}[\chi_m^s(\mbf{x}-\mbf{x}_0|s)] &=-\frac{Es}{8 \pi} {r^*}^2 \big(\ln r^* - 1/2 \big)-\frac{E s h}{2 \pi} r^*\sin\varphi^* \ln r^* -\frac{E s h^2}{2 \pi} \ln r^*,
\end{align}  
\end{subequations}
where we introduced polar coordinates $(r,\varphi)$ and $(r^*,\varphi^*)$  centered at the disclination ($r=\sqrt{(x-x_0)^2+(y-y_0)^2}$, $\varphi=\arctan[(y-y_0)/(x-x_0)]$) and the position of its image ($r^*=\sqrt{(x-x_0^*)^2+(y-y_0^*)^2}$, $\varphi^*=\arctan[(y-y_0^*)/(x-x_0^*)]$), respectively. It is straightforward to check that the stress fields resulting from the Airy stress function $\chi^\text{tot}=\chi_m^s+\mathcal{I}[\chi_m^s]$ satisfy the boundary conditions along the edge (see Fig.~\ref{Fig:ImageIllus}b). Just like in electrostatics, the image of the image disclination is the original disclination itself, i.e. $\mathcal{I}\big[\mathcal{I}[\chi_m^s(\mbf{x}-\mbf{x}_0|s)]\big]=\chi_m^s(\mbf{x}-\mbf{x}_0|s)$.

The Airy stress function for the disclination $\chi_m^s$ can be used to systematically construct the Airy stress functions for all other elastic multipoles, as  discussed in detail in the companion paper~\cite{sarkar2019elastic}. Here, we repeat this procedure to construct the images for all other elastic multipoles. First, we show how to construct an image for a topological dipole (dislocation), which is formed at $\mbf{x}_0=(x_0,y_0)=(x_0,y_b-h)$, when two opposite disclinations with charges $\pm s$ are located at $\mbf{x}_\pm=\mbf{x}_0\pm a (\cos\theta,\sin\theta)$, where the angle $\theta$ describes the orientation of dislocation. Thus the Airy stress functions $\chi_d^s(\mbf{x}-\mbf{x}_0)$ for the dislocation and $\mathcal{I}[\chi_d^s(\mbf{x}-\mbf{x}_0)]$ for its image   at position $\mbf{x}^*_0=(x_0^*,y_0^*)=(x_0,y_b+h)$  can be expressed as
\begin{subequations}
  \begin{align}
  \chi_d^s(\mbf{x}-\mbf{x}_0|sa,\theta) &= \chi_m^s(\mbf{x}-\mbf{x}_+|s)+ \chi_m^s(\mbf{x}-\mbf{x}_-|-s)\ \xlongrightarrow{{a\to 0}}\ 
  -\frac{E s a}{2\pi}
  r \ln r \cos (\varphi-\theta) ,\\
\mathcal{I}[\chi_d^s(\mbf{x}-\mbf{x}_0|sa,\theta)] &= \mathcal{I}[\chi_m^s(\mbf{x}-\mbf{x}_+|s)]+ \mathcal{I}[\chi_m^s(\mbf{x}-\mbf{x}_-|-s)], \nonumber \\
  \mathcal{I}[\chi_d^s(\mbf{x}-\mbf{x}_0|sa,\theta)] & \xlongrightarrow{{a\to 0}} +\frac{E s a}{2\pi}
  r^* \ln r^* \cos (\varphi^*-\theta)  + \frac{Esah}{\pi} \left[\frac{1}{2}\sin(2 \varphi^* + \theta) +\sin \theta \left(\ln r^* -\frac{1}{2}\right)\right]\nonumber\\
  &\phantom{\xlongrightarrow{{a\to 0}}+\frac{E s a}{2\pi}
  r^* \ln r^*\cos (\varphi^*-\theta) } \  +\frac{Esah^2}{\pi} \frac{\cos(\varphi^* + \theta)}{r^*},
  \end{align}
\end{subequations}
where we again used polar coordinates $(r, \varphi)$ centered at the dislocation and $(r^*, \varphi^*)$ centered at its image.

The Airy stress functions for a quadrupole and higher-order multipoles and their images can be constructed similarly. The multipole $\mbf{Q}_n^s$ of order $n$ is constructed by placing $n$ positive and $n$ negative disclinations symmetrically around $\mbf{x}_0=(x_0,y_0)=(x_0,y_b-h)$, such that disclinations of charges $s_i=s(-1)^i$ are placed at positions $\mbf{x}_i=\mbf{x}_0+a \big(\cos(\theta+i\pi/n),\sin(\theta+i\pi /n)\big)$, where $i~\in~\{0,1,\ldots,2n-1\}$ and the angle $\theta$ describes the orientation of elastic multipole. The Airy stress functions for such multipoles $\chi_{n}^{s}$ and their images $\mathcal{I}[\chi_{n}^{s}]$ at positions $\mbf{x}^*_0=(x_0^*,y_0^*)=(x_0,y_b+h)$ are expressed in polar coordinates as
\begin{subequations}
  \begin{align}
      \chi_{n}^{s}(\mbf{x}-\mbf{x}_0|sa^n,\theta) &=\sum_{i=0}^{2n-1}  \chi_m^s(\mbf{x}-\mbf{x}_i|s_i) \ \xlongrightarrow{{a\to 0}}\ +\frac{E s a^n}{4(n-1) \pi} \frac{ \cos\big(n (\varphi-\theta)\big)}{r^{n-2}},\\
      \mathcal{I}[\chi_{n}^{s}(\mbf{x}-\mbf{x}_0|sa^n,\theta)] &=\sum_{i=0}^{2n-1} \mathcal{I}[\chi_m^s(\mbf{x}-\mbf{x}_i|s_i)], \nonumber \\
      \mathcal{I}[\chi_{n}^{s}(\mbf{x}-\mbf{x}_0|sa^n,\theta)] & \xlongrightarrow{{a\to 0}}\ +\frac{E s a^n}{4(n-1) \pi} \frac{\left[(n-1) \cos\big(n \varphi^* + n \theta\big) - n \cos\big((n-2)\varphi^* +n \theta\big) \right]}{{r^*}^{n-2}}\nonumber \\
      & \quad \quad \quad + \frac{E s a^n h}{2(n-1) \pi} \frac{\left[(n-1) \sin\big((n+1)\varphi^* + n \theta\big) - (2n-1) \sin\big((n-1)\varphi^* +n \theta\big) \right]}{{r^*}^{n-1}} \nonumber \\
      & \quad \quad \quad +\frac{E s a^n h^2}{\pi} \frac{\cos\big(n \varphi^* + n \theta\big)}{{r^*}^{n}}.
  \end{align}
\end{subequations}

In 2D elasticity, there is another type of monopole besides disclinations. This is a non-topological monopole with charge $p$ that describes local contraction of material at point $\mbf{x}_0=(x_0,y_0)=(x_0,y_b-h)$ and the corresponding Airy stress function $\chi_m^p$ can be obtained as the solution of equation $\Delta \Delta \chi_m^p = E p \Delta_0 \delta (\mbf{x}-\mbf{x}_0)$, where $\Delta_0$ corresponds to the Laplace operator with respect to $\mbf{x}_0$.~\cite{Moshe2,sarkar2019elastic} The Airy stress functions $\chi_m^p$ for this non-topological monopole and $\mathcal{I}[\chi_m^p]$ for its image at position $\mbf{x}^*_0=(x_0^*,y_0^*)=(x_0,y_b+h)$ can thus be obtained from the Airy stress function $\chi_m^s$ for disclination as
\begin{subequations}
  \begin{align}
    \chi_m^p(\mbf{x}-\mbf{x}_0|p)&=\Delta_0 \chi_m^s(\mbf{x}-\mbf{x}_0|p) = +\frac{E p}{4 \pi} \left(1+2 \ln r\right),\\  
    \mathcal{I}[\chi_m^p(\mbf{x}-\mbf{x}_0|p)]&=\Delta_0\mathcal{I}[ \chi_m^s(\mbf{x}-\mbf{x}_0|p)]= +\frac{Ep}{4 \pi} \left[1-2 \ln r^* - 2 \cos(2 \varphi^*) \right] + \frac{E p h}{\pi} \frac{\sin \varphi^*}{r^*}.
  \end{align}
\end{subequations}

The Airy stress functions for a dipole, a quadrupole, and higher-order multipoles that arise from non-topological monopoles $p$ and their images can be constructed similarly as  for higher-order multipoles arising from disclinations. The multipole $\mbf{Q}_n^p$ of order $n$ is constructed by placing $n$ positive and $n$ negative non-topological monopoles symmetrically around $\mbf{x}_0=(x_0,y_0)=(x_0,y_b-h)$, such that non-topological monopoles of charges $p_i=p(-1)^i$ are placed at positions $\mbf{x}_i=\mbf{x}_0+a \big(\cos(\theta+i\pi/n),\sin(\theta+i\pi /n)\big)$, where $i~\in~\{0,1,\ldots,2n-1\}$ and the angle $\theta$ describes the orientation of multipole. Airy stress functions $\chi_{n}^{p}$ for such multipoles  and $\mathcal{I}[\chi_{n}^{p}]$ for their images at positions $\mbf{x}^*_0=(x_0^*,y_0^*)=(x_0,y_b+h)$ are expressed in polar coordinates as
\begin{subequations}
  \begin{align}
      \chi_{n}^{p}(\mbf{x}-\mbf{x}_0|pa^n,\theta) &=\sum_{i=0}^{2n-1}  \chi_n^p(\mbf{x}-\mbf{x}_i|p_i) \ \xlongrightarrow{{a\to 0}}\ -\frac{E p a^n}{\pi}\  \frac{ \cos\big(n (\varphi-\theta)\big)}{r^{n}},\\
      \mathcal{I}[\chi_{n}^{p}(\mbf{x}-\mbf{x}_0|pa^n,\theta)] &=\sum_{i=0}^{2n-1} \mathcal{I}[\chi_m^p(\mbf{x}-\mbf{x}_i|p_i)], \nonumber \\
      \mathcal{I}[\chi_{n}^{p}(\mbf{x}-\mbf{x}_0|pa^n,\theta)] & \xlongrightarrow{{a\to 0}}\ \frac{E p a^n}{\pi}\  \frac{\left[(n+1) \cos\big(n \varphi^* + n \theta\big) - n \cos\big((n+2)\varphi^* +n \theta\big) \right]}{{r^*}^{n}}\nonumber \\
      & \quad \quad \quad + \frac{E p a^n h}{\pi}\  \frac{2n \sin\big((n+1)\varphi^* + n \theta\big)}{{r^*}^{n+1}}.
  \end{align}
\end{subequations}
 
 Note that the elastic multipoles described above are related to the terms in the Michell solution~\cite{Michell} for the biharmonic equation, which is discussed in detail in the companion paper~\cite{sarkar2019elastic}. The operator $\mathcal{I}$ for constructing the Airy stress functions of images can be written in a compact form  as was previously shown by Adeerogba~\cite{Adeerogba}
\begin{equation}
\label{eq:AdeerogbaOpFree}
\mathcal{I}\big[\chi_0(x,y)\big]=-\left(1-2(y-y_b)\frac{\partial}{\partial y}+(y-y_b)^2\Delta\right)\chi_0(x,2y_b-y).
\end{equation}

\section{Elastic multipole method with image charges}\label{sec:Image}

Using the concept of image charges described  in Section~\ref{sec:images} and procedures that were presented in the companion paper~\cite{sarkar2019elastic}, we describe a method for calculating the linear deformation of circular inclusions and holes embedded either in a semi-infinite ($y<y_b$) elastic matrix,  in an infinite elastic strip ($y_a<y<y_b$), or in an elastic disk with radius $R$. External load induces elastic multipoles at the centers of  inclusions and holes, which further induce image multipoles in order to satisfy boundary condition at the edges. The amplitudes of induced multipoles are obtained from boundary conditions (continuity of tractions and displacements) between different materials in the elastic matrix.

In the following subsections, we describe the method for the general case where  circular inclusions can have different sizes and material properties (holes correspond to zero shear modulus). Note that our method applies to the deformation of cylindrical holes and inclusions embedded in a thin plate (plane stress) as well as of cylindrical inclusions and holes embedded in an infinitely thick elastic matrix (plane strain) by appropriately setting the values of Kolosov's constants~\cite{Barber}. In Section~\ref{sec:flat_edge_traction}, we describe the method for a semi-infinite elastic matrix ($y<y_b$) subject to uniform tractions along the boundary. In Section~\ref{sec:strip}, we adapt this method for the case of an infinite elastic strip ($y_a<y<y_b$) subject to uniform tractions along the boundary. In Section~\ref{sec:flat_edge_displacements}, we use our method to describe the deformation of a semi-infinite elastic matrix ($y<y_b$) subject to displacement control boundary conditions. Finally, in Section~\ref{sec:disk}, we describe the deformation of a circular disk subject to either hydrostatic stress, no-slip (displacement control), and slip boundary conditions. In all of these sections, we compare the results of our method with linear finite element simulations and  experiments.

\subsection{Inclusions in a semi-infinite elastic matrix with uniform tractions along the outer boundary of the matrix}
\label{sec:flat_edge_traction}
\subsubsection{Method}

\begin{figure}[!t]
  \centering
  \includegraphics[scale=1]{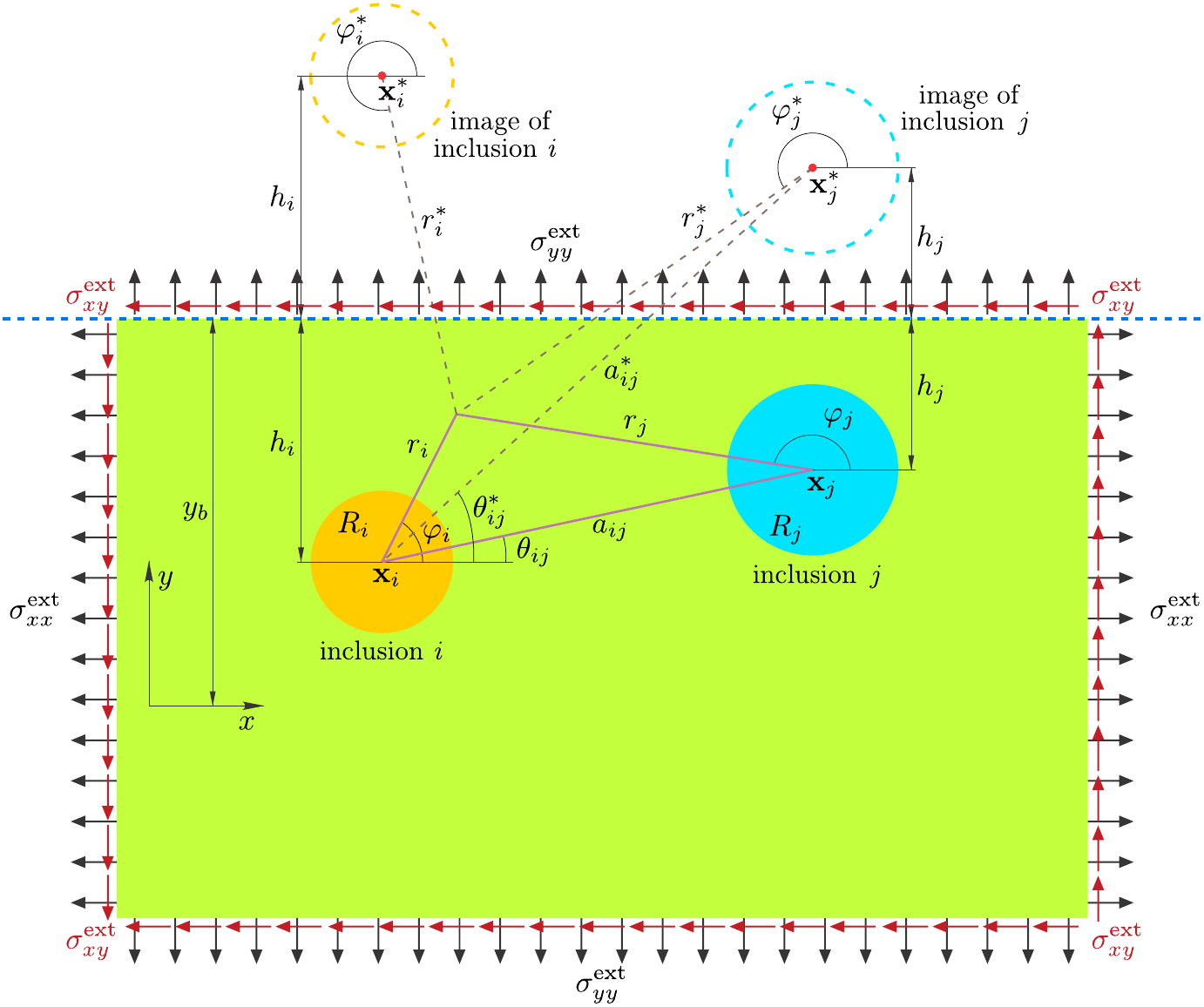}
  \caption{Illustration of external loading ($\sigma_{xx}^\text{ext}$, $\sigma_{yy}^\text{ext}$, $\sigma_{xy}^\text{ext}$) for the semi-infinite structure ($y<y_b$) with circular inclusions (colored disks) with uniform tractions along the edge (dashed blue line). Images of inclusions are represented with dashed circles. The schematic describes polar coordinates ($r_i,\varphi_i$) and ($r_j,\varphi_j$) relative to the centers $\mbf{x}_i$ of the $i^{\text{th}}$ inclusion with radius $R_i$ and $\mbf{x}_j$ of the $j^{\text{th}}$ inclusion with radius $R_j$, respectively. Similarly, we define polar coordinates ($r_i^*,\varphi_i^*$) and ($r_j^*,\varphi_j^*$) relative to the centers $\mbf{x}_i^*$ of the image of  the $i^{\text{th}}$ inclusion and $\mbf{x}_j^*$ of the image of the $j^{\text{th}}$ inclusion, respectively. Here, $a_{ij}$ ($a_{ij}^*$) is the separation distance between the centers of the $i^{\text{th}}$ inclusion and the $j^{\text{th}}$ inclusion (the image of  the $j^{\text{th}}$ inclusion) and $\theta_{ij}$ ($\theta_{ij}^*$) is the angle between the line joining their centers and the $x$-axis.
  }
\label{Fig:schematic2}
\end{figure}

Let us consider a 2D semi-infinite elastic matrix with the Young's modulus $E_0$ and Poisson's ratio $\nu_0$ filling the half-space $y<y_b$ subject to external stresses ($\sigma_{xx}^\textrm{ext}$, $\sigma_{yy}^\textrm{ext}$, $\sigma_{xy}^\textrm{ext}$). Embedded in the matrix are $N$ circular inclusions with radii $R_i$ centered at positions $\mbf{x}_i=(x_i,y_i)=(x_i,y_b-h_i)$ with Young's moduli $E_i$ and Poisson's ratios $\nu_i$, where $i\in\{1,\ldots, N\}$ and $h_i$ is the distance from the center of the $i^\text{th}$ inclusion to the traction controlled edge at $y=y_b$ (see Fig.~\ref{Fig:schematic2}). Holes are described with the zero Young's modulus ($E_i=0$). 

External stress, represented with the Airy stress function
\begin{equation}
    \chi_\textrm{ext}(x,y)=\frac{1}{2}\sigma_{xx}^\textrm{ext} y^2 + \frac{1}{2}\sigma_{yy}^\textrm{ext} x^2 - \sigma_{xy}^\textrm{ext} xy,
    \label{eq:ChiExt}
\end{equation}
induces non-topological monopoles, non-topological dipoles, quadrupoles  and higher-order multipoles  at the centers of inclusions, while topological defects (disclination, dislocation) cannot be induced, as was discussed in the companion paper~\cite{sarkar2019elastic}. Thus the Airy
stress function $ \chi_{\text{out}}\big(r_i,\varphi_i|\mbf{a}_{i,\text{out}}\big)$ outside the $i^{\textrm{th}}$ inclusion due to the induced multipoles can be expanded as
\begin{equation}
  \label{eq:AiryInducedOut}
  \begin{split}
  \chi_{\text{out}}\big(r_i,\varphi_i|\mbf{a}_{i,\text{out}}\big) = &A_{i,0} R_i^2 \ln \left(\frac{r_i}{R_i}\right) + \sum_{n=1}^\infty R_i^2 \left[ A_{i,n} \left(\frac{r_i}{R_i}\right)^{-n}\cos(n\varphi_i) +B_{i,n}
  \left(\frac{r_i}{R_i}\right)^{-n}\sin(n \varphi_i)\right]\\
  &\quad \quad \quad \quad \quad\quad\quad  +\sum_{n=2}^{\infty} R_i^2 \left[ C_{i,n} \left(\frac{r_i}{R_i}\right)^{-n+2}\cos (n
  \varphi_i) + D_{i,n}
  \left(\frac{r_i}{R_i}\right)^{-n+2}\sin (n \varphi_i)\right],
  \end{split}
\end{equation}
where the origin of polar coordinates $(r_i = \sqrt{(x-x_i)^2+(y-y_i)^2},\varphi_i = \arctan[(y-y_i)/(x-x_i)])$ is at the center $\mbf{x}_i=(x_i,y_i)=(x_i,y_b-h_i)$ of the $i^\text{th}$ inclusion with radius $R_i$ and  the set of amplitudes of induced multipoles is $\mbf{a}_{i,\text{out}}=\{A_{i,0}, A_{i,1},\dots,B_{i,1}, B_{i,2},\dots, C_{i,2}, C_{i,3},\dots, D_{i,2}, D_{i,3},\dots\}$. The induced elastic multipoles at the center of the $i^\text{th}$ inclusion then further induce image multipoles at position $\mbf{x}_i^*=(x_i^*,y_i^*)=(x_i,y_b+h_i)$ to satisfy  boundary conditions at the edge ($\sigma_{yy}(x,y_b)=\sigma_{yy}^\textrm{ext}$,  $\sigma_{xy}(x,y_b)=\sigma_{xy}^\textrm{ext}$). The Airy stress functions of image multipoles due to the $i^\text{th}$ inclusion can be expanded as
\begin{equation}
  \label{eq:AiryInducedOutImage2}
  \begin{split}
  \mathcal{I}[\chi_{\text{out}}\big(r_i,\varphi_i|\mbf{a}_{i,\text{out}}\big)] = & \chi^*_{\text{out}}\big(r_i^*,\varphi_i^*|\mbf{a}^*_{i,\text{out}}\big),\\
 \chi^*_{\text{out}}\big(r_i^*,\varphi_i^*|\mbf{a}^*_{i,\text{out}}\big)= &A_{i,0}^* R_i^2 \ln \left(\frac{r_i^*}{R_i}\right) + \sum_{n=1}^\infty R_i^2 \left[ A_{i,n}^{*} \left(\frac{r_i^*}{R_i}\right)^{-n}\cos(n\varphi_i^*) +B_{i,n}^{*}
  \left(\frac{r_i^*}{R_i}\right)^{-n}\sin(n \varphi_i^*)\right]\\
  &+ B_{i,0}^{*} R_i^2 \varphi_i^* \quad \, \ +\sum_{n=2}^{\infty} R_i^2 \left[ C_{i,n}^{*} \left(\frac{r_i^*}{R_i}\right)^{-n+2}\cos (n
  \varphi_i^*) + D_{i,n}^{*}
  \left(\frac{r_i^*}{R_i}\right)^{-n+2}\sin (n \varphi_i^*)\right],
  \end{split}
\end{equation}
\noindent where the origin of polar coordinates $(r_i^* = \sqrt{(x-x_i^*)^2+(y-y_i^*)^2},\varphi_i^* = \arctan[(y-y_i^*)/(x-x_i^*)])$ is at the center $\mbf{x}_i^*$ of the image of the $i^\text{th}$ inclusion. Note that in the above expansion in Eq.~(\ref{eq:AiryInducedOutImage2}) we included the term proportional to $B_{i,0}^{*} \varphi_i^*$, which is absent in the expansion for the Airy stress function $\chi_{\text{out}}\big(r_i,\varphi_i|\mbf{a}_{i,\text{out}}\big)$ of  the $i^\text{th}$ inclusion in Eq.~(\ref{eq:AiryInducedOut}). This term is not allowed for inclusions located inside the elastic matrix because the Airy stress function $\chi_{\text{out}}\big(r_i,\varphi_i|\mbf{a}_{i,\text{out}}\big)$ must be continuous as the angle $\varphi_i$ spans the range of values from $0$ to $2 \pi$. On the other hand, the images of inclusions are located outside the elastic matrix (see Fig.~\ref{Fig:schematic2}) and thus the angle $\varphi_i^*$ cannot attain  values in the whole range from $0$ to $2 \pi$. The amplitudes  $\mbf{a}_{i,\text{out}}^{*}=\{A_{i,0}^{*}, A_{i,1}^{*},\dots,B_{i,0}^{*}, B_{i,1}^{*},\dots, C_{i,2}^{*}, C_{i,3}^{*},\dots, D_{i,2}^{*}, D_{i,3}^{*},\dots\}$ of image multipoles are related to the amplitudes  $\mbf{a}_{i,\text{out}}$ of induced multipoles for the $i^\text{th}$ inclusion as 
\begin{align}
\begin{split}
    A_{i,n}^{*} &= \begin{cases}
    -A_{i,0}, & n=0, \\
    -(n+1)A_{i,n}-(n+2)C_{i,n+2} -2(n-1)\frac{h_i}{R_i}B_{i,n-1}-2(2n+1)\frac{h_i}{R_i}D_{i,n+1} +4(n-1)\frac{h_i^2}{R_i^2}C_{i,n}, &  n\geq 1,
    \end{cases}\\
    B_{i,n}^{*} &= \begin{cases}
    0,  & n=0,\\
    +2B_{i,1} +3D_{i,3} +2\frac{h_i}{R_i}A_{i,0} -6\frac{h_i}{R_i}C_{i,2}, & n = 1,\\
    +(n+1)B_{i,n} +(n+2)D_{i,n+2} -2(n-1)\frac{h_i}{R_i}A_{i,n-1} -2(2n+1)\frac{h_i}{R_i}C_{i,n+1} -4(n-1)\frac{h_i^2}{R_i^2}D_{i,n}, & n\geq 2,
    \end{cases}\\
    C_{i,n}^{*} &= \begin{cases}
    +C_{i,2} - A_{i,0}, & n = 2,\\
    +(n-1)C_{i,n}+(n-2)A_{i,n-2}+2(n-2) \frac{h_i}{R_i} D_{i,n-1}, {\hskip 60mm} & n\geq 3,
    \end{cases}\\
    D_{i,n}^{*} &=\begin{cases} -(n-1)D_{i,n}-(n-2)B_{i,n-2}+2(n-2) \frac{h_i}{R_i}C_{i,n-1},  {\hskip 61mm} & n\geq 2, \end{cases}
\end{split}
\end{align}
where we used  results for the image multipoles derived in Section~\ref{sec:images}.
The total Airy stress function outside all inclusions can then be written as 
\begin{equation}
\label{eq:AiryOut}
\chi^{\text{tot}}_{\text{out}}\big(x,y|\mbf{a}_{\text{out}}\big)=\chi_{\text{ext}}(x,y)+\sum_{i=1}^N \chi_{\text{out}}\big(r_i(x,y),\varphi_i(x,y)|\mbf{a}_{i,\text{out}}\big)+ \sum_{i=1}^N \mathcal{I}\left[\chi_{\text{out}}\big(r_i(x,y),\varphi_i(x,y)|\mbf{a}_{i,\text{out}}\big)\right],
\end{equation}
where the first term is due to external stress and the two summations describe contributions due to induced multipoles at the centers of inclusions and due to their images. The set of amplitudes of induced multipoles for all inclusions is defined as $\mbf{a}_{\text{out}}=\{\mbf{a}_{1,\text{out}}, \ldots, \mbf{a}_{N,\text{out}}\}$.

Similarly, we expand the induced Airy stress function inside  the $i^\text{th}$ inclusion as~\cite{sarkar2019elastic}
\begin{equation}
  \label{eq:AiryInducedIn}
  \begin{split}
  \chi_{\text{in}}\big(r_i,\varphi_i|\mbf{a}_{i,\text{in}}\big)=& \phantom{+c_{i,0} r_i^2+} \ \  \sum_{n=2}^{\infty} R_i^2 \left[a_{i,n} \left(\frac{r_i}{R_i}\right)^n\cos(n\varphi_i)+ b_{i,n}
  \left(\frac{r_i}{R_i}\right)^n\sin(n\varphi_i)\right]\\
  &+c_{i,0} r_i^2 +\sum_{n=1}^{\infty} R_i^2 \left[c_{i,n} \left(\frac{r_i}{R_i}\right)^{n+2}\cos
  (n\varphi_i) + d_{i,n}
  \left(\frac{r_i}{R_i}\right)^{n+2}\sin (n\varphi_i) \right],
  \end{split}
\end{equation}
where we kept only terms that generate finite stresses at the center of inclusion and we omitted constant and linear terms $\{1,r_i \cos\varphi_i, r_i \sin \varphi_i\}$ that correspond to zero stresses. The set of amplitudes of induced multipoles is represented as $\mbf{a}_{i,\text{in}}=\{a_{i,2}, a_{i,3},\dots,b_{i,2}, b_{i,3},\dots, c_{i,0}, c_{i,1},\dots, d_{i,1}, d_{i,2},\dots\}$. The total Airy stress function inside the $i^\text{th}$ inclusion is thus
\begin{equation}
    \chi^{\text{tot}}_{\text{in}}\big(x,y|\mbf{a}_{i,\text{in}}\big)=\chi_{\text{ext}}(x,y)+\chi_{\text{in}}\big(r_i(x,y),\varphi_i(x,y)|\mbf{a}_{i,\text{in}}\big),
    \label{eq:AiryIn}    
\end{equation}
where the first term is due to external stress and the second term is due to induced multipoles.

The amplitudes of induced multipoles $\mbf{a}_{i,\text{out}}$ and $\mbf{a}_{i,\text{in}}$ are obtained by satisfying  boundary conditions that tractions and displacements are continuous across the circumference of each inclusion~\cite{sarkar2019elastic}
\begin{subequations}
  \begin{align}
    \sigma_{\text{in},rr}^{\text{tot}}\big(r_i=R_i,\varphi_i|\mbf{a}_{i,\text{in}}\big)&=\sigma_{\text{out},rr}^{\text{tot}}\big(r_i=R_i,\varphi_i|\mbf{a}_{\text{out}}\big), \label{eq:BC:sigmaRR}\\
    \sigma_{\text{in},r\varphi}^{\text{tot}}\big(r_i=R_i,\varphi_i|\mbf{a}_{i,\text{in}}\big)&=\sigma_{\text{out},r\varphi}^{\text{tot}}\big(r_i=R_i,\varphi_i|\mbf{a}_{\text{out}}\big), \label{eq:BC:sigmaRF}\\
    u_{\text{in},r}^{\text{tot}}\big(r_i=R_i,\varphi_i|\mbf{a}_{i,\text{in}}\big)&=u_{\text{out},r}^{\text{tot}}\big(r_i=R_i,\varphi_i|\mbf{a}_{\text{out}}\big), \label{eq:BC:uR}\\
    u_{\text{in},\varphi}^{\text{tot}}\big(r_i=R_i,\varphi_i|\mbf{a}_{i,\text{in}}\big)&=u_{\text{out},\varphi}^{\text{tot}}\big(r_i=R_i,\varphi_i|\mbf{a}_{\text{out}}\big),\label{eq:BC:uF}
   \end{align}
   \label{eq:BC}%
\end{subequations}
where stresses and displacements are obtained from the total Airy stress functions $\chi^{\text{tot}}_{\text{in}}\big(x,y|\mbf{a}_{i,\text{in}}\big)$ inside the $i^\text{th}$ inclusion (see Eq.~(\ref{eq:AiryIn})) and $\chi^{\text{tot}}_{\text{out}}\big(x,y|\mbf{a}_{\text{out}}\big)$ outside all inclusions (see Eq.~(\ref{eq:AiryOut})). In the boundary conditions for  the $i^\text{th}$ inclusion in the above Eq.~(\ref{eq:BC}), we can easily take into account contributions due to the induced multipoles $\mbf{a}_{i,\text{in}}$ and $\mbf{a}_{i,\text{out}}$ in this inclusion and due to external stresses $\sigma_{xx}^\textrm{ext}$, $\sigma_{yy}^\textrm{ext}$, and $\sigma_{xy}^\textrm{ext}$ after rewriting the corresponding Airy stress function $\chi_\textrm{ext}(x,y)$ in Eq.~(\ref{eq:ChiExt}) in polar coordinates centered at the  $i^\text{th}$ inclusion as
\begin{equation}
\chi_\textrm{ext}(r_i,\varphi_i)=\frac{1}{4}(\sigma_{xx}^\textrm{ext}+\sigma_{yy}^\textrm{ext})r_i^2  - \frac{1}{4}(\sigma_{xx}^\textrm{ext}-\sigma_{yy}^\textrm{ext}) r_i^2 \cos(2 \varphi_i) - \frac{1}{2}\sigma_{xy}^\textrm{ext} r_i^2 \sin(2 \varphi_i).
\label{eq:ChiExtI}
\end{equation}
This can be done with the help of Table~\ref{table:MichellStressDisplacement}, which shows the values of stresses and displacements corresponding to each term in the Michell solution~\cite{Barber}.
\begingroup
\setlength{\tabcolsep}{2pt}
\begin{table}[!b]
{\centering
\caption{Stresses $\sigma_{ij}$ and displacements $u_i$ corresponding to different terms for the Airy stress function $\chi$ in the Michell solution~\cite{Barber}. The value of Kolosov's constant for plane stress is $\kappa=(3-\nu)/(1+\nu)$ and for plane strain is  $\kappa=3-4\nu$. Here, $\mu$ is the shear modulus and $\nu$ is the Poisson's ratio. \label{table:MichellStressDisplacement}}
\footnotesize
\def\arraystretch{1.1}
\begin{tabular}{|c|c|c|c|c|}
\hline
$\chi$ & $\sigma_{rr}$ & $\sigma_{r\varphi}$ &  $\sigma_{\varphi\varphi}$ & $2\mu \begin{pmatrix}u_{r}\\u_{\varphi}\end{pmatrix}$ \\
\hline
$r^2$ & $2$ & $0$ & $2$ & $r \, \begin{pmatrix}\kappa-1\\
                                0\end{pmatrix}$ \\
$\ln r$ & $r^{-2}$ & $0$ & -$r^{-2}$ & $r^{-1}\, \begin{pmatrix}-1\\
                                0\end{pmatrix}$\\
$r^{n+2}\cos (n\varphi)$ & $-(n+1)(n-2)r^n\cos (n\varphi)$ & $n(n+1)r^n\sin (n\varphi)$ & $(n+1)(n+2)r^n\cos (n\varphi)$ & $r^{n+1}\, \begin{pmatrix}(\kappa-n-1)\cos (n\varphi) \\
                            (\kappa+n+1)\sin (n\varphi) \end{pmatrix}$\\
$r^{n+2}\sin (n\varphi)$ & $-(n+1)(n-2)r^n\sin (n\varphi)$ & $-n(n+1)r^n\cos (n\varphi)$ & $(n+1)(n+2)r^n\sin (n\varphi)$ & $r^{n+1}\, \begin{pmatrix}(\kappa-n-1)\sin (n\varphi) \\
                    -(\kappa+n+1)\cos (n\varphi)\end{pmatrix}$\\
$r^{-n+2}\cos (n\varphi)$ &$-(n+2)(n-1)r^{-n}\cos (n\varphi)$& $-n(n-1)r^{-n}\sin (n\varphi)$ & $(n-1)(n-2)r^{-n}\cos (n\varphi)$ &$r^{-n+1}\,\begin{pmatrix}(\kappa+n-1)\cos (n\varphi) \\
                        -(\kappa-n+1)\sin (n\varphi) \end{pmatrix}$\\
$r^{-n+2}\sin (n\varphi)$ & $-(n+2)(n-1)r^{-n}\sin (n\varphi)$ & $n(n-1)r^{-n}\cos (n\varphi)$ & $(n-1)(n-2)r^{-n}\sin (n\varphi)$ & $r^{-n+1}\, \begin{pmatrix}(\kappa+n-1)\sin (n\varphi) \\
                    (\kappa-n+1)\cos (n\varphi)\end{pmatrix}$\\
$r^{n}\cos (n\varphi)$ & $-n(n-1)r^{n-2}\cos (n\varphi)$ & $n(n-1)r^{n-2}\sin (n\varphi)$ & $n(n-1)r^{n-2}\cos (n\varphi)$ & $r^{n-1}\, \begin{pmatrix}-n \cos (n\varphi) \\
                        n \sin (n\varphi)\end{pmatrix}$\\
$r^{n}\sin (n\varphi)$ & $-n(n-1)r^{n-2}\sin (n\varphi)$ & $-n(n-1)r^{n-2}\cos (n\varphi)$ & $n(n-1)r^{n-2}\sin (n\varphi)$ &  $r^{n-1} \begin{pmatrix}-n \sin (n\varphi)\\
                        -n \cos (n\varphi)\end{pmatrix}$\\
$r^{-n}\cos (n\varphi)$ & $-n(n+1)r^{-n-2}\cos (n\varphi)$ & $-n(n+1)r^{-n-2}\sin (n\varphi)$ & $n(n+1)r^{-n-2}\cos (n\varphi)$ & $r^{-n-1}\, \begin{pmatrix}n \cos (n\varphi) \\
                        n \sin (n\varphi)\end{pmatrix}$\\
$r^{-n}\sin (n\varphi)$ & $-n(n+1)r^{-n-2}\sin (n\varphi)$ & $n(n+1)r^{-n-2}\cos (n\varphi)$ & $n(n+1)r^{-n-2}\sin (n\varphi)$ & $r^{-n-1}\,\begin{pmatrix}n \sin (n\varphi)\\-n \cos (n\varphi)\end{pmatrix}$\\
\hline
\end{tabular}
}
\end{table}
\endgroup
In order to calculate the contributions from  induced multipoles in other inclusions as well as all the contributions from  image multipoles from the $i^\text{th}$ and other inclusions, we need to expand  expressions for the Airy stress functions $\chi_{\text{out}}\big(r_j,\varphi_j|\mbf{a}_{j,\text{out}}\big)$ and $\mathcal{I}[\chi_{\text{out}}\big(r_j,\varphi_j|\mbf{a}_{j,\text{out}}\big)]$ around the center $\mbf{x}_i$ of  the $i^\text{th}$ inclusion. Polar coordinates $(r_j,\varphi_j)$ centered at the $j^\text{th}$ inclusion can be expressed in terms of polar coordinates $(r_i,\varphi_i)$ centered at the $i^\text{th}$ inclusion as
\begin{equation}
\begin{split}
r_j(r_i,\varphi_i)&=\sqrt{r_i^2+a_{ij}^2-2 r_i a_{ij} \cos(\varphi_i - \theta_{ij})},\\
\varphi_j(r_i,\varphi_i)& =\pi +\theta_{ij} - \arctan\left[\frac{r_i \sin(\varphi_i  - \theta_{ij})}{\big(a_{ij}-r_i \cos(\varphi_i  - \theta_{ij})\big)}\right],
\end{split}
\end{equation}
where $a_{ij}=\sqrt{(x_i-x_j)^2+(y_i-y_j)^2}$ is the distance between the centers of  the $i^{\text{th}}$ and $j^{\text{th}}$ inclusions and $\theta_{ij} = \arctan [(y_j-y_i)/(x_j-x_i)]$ is the angle between the line joining the centers of inclusions and the
$x$-axis, as shown in Fig.~\ref{Fig:schematic2}. Similarly,  polar coordinates $(r_j^*,\varphi_j^*)$ centered at the image of the $j^\text{th}$ inclusion can be expressed in terms of polar coordinates $(r_i,\varphi_i)$ centered at the $i^\text{th}$ inclusion as
\begin{equation}
\begin{split}
r_j^*(r_i,\varphi_i)&=\sqrt{r_i^2+{a_{ij}^*}^2-2 r_i a_{ij}^* \cos(\varphi_i - \theta_{ij}^*)},\\
\varphi_j^*(r_i,\varphi_i)& =\pi +\theta_{ij}^* - \arctan\left[\frac{r_i \sin(\varphi_i  - \theta_{ij}^*)}{\big(a_{ij}^*-r_i \cos(\varphi_i  - \theta_{ij}^*)\big)}\right],
\end{split}
\end{equation}
where $a^*_{ij}=\sqrt{(x_i-x_j^*)^2+(y_i-y_j^*)^2}$ is the distance between the centers of the $i^{\text{th}}$ inclusion and the image of  the $j^{\text{th}}$ inclusion and $\theta^*_{ij}=\arctan [(y_j^*-y_i)/(x_j^*-x_i)]$ is the angle between the line joining these centers and the
$x$-axis as shown in Fig.~\ref{Fig:schematic2}.
The Airy stress functions $\chi_{\text{out}}\big(r_j,\varphi_j|\mbf{a}_{j,\text{out}}\big)$ and $\mathcal{I}[\chi_{\text{out}}\big(r_j,\varphi_j|\mbf{a}_{j,\text{out}}\big)]$ due to induced multipoles centered at the $j^{\text{th}}$ inclusion and due to their image multipoles, respectively, can be expanded in Taylor series around the center of the $i^{\text{th}}$ inclusion as~\cite{Green,sarkar2019elastic}
\begin{subequations}
\begin{align}
\begin{split}
\label{eq:ChiJOutI}
\chi_\text{out}\big(r_j(r_i, \varphi_i),\varphi_j(r_i, \varphi_i)|\mbf{a}_{j,\textrm{out}}\big)=&\sum_{n=2}^\infty R_j^2 \frac{r_i^n}{a_{ij}^n} \Big[\cos(n \varphi_i) f^n_c\big(R_j/a_{ij},\theta_{ij}|\mbf{a}_{j,\textrm{out}}\big)+\sin(n \varphi_i) f^n_s\big(R_j/a_{ij},\theta_{ij}|\mbf{a}_{j,\textrm{out}}\big)\Big]\\
&\hspace{-1cm}
+\sum_{n=0}^\infty R_j^2\frac{r_i^{n+2}}{a_{ij}^{n+2}} \Big[\cos(n \varphi_i) g^n_c\big(R_j/a_{ij},\theta_{ij}|\mbf{a}_{j,\textrm{out}}\big)+\sin(n \varphi_i) g^n_s\big(R_j/a_{ij},\theta_{ij}|\mbf{a}_{j,\textrm{out}}\big)\Big],
\end{split}\\
\begin{split}
\label{eq:ImageChiJOutI}
\mathcal{I}\left[\chi_\text{out}\big(r_j(r_i, \varphi_i),\varphi_j(r_i, \varphi_i)|\mbf{a}_{j,\textrm{out}}\big)\right]=&\sum_{n=2}^\infty R_j^2 \frac{r_i^n}{a_{ij}^{*n}} \Big[\cos(n \varphi_i) f^n_c\big(R_j/a^*_{ij},\theta^*_{ij}|\mbf{a}_{j,\textrm{out}}^{*}\big)+\sin(n \varphi_i) f^n_s\big(R_j/a_{ij}^*,\theta_{ij}^*|\mbf{a}_{j,\textrm{out}}^{*}\big)\Big]\\
&\hspace{-1cm}
+\sum_{n=0}^\infty R_j^2\frac{r_i^{n+2}}{a_{ij}^{*n+2}} \Big[\cos(n \varphi_i) g^n_c\big(R_j/a_{ij}^*,\theta_{ij}^*|\mbf{a}_{j,\textrm{out}}^{*}\big)+\sin(n \varphi_i) g^n_s\big(R_j/a_{ij}^*,\theta_{ij}^*|\mbf{a}_{j,\textrm{out}}^{*}\big)\Big],
\end{split}
\end{align}
\label{eq:AiryOutTaylorExpansion}%
\end{subequations}
\noindent where we omitted constant and linear terms $\{1, r_i \cos\varphi_i, r_i \sin \varphi_i\}$ that correspond to zero stresses and we introduced  functions
\begin{subequations}
    \begin{align}
f^n_c\left(R_j/a_{ij},\theta_{ij}|\mbf{a}_{j,\textrm{out}}\right)&=\sum_{m=0}^\infty  \left(\frac{R_j^m}{a_{ij}^m} \Big[A_{j,m} \mathcal{A}_n^m(\theta_{ij})+B_{j,m} \mathcal{B}_n^m(\theta_{ij})\Big]+\frac{R_j^{m-2}}{a_{ij}^{m-2}}\Big[C_{j,m} \mathcal{C}_n^m(\theta_{ij})+D_{j,m} \mathcal{D}_n^m(\theta_{ij})\Big]\right),\\
f^n_s\left(R_j/a_{ij},\theta_{ij}|\mbf{a}_{j,\textrm{out}}\right)&=\sum_{m=0}^\infty  \left(\frac{R_j^m}{a_{ij}^m} \Big[A_{j,m} \mathcal{B}_n^m(\theta_{ij})-B_{j,m} \mathcal{A}_n^m(\theta_{ij})\Big]+\frac{R_j^{m-2}}{a_{ij}^{m-2}}\Big[C_{j,m} \mathcal{D}_n^m(\theta_{ij})-D_{j,m} \mathcal{C}_n^m(\theta_{ij})\Big]\right),\\
g^n_c\left(R_j/a_{ij},\theta_{ij}|\mbf{a}_{j,\textrm{out}}\right)&=\sum_{m=2}^\infty\frac{R_j^{m-2}}{a_{ij}^{m-2}}\Big[C_{j,m} \mathcal{E}_n^m(\theta_{ij})+D_{j,m} \mathcal{F}_n^m(\theta_{ij})\Big],\\
g^n_s\left(R_j/a_{ij},\theta_{ij}|\mbf{a}_{j,\textrm{out}}\right)&=\sum_{m=2}^\infty\frac{R_j^{m-2}}{a_{ij}^{m-2}}\Big[C_{j,m} \mathcal{F}_n^m(\theta_{ij})-D_{j,m} \mathcal{E}_n^m(\theta_{ij})\Big].
    \end{align}
   \label{eq:TaylorExpansion}%
\end{subequations}
 In the above Eq.~(\ref{eq:TaylorExpansion}), we set $C_{j,0}=D_{j,0}=C_{j,1}=D_{j,1}=0$ and  introduced coefficients $\mathcal{A}_n^m(\theta_{ij})$, $\mathcal{B}_n^m(\theta_{ij})$, $\mathcal{C}_n^m(\theta_{ij})$, $\mathcal{D}_n^m(\theta_{ij})$, $\mathcal{E}_n^m(\theta_{ij})$, and $\mathcal{F}_n^m(\theta_{ij})$ that are summarized in Table~\ref{tab:CoefficientsFG}.
\begin{table}[!t]
{
\centering
\caption{Coefficients for the expansion of the Airy stress functions $\chi_\text{out}\big(r_j(r_i, \varphi_i),\varphi_j(r_i, \varphi_i)|\mbf{a}_{j,\textrm{out}}\big)$ and $\mathcal{I}\left[\chi_\text{out}\big(r_j(r_i, \varphi_i),\varphi_j(r_i, \varphi_i)|\mbf{a}_{j,\textrm{out}}\big)\right]$ in Eqs.~(\ref{eq:AiryOutTaylorExpansion}-\ref{eq:TaylorExpansion}).}
\label{tab:CoefficientsFG}
\def\arraystretch{2}
\begin{tabular}{|l|l|l|}
\hline
$n \ge 2$&  $\mathcal{A}_n^0(\theta_{ij})=-\frac{1}{n} \cos(n\theta_{ij})$ & $\mathcal{B}_n^0(\theta_{ij})=-\frac{1}{n} \sin(n\theta_{ij})$\\ 
\hline
$n \ge 2, m \ge 1$ & $\mathcal{A}_n^m(\theta_{ij})=(-1)^m {{m+n-1}\choose{n}} \cos\big((m+n)\theta_{ij}\big)$ &  $\mathcal{B}_n^m(\theta_{ij})=(-1)^m {{m+n-1}\choose{n}} \sin\big((m+n)\theta_{ij}\big)$ \\
\hline
$n \ge 0, m \ge 2$ & $\mathcal{C}_n^m(\theta_{ij})=(-1)^m {{m+n-2}\choose{n}} \cos\big((m+n)\theta_{ij}\big)$ & $\mathcal{D}_n^m(\theta_{ij})=(-1)^m {{m+n-2}\choose{n}} \sin\big((m+n)\theta_{ij}\big)$\\
\hline
$n \ge 0, m \ge 2$ &$\mathcal{E}_n^m(\theta_{ij})=(-1)^{m-1} {{m+n-1}\choose{n+1}} \cos\big((m+n)\theta_{ij}\big)$ & $\mathcal{F}_n^m(\theta_{ij})=(-1)^{m-1} {{m+n-1}\choose{n+1}} \sin\big((m+n)\theta_{ij}\big)$\\
\hline
\end{tabular}
}
\end{table}

Next, we calculate stresses and displacements at the circumference of the $i^\text{th}$ inclusion by using expressions for the Airy stress functions \externalStress{$\chi_\textrm{ext}(r_i,\varphi_i)$} in Eq.~(\ref{eq:ChiExtI}) due to external stresses, \inclusionI{$\chi_{\text{in}}\big(r_i,\varphi_i|\mbf{a}_{i,\text{in}}\big)$} and \inclusionI{$\chi_{\text{out}}\big(r_i,\varphi_i|\mbf{a}_{i,\text{out}}\big)$} in Eqs.~(\ref{eq:AiryInducedIn}) and Eq.~(\ref{eq:AiryInducedOut}) due to induced multipoles for the $i^\text{th}$ inclusion, \inclusionJ{$\chi_{\text{out}}\big(r_j,\varphi_j|\mbf{a}_{j,\text{out}}\big)$} in Eq.~(\ref{eq:ChiJOutI}) due to induced multipoles  for the $j^\text{th}$ inclusion ($j\ne i$), and \imageInclusion{$\mathcal{I}[\chi_{\text{out}}\big(r_j,\varphi_j|\mbf{a}_{j,\text{out}}\big)]$} in Eq.~(\ref{eq:ImageChiJOutI}) due to image multipoles  for the $j^\text{th}$ inclusion ($j=1,\ldots, N$). With the help of Table~\ref{table:MichellStressDisplacement}, which shows how to convert each term of the Airy stress function to stresses and displacements, we obtain
\begin{subequations}
\small
\begin{align}
\sigma_{\text{in},rr}^{\text{tot}}\big(r_i=R_i,\varphi_i|\mbf{a}_{i,\text{in}}\big)&=\externalStress{\frac{1}{2}(\sigma_{xx}^\textrm{ext} + \sigma_{yy}^\textrm{ext}) + \frac{1}{2}(\sigma_{xx}^\textrm{ext} - \sigma_{yy}^\textrm{ext}) \cos(2 \varphi_i) + \sigma_{xy}^\textrm{ext} \sin (2 \varphi_i)}+ \inclusionI{2 c_{i,0}} \nonumber\\
      &\hspace{-2.0cm}\inclusionI{-\sum_{n=1}^\infty \Big[n (n-1) \big(a_{i,n} \cos (n \varphi_i) + b_{i,n} \sin (n \varphi_i) \big)+(n+1)(n-2)\big(c_{i,n} \cos (n \varphi_i) + d_{i,n} \sin (n \varphi_i) \big) \Big]},\\
\sigma_{\text{out},rr}^{\text{tot}}\big(r_i=R_i,\varphi_i|\mbf{a}_{\text{out}}\big)&=  \externalStress{\frac{1}{2}(\sigma_{xx}^\textrm{ext} + \sigma_{yy}^\textrm{ext}) + \frac{1}{2}(\sigma_{xx}^\textrm{ext} - \sigma_{yy}^\textrm{ext}) \cos(2 \varphi_i) + \sigma_{xy}^\textrm{ext} \sin (2 \varphi_i)} + \inclusionI{A_{i,0}}\nonumber \\
    &\hspace{-2.0cm}\inclusionI{-\sum_{n=1}^\infty \Big[n (n+1) \big(A_{i,n} \cos (n \varphi_i) + B_{i,n} \sin (n \varphi_i) \big)+(n+2)(n-1)\big(C_{i,n} \cos (n \varphi_i) + D_{i,n} \sin (n \varphi_i) \big) \Big]}\nonumber\\
    &\hspace{-2.0cm}\inclusionJ{-\sum_{j\ne i}\sum_{n=2}^\infty \frac{R_j^2 R_i^{n-2} }{a_{ij}^n} n (n-1)  \Big[\cos(n \varphi_i) f^n_c\big(R_j/a_{ij},\theta_{ij}|\mbf{a}_{j,\textrm{out}}\big)+\sin(n \varphi_i) f^n_s\big(R_j/a_{ij},\theta_{ij}|\mbf{a}_{j,\textrm{out}}\big)\Big]}\nonumber \\
    &\hspace{-2.0cm}\inclusionJ{-\sum_{j\ne i}\sum_{n=0}^\infty \frac{R_j^2 R_i^{n}}{a_{ij}^{n+2}}  (n+1) (n-2) \Big[\cos(n \varphi_i) g^n_c\big(R_j/a_{ij},\theta_{ij}|\mbf{a}_{j,\textrm{out}}\big)+\sin(n \varphi_i) g^n_s\big(R_j/a_{ij},\theta_{ij}|\mbf{a}_{j,\textrm{out}}\big)\Big]}\nonumber \\
    &\hspace{-2.0cm}\imageInclusion{-\sum_{j=1}^N\sum_{n=2}^\infty \frac{R_j^2 R_i^{n-2} }{a_{ij}^{*n}} n (n-1)  \Big[\cos(n \varphi_i) f^n_c\big(R_j/a_{ij}^*,\theta_{ij}^*|\mbf{a}_{j,\textrm{out}}^{*}\big)+\sin(n \varphi_i) f^n_s\big(R_j/a_{ij}^*,\theta_{ij}^*|\mbf{a}_{j,\textrm{out}}^{*}\big)\Big]}\nonumber \\
    &\hspace{-2.0cm}\imageInclusion{-\sum_{j=1}^N\sum_{n=0}^\infty \frac{R_j^2 R_i^{n}}{a_{ij}^{*n+2}}  (n+1) (n-2) \Big[\cos(n \varphi_i) g^n_c\big(R_j/a_{ij}^*,\theta_{ij^*}|\mbf{a}_{j,\textrm{out}}^{*}\big)+\sin(n \varphi_i) g^n_s\big(R_j/a_{ij}^*,\theta_{ij}^*|\mbf{a}_{j,\textrm{out}}^{*}\big)\Big]},\\
\sigma_{\text{in},r\varphi}^{\text{tot}}\big(r_i=R_i,\varphi_i|\mbf{a}_{i,\text{in}}\big)&= \externalStress{-\frac{1}{2}(\sigma_{xx}^\textrm{ext} - \sigma_{yy}^\textrm{ext}) \sin(2 \varphi_i) + \sigma_{xy}^\textrm{ext} \cos (2 \varphi_i)} \nonumber\\
      &\hspace{-2.0cm}\inclusionI{+\sum_{n=1}^\infty \Big[n (n-1) \big(a_{i,n} \sin (n \varphi_i) - b_{i,n} \cos (n \varphi_i) \big)+n(n+1)\big(c_{i,n} \sin (n \varphi_i) - d_{i,n} \cos (n \varphi_i) \big) \Big]},\\
\sigma_{\text{out},r\varphi}^{\text{tot}}\big(r_i=R_i,\varphi_i|\mbf{a}_{\text{out}}\big)&= \externalStress{ -\frac{1}{2}(\sigma_{xx}^\textrm{ext} - \sigma_{yy}^\textrm{ext}) \sin(2 \varphi_i) + \sigma_{xy}^\textrm{ext} \cos (2 \varphi_i)}\nonumber \\
    &\hspace{-2.0cm}\inclusionI{-\sum_{n=1}^\infty \Big[n (n+1) \big(A_{i,n} \sin (n \varphi_i) - B_{i,n} \cos (n \varphi_i) \big)+n(n-1)\big(C_{i,n} \sin (n \varphi_i) - D_{i,n} \cos (n \varphi_i) \big) \Big]}\nonumber\\
    &\hspace{-2.0cm}\inclusionJ{+\sum_{j\ne i}\sum_{n=2}^\infty \frac{R_j^2 R_i^{n-2} }{a_{ij}^n} n (n-1)  \Big[\sin(n \varphi_i) f^n_c\big(R_j/a_{ij},\theta_{ij}|\mbf{a}_{j,\textrm{out}}\big)-\cos(n \varphi_i) f^n_s\big(R_j/a_{ij},\theta_{ij}|\mbf{a}_{j,\textrm{out}}\big)\Big]}\nonumber \\
    &\hspace{-2.0cm}\inclusionJ{+\sum_{j\ne i}\sum_{n=0}^\infty \frac{R_j^2 R_i^{n}}{a_{ij}^{n+2}}  n(n+1) \Big[\sin(n \varphi_i) g^n_c\big(R_j/a_{ij},\theta_{ij}|\mbf{a}_{j,\textrm{out}}\big)-\cos(n \varphi_i) g^n_s\big(R_j/a_{ij},\theta_{ij}|\mbf{a}_{j,\textrm{out}}\big)\Big]}\nonumber \\
    &\hspace{-2cm}\imageInclusion{+\sum_{j=1}^N\sum_{n=2}^\infty \frac{R_j^2 R_i^{n-2} }{a_{ij}^{*n}} n (n-1)  \Big[\sin(n \varphi_i) f^n_c\big(R_j/a_{ij}^*,\theta_{ij}^*|\mbf{a}_{j,\textrm{out}}^{*}\big)-\cos(n \varphi_i) f^n_s\big(R_j/a_{ij}^*,\theta_{ij}^*|\mbf{a}_{j,\textrm{out}}^{*}\big)\Big]}\nonumber \\
    &\hspace{-2cm}\imageInclusion{+\sum_{j=1}^N\sum_{n=0}^\infty \frac{R_j^2 R_i^{n}}{a_{ij}^{*n+2}}  n(n+1) \Big[\sin(n \varphi_i) g^n_c\big(R_j/a_{ij}^*,\theta_{ij}^*|\mbf{a}_{j,\textrm{out}}^{*}\big)-\cos(n \varphi_i) g^n_s\big(R_j/a_{ij}^*,\theta_{ij}^*|\mbf{a}_{j,\textrm{out}}^{*}\big)\Big]}, \\
\frac{2\mu_i}{R_i}\,  u_{\text{in},r}^{\text{tot}}\big(r_i=R_i,\varphi_i|\mbf{a}_{i,\text{in}}\big)&= \externalStress{-\frac{1}{4}(\sigma_{xx}^\textrm{ext} + \sigma_{yy}^\textrm{ext})(1-\kappa_i) + \frac{1}{2}(\sigma_{xx}^\textrm{ext} - \sigma_{yy}^\textrm{ext}) \cos(2 \varphi_i) + \sigma_{xy}^\textrm{ext}\sin (2 \varphi_i)} +\inclusionI{c_{i,0} (\kappa_i-1)} \nonumber\\
      &\hspace{-2.0cm}\inclusionI{-\sum_{n=1}^\infty \Big[n \big(a_{i,n} \cos (n \varphi_i) + b_{i,n} \sin (n \varphi_i) \big)+(n+1-\kappa_i)\big(c_{i,n} \cos (n \varphi_i) + d_{i,n} \sin (n \varphi_i) \big) \Big]},\\
\frac{2\mu_0}{R_i}\, u_{\text{out},r}^{\text{tot}}\big(r_i=R_i,\varphi_i|\mbf{a}_{\text{out}}\big)&= \externalStress{ -\frac{1}{4}(\sigma_{xx}^\textrm{ext} + \sigma_{yy}^\textrm{ext})(1-\kappa_0) + \frac{1}{2}(\sigma_{xx}^\textrm{ext} - \sigma_{yy}^\textrm{ext}) \cos(2 \varphi_i) + \sigma_{xy}^\textrm{ext} \sin (2 \varphi_i)} \inclusionI{- A_{i,0}}\nonumber \\
    &\hspace{-2.0cm}\inclusionI{+\sum_{n=1}^\infty \Big[n \big(A_{i,n} \cos (n \varphi_i) + B_{i,n} \sin (n \varphi_i) \big)+(n-1+\kappa_0)\big(C_{i,n} \cos (n \varphi_i) + D_{i,n} \sin (n \varphi_i) \big) \Big]}\nonumber\\
    &\hspace{-2.0cm}\inclusionJ{-\sum_{j\ne i}\sum_{n=2}^\infty \frac{R_j^2 R_i^{n-2} }{a_{ij}^n} n  \Big[\cos(n \varphi_i) f^n_c\big(R_j/a_{ij},\theta_{ij}|\mbf{a}_{j,\textrm{out}}\big)+\sin(n \varphi_i) f^n_s\big(R_j/a_{ij},\theta_{ij}|\mbf{a}_{j,\textrm{out}}\big)\Big]}\nonumber \\
    &\hspace{-2.0cm}\inclusionJ{-\sum_{j\ne i}\sum_{n=0}^\infty \frac{R_j^2 R_i^{n}}{a_{ij}^{n+2}}  (n+1-\kappa_0) \Big[\cos(n \varphi_i) g^n_c\big(R_j/a_{ij},\theta_{ij}|\mbf{a}_{j,\textrm{out}}\big)+\sin(n \varphi_i) g^n_s\big(R_j/a_{ij},\theta_{ij}|\mbf{a}_{j,\textrm{out}}\big)\Big]}\nonumber \\
    &\hspace{-2cm}\imageInclusion{-\sum_{j=1}^N\sum_{n=2}^\infty \frac{R_j^2 R_i^{n-2} }{a_{ij}^{*n}} n  \Big[\cos(n \varphi_i) f^n_c\big(R_j/a_{ij}^*,\theta_{ij}^*|\mbf{a}_{j,\textrm{out}}^{*}\big)+\sin(n \varphi_i) f^n_s\big(R_j/a_{ij}^*,\theta_{ij}^*|\mbf{a}_{j,\textrm{out}}^{*}\big)\Big]}\nonumber \\
    &\hspace{-2cm}\imageInclusion{-\sum_{j=1}^N\sum_{n=0}^\infty \frac{R_j^2 R_i^{n}}{a_{ij}^{*n+2}}  (n+1-\kappa_0) \Big[\cos(n \varphi_i) g^n_c\big(R_j/a_{ij}^*,\theta_{ij}^*|\mbf{a}_{j,\textrm{out}}^{*}\big)+\sin(n \varphi_i) g^n_s\big(R_j/a_{ij}^*,\theta_{ij}^*|\mbf{a}_{j,\textrm{out}}^{*}\big)\Big]},\\
\frac{2\mu_i}{R_i}\, u_{\text{in},\varphi}^{\text{tot}}\big(r_i=R_i,\varphi_i|\mbf{a}_{i,\text{in}}\big)&= \externalStress{-\frac{1}{2}(\sigma_{xx}^\textrm{ext} - \sigma_{yy}^\textrm{ext}) \sin(2 \varphi_i) + \sigma_{xy}^\textrm{ext} \cos (2 \varphi_i)} \nonumber\\
      &\hspace{-2.0cm}\inclusionI{+\sum_{n=1}^\infty \Big[n  \big(a_{i,n} \sin (n \varphi_i) - b_{i,n} \cos (n \varphi_i) \big)+(n+1+\kappa_i)\big(c_{i,n} \sin (n \varphi_i) - d_{i,n} \cos (n \varphi_i) \big) \Big]},\\
\frac{2\mu_0}{R_i}\,u_{\text{out},\varphi}^{\text{tot}}\big(r_i=R_i,\varphi_i|\mbf{a}_{\text{out}}\big)&= \externalStress{ -\frac{1}{2}(\sigma_{xx}^\textrm{ext} - \sigma_{yy}^\textrm{ext}) \sin(2 \varphi_i) + \sigma_{xy}^\textrm{ext} \cos (2 \varphi_i)}\nonumber \\
    &\hspace{-2.0cm}\inclusionI{+\sum_{n=1}^\infty \Big[n  \big(A_{i,n} \sin (n \varphi_i) - B_{i,n} \cos (n \varphi_i) \big)+(n-1-\kappa_0)\big(C_{i,n} \sin (n \varphi_i) - D_{i,n} \cos (n \varphi_i) \big) \Big]}\nonumber\\
    &\hspace{-2.0cm}\inclusionJ{+\sum_{j\ne i}\sum_{n=2}^\infty \frac{R_j^2 R_i^{n-2} }{a_{ij}^n} n \Big[\sin(n \varphi_i) f^n_c\big(R_j/a_{ij},\theta_{ij}|\mbf{a}_{j,\textrm{out}}\big)-\cos(n \varphi_i) f^n_s\big(R_j/a_{ij},\theta_{ij}|\mbf{a}_{j,\textrm{out}}\big)\Big]}\nonumber \\
    &\hspace{-2.0cm}\inclusionJ{+\sum_{j\ne i}\sum_{n=0}^\infty \frac{R_j^2 R_i^{n}}{a_{ij}^{n+2}}  (n+1+\kappa_0) \Big[\sin(n \varphi_i) g^n_c\big(R_j/a_{ij},\theta_{ij}|\mbf{a}_{j,\textrm{out}}\big)-\cos(n \varphi_i) g^n_s\big(R_j/a_{ij},\theta_{ij}|\mbf{a}_{j,\textrm{out}}\big)\Big]}\nonumber \\
    &\hspace{-2cm}\imageInclusion{+\sum_{j=1}^N\sum_{n=2}^\infty \frac{R_j^2 R_i^{n-2} }{a_{ij}^{*n}} n \Big[\sin(n \varphi_i) f^n_c\big(R_j/a_{ij}^*,\theta_{ij}^*|\mbf{a}_{j,\textrm{out}}^{*}\big)-\cos(n \varphi_i) f^n_s\big(R_j/a_{ij}^*,\theta_{ij}^*|\mbf{a}_{j,\textrm{out}}^{*}\big)\Big]}\nonumber \\
    &\hspace{-2cm}\imageInclusion{+\sum_{j=1}^N\sum_{n=0}^\infty \frac{R_j^2 R_i^{n}}{a_{ij}^{*n+2}}  (n+1+\kappa_0) \Big[\sin(n \varphi_i) g^n_c\big(R_j/a_{ij}^*,\theta_{ij}^*|\mbf{a}_{j,\textrm{out}}^{*}\big)-\cos(n \varphi_i) g^n_s\big(R_j/a_{ij}^*,\theta_{ij}^*|\mbf{a}_{j,\textrm{out}}^{*}\big)\Big]}.
\end{align}
\label{eq:BoundaryTractionsDisplacements}%
\end{subequations}
Colors in the above equations correspond to the Airy stress functions \externalStress{$\chi_\textrm{ext}(r_i,\varphi_i)$}, \inclusionI{$\chi_{\text{in}}\big(r_i,\varphi_i|\mbf{a}_{i,\text{in}}\big)$}, \inclusionI{$\chi_{\text{out}}\big(r_i,\varphi_i|\mbf{a}_{i,\text{out}}\big)$},  \inclusionJ{$\chi_{\text{out}}\big(r_j,\varphi_j|\mbf{a}_{j,\text{out}}\big)$}, and \imageInclusion{$\mathcal{I}\left[\chi_{\text{out}}\big(r_j,\varphi_j|\mbf{a}_{j,\text{out}}\big)\right]$}. e introduced the shear modulus $\mu_i=E_i/[2(1+\nu_i)]$ and  the Kolosov's constant $\kappa_i$ for the $i^\text{th}$ inclusion, where the value of Kolosov's constant is $\kappa_i=(3-\nu_i)/(1+\nu_i)$ for plane stress and $\kappa_i=3-4\nu_i$ for plane strain conditions~\cite{Barber}. Similarly, we define the shear modulus $\mu_0=E_0/[2(1+\nu_0)]$ and the Kolosov's constant $\kappa_0$ for the elastic matrix.

\begin{figure}[!t]
  \centering
  \includegraphics[scale=1]{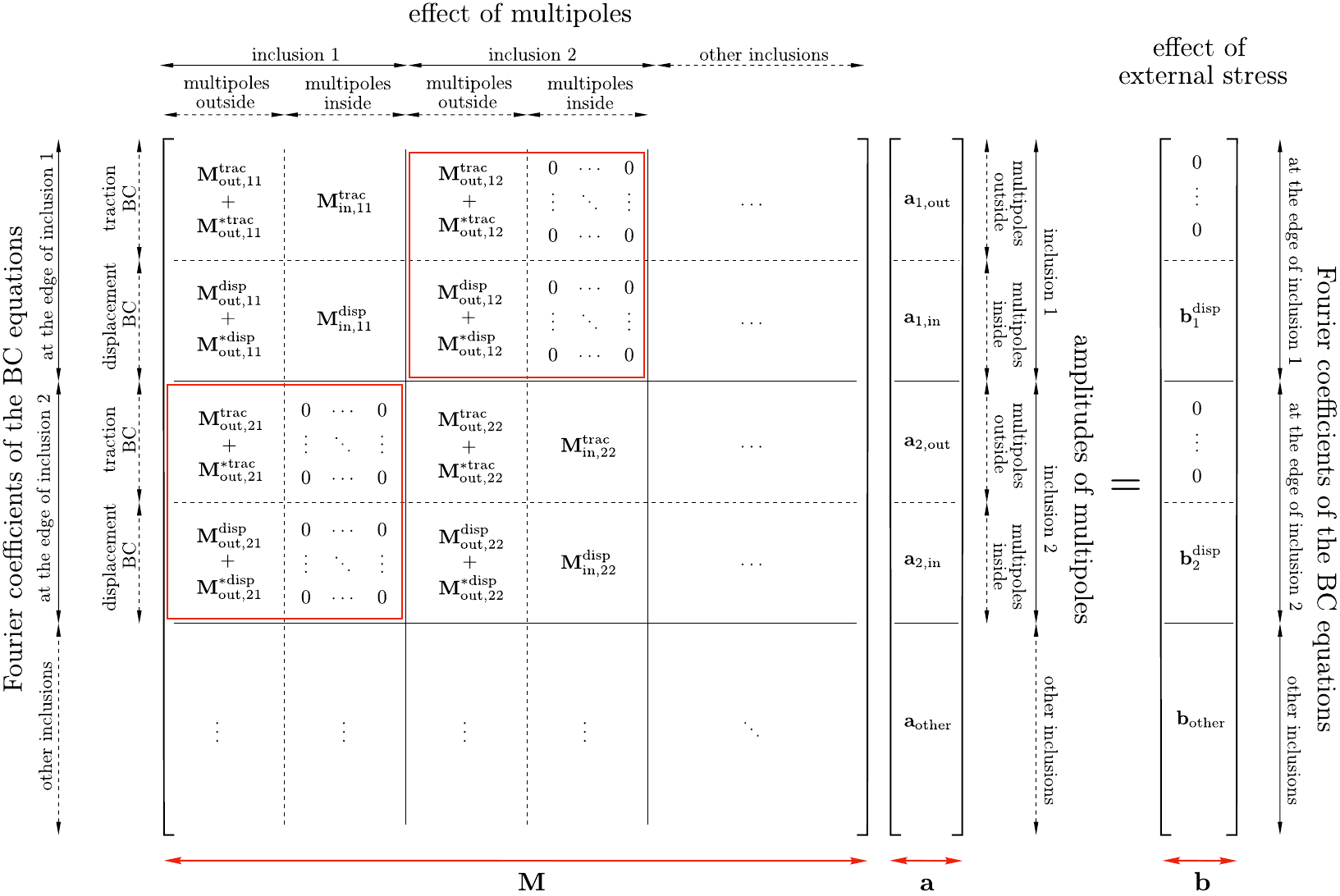}
  \caption{Structure of the system of Eqs.~(\ref{eq:matrix}) for the amplitudes of induced multipoles $\mbf{a}_{i,\textrm{out}}$, $ \mbf{a}_{i,\textrm{in}}$ for inclusions $i\in\{1,\ldots, N\}$. The matrix $\mathbf{M}$ is  divided into $4 N^2$ blocks, where blocks $\mbf{M}_{\text{in},ij}^\text{trac}$ and $\mbf{M}_{\text{out},ij}^\text{trac} + \mbf{M}_{\text{out},ij}^{*\text{trac}}$ correspond to the boundary conditions for tractions around the circumference of  the $i^\text{th}$ inclusion in Eq.~(\ref{eq:BC:sigmaRR}, \ref{eq:BC:sigmaRF}), and blocks $\mbf{M}_{\text{in},ij}^\text{disp}$ and $\mbf{M}_{\text{out},ij}^\text{disp}+\mbf{M}_{\text{out},ij}^{*\text{disp}}$ correspond to the boundary conditions for displacements around the circumference of  the $i^\text{th}$ inclusion in Eq.~(\ref{eq:BC:uR}, \ref{eq:BC:uF}). The red boxes mark  blocks with $i \ne j$ that account for the  interactions between different inclusions. The effect of external stresses is contained in vectors $\mbf{b}_i^\text{disp}$. See text for the detailed description of elements represented in this system of equations.}
\label{Fig:matrix2}
\end{figure}

The boundary conditions in Eq.~(\ref{eq:BC}) have to be satisfied at every point ($\varphi_i$) on the circumference of the $i^\text{th}$ inclusion. Thus the coefficients of Fourier modes $\{1, \cos(n \varphi_i), \sin (n \varphi_i)\}$ have to match in the expansions of tractions and displacements in Eq.~(\ref{eq:BoundaryTractionsDisplacements}). This enables us to construct a matrix equation for the set of induced multipoles $\{\mbf{a}_{i,\textrm{out}}, \mbf{a}_{i,\textrm{in}}\}$ in the form (see also Fig.~\ref{Fig:matrix2})
 \begin{equation}
     \begin{pmatrix}
     \mbf{M}_{\text{out},ij}^\text{trac} + \mbf{M}_{\text{out},ij}^{*\text{trac}}, & \mbf{M}_{\text{in},ij}^\text{trac} \\
    \mbf{M}_{\text{out},ij}^\text{disp} + \mbf{M}_{\text{out},ij}^{*\text{disp}},  & \mbf{M}_{\text{in},ij}^\text{disp} \\
     \end{pmatrix}
     \begin{pmatrix}
     \mbf{a}_{j,\textrm{out}} \\
     \mbf{a}_{j,\textrm{in}} \\
     \end{pmatrix}
     =
     \begin{pmatrix}
     \mbf{0} \\
     \mbf{b}_i^\text{disp}\\
     \end{pmatrix},
     \label{eq:matrix}
 \end{equation}
where the summation over inclusions $j$ is implied. The top and bottom rows of matrix $\mbf{M}$ in the above equation are obtained from the boundary conditions in Eq.~(\ref{eq:BC}) for tractions (superscript `trac') and displacements (superscript `disp'), respectively. The left and right columns of matrix $\mbf{M}$ describe the effect of  induced multipoles $\mbf{a}_{j,\textrm{out}}$ and $\mbf{a}_{j,\textrm{in}}$, respectively. Matrices $\mbf{M}_{ij}^*$ describe interactions between the $i^\text{th}$ inclusion and the image of  the $j^\text{th}$ inclusion;  these are new compared to the companion paper~\cite{sarkar2019elastic}. Note that each inclusion interacts with its  image as well as with images of other inclusions. The entries in matrices $\mbf{M}_{\text{out},ii}^\text{trac}$, $\mbf{M}_{\text{out},ii}^{*\text{trac}}$, and $\mbf{M}_{\text{in},ii}^\text{trac}$ for the $i^\text{th}$ inclusion are numbers that depend on the degrees of induced multipoles. The entries in matrices $\mbf{M}_{\text{out},ii}^\text{disp}$, $\mbf{M}_{\text{out},ii}^{*\text{disp}}$, and $\mbf{M}_{\text{in},ii}^\text{disp}$ for the $i^\text{th}$ inclusion depend on the degrees  of induced multipoles, the radius of inclusion $R_i$ and the material properties of the inclusion~($\mu_i,\kappa_i$) and elastic matrix~($\mu_0,\kappa_0$). Additionally, the entries in matrices $\mbf{M}_{\text{out},ii}^{*\text{trac}}$ and $\mbf{M}_{\text{out},ii}^{*\text{disp}}$ also depend on the distance $h_i$ from the center of  the $i^\text{th}$ inclusion to the boundary. Matrices $\mbf{M}_{\text{out},ij}^\text{trac}$ and $\mbf{M}_{\text{out},ij}^\text{disp}$ encode the interactions between inclusions $i$ and $j$. The entries in these matrices depend on the degrees of induced multipoles, the radii $R_i$ and $R_j$ of inclusions, the angle $\theta_{ij}$, and the separation distance $a_{ij}$ between inclusions (see Fig.~\ref{Fig:schematic2}). In addition to that, the entries in matrix $\mbf{M}_{\text{out},ij}^\text{disp}$ also depend on the material properties of the elastic matrix ($\mu_0,\kappa_0$). 
Matrices $\mbf{M}_{\text{out},ij}^{*\text{trac}}$ and $\mbf{M}_{\text{out},ij}^{*\text{disp}}$ describe the interaction between the $i^\text{th}$ inclusion and the image of the $j^\text{th}$ inclusion.  These matrices depend on the degrees of image multipoles, the radii $R_i$ and $R_j$ of inclusions, the distance $h_j$ from the center of  the $j^{th}$ inclusion to the boundary, the angle $\theta_{ij}^\text{*}$, and the separation distance $a_{ij}^*$ between the $i^\text{th}$ inclusion and the image of the $j^\text{th}$ inclusion (see Fig.~\ref{Fig:schematic2}). Matrices $\mbf{M}_{\text{out},ij}^{*\text{disp}}$ additionally depend on the material properties of the elastic matrix~($\mu_0,\kappa_0$).
Note that the other matrices are zero, i.e. $\mbf{M}_{\text{in},ij}^\text{trac}=\mbf{M}_{\text{in},ij}^\text{disp}=0$. The entries in vector $\mbf{b}_i^\text{disp}$ depend on the magnitude of external stresses ($\sigma_{xx}^\text{ext}$, $\sigma_{yy}^\text{ext}$, $\sigma_{xy}^\text{ext}$), the degrees of induced multipoles, the radius of inclusion $R_i$, and the material properties of the inclusion ($\mu_i,\kappa_i$) and elastic matrix ($\mu_0,\kappa_0$). Note that in $\mbf{b}_i^\text{disp}$ the only nonzero entries are the ones that correspond to Fourier modes $1$, $\cos(2 \varphi_i)$, and  $\sin(2 \varphi_i)$.

To numerically solve the system of equations for induced multipoles in Eq.~(\ref{eq:matrix}) we truncate the multipole expansion at degree $n_\text{max}$. For each inclusion $i$, there are $4 n_\text{max}-1$ unknown amplitudes of multipoles  $\mbf{a}_{i,\text{out}}=\{A_{i,0}, A_{i,1},\dots, A_{i,n_\text{max}}, B_{i,1}, B_{i,2},\dots, B_{i,n_\text{max}}, C_{i,2}, C_{i,3},\dots, C_{i,n_\text{max}}, D_{i,2}, D_{i,3},\dots, D_{i,n_\text{max}} \}$ and  $4 n_\text{max}-1$ unknown amplitudes of multipoles $\mbf{a}_{i,\text{in}}=\{a_{i,2}, a_{i,3},\dots,a_{i,n_\text{max}},b_{i,2}, b_{i,3},\dots,b_{i,n_\text{max}}, c_{i,0}, c_{i,1},\dots,c_{i,n_\text{max}}, d_{i,1}, d_{i,2},\dots, d_{i,n_\text{max}}\}$. Furthermore, we truncate the series for the Airy stress function of image multipoles $\mathcal{I}\left[\chi_{\text{out}}\big(r_i,\varphi_i|\mbf{a}_{i,\text{out}}\big)\right]$ in Eq.~(\ref{eq:AiryInducedOutImage2}) as well as the Taylor expansions for the Airy stress functions $\chi_{\text{out}}\big(r_j(r_i,\varphi_i),\varphi_j(r_i,\varphi_i)|\mbf{a}_{j,\text{out}}\big)$ and $\mathcal{I}\left[\chi_{\text{out}}\big(r_j(r_i,\varphi_i),\varphi_j(r_i,\varphi_i)|\mbf{a}_{j,\text{out}}\big)\right]$ in Eq.~(\ref{eq:AiryOutTaylorExpansion}) at the same order $n_\text{max}$. By matching the coefficients of Fourier modes $\{1, \cos\varphi_i, \sin \varphi_i, \ldots, \cos(n_\text{max} \varphi_i),\sin(n_\text{max} \varphi_i)\}$ in the expansions for tractions and displacements in Eq.~(\ref{eq:BoundaryTractionsDisplacements}) around the circumference of  the $i^\text{th}$ inclusion, we in principle get $2(2n_\text{max}+1)$ equations from tractions and $2(2n_\text{max}+1)$ equations from displacements. However, the zero Fourier modes for $\sigma_{r\varphi}$ and $u_\varphi$ are equal to zero. Furthermore, the coefficients of Fourier modes $\cos \varphi_i$ and $\sin \varphi_i$ are identical for each of the $\sigma_{rr}$, $\sigma_{r\varphi}$, $u_r$, and $u_\varphi$ in Eq.~(\ref{eq:BoundaryTractionsDisplacements}). By removing the equations, which do not provide any new information, the dimensions of matrices $\mathbf{M}_{\text{out},ij}^{\text{trac}}$, $\mathbf{M}_{\text{out},ij}^{*\text{trac}}$, $\mathbf{M}_{\text{in},ij}^{\text{trac}}$, $\mathbf{M}_{\text{out},ij}^{\text{disp}}$, $\mathbf{M}_{\text{out},ij}^{*\text{disp}}$, and $\mathbf{M}_{\text{in},ij}^{\text{disp}}$, become $(4 n_\text{max}-1)\times(4 n_\text{max}-1)$. Thus Eq.~(\ref{eq:matrix}) describes the system of $N(8 n_\text{max} -2)$ equations for the amplitudes of induced multipoles $\{\mbf{a}_{1,\text{out}}, \mbf{a}_{1,\text{in}},\ldots,\mbf{a}_{N,\text{out}}, \mbf{a}_{N,\text{in}}\}$. The solution of this system of equations gives amplitudes of induced multipoles, which are linear functions of applied loads $\sigma_{xx}^\text{ext}$, $\sigma_{yy}^\text{ext}$, and $\sigma_{xy}^\text{ext}$.
These amplitudes are then used to obtain the Airy stress functions $\chi^{\text{tot}}_{\text{in}}\big(x,y|\mbf{a}_{i,\text{in}}\big)$ in Eq.~(\ref{eq:AiryIn}) inside inclusions and $\chi^{\text{tot}}_{\text{out}}\big(x,y|\mbf{a}_{\text{out}}\big)$ in Eq.~(\ref{eq:AiryOut}) outside all inclusions, which enables us to calculate stresses and displacements everywhere in the structure. The accuracy of obtained results depends on the number $n_\text{max}$ for the maximum degree of induced multipoles, where larger number $n_\text{max}$ yields more accurate results. In the next  sections, we compare  results obtained with the method described above  with  linear finite element simulations and experiments.

\par
\begin{figure}[!t]
  \centering
  \includegraphics[scale=1]{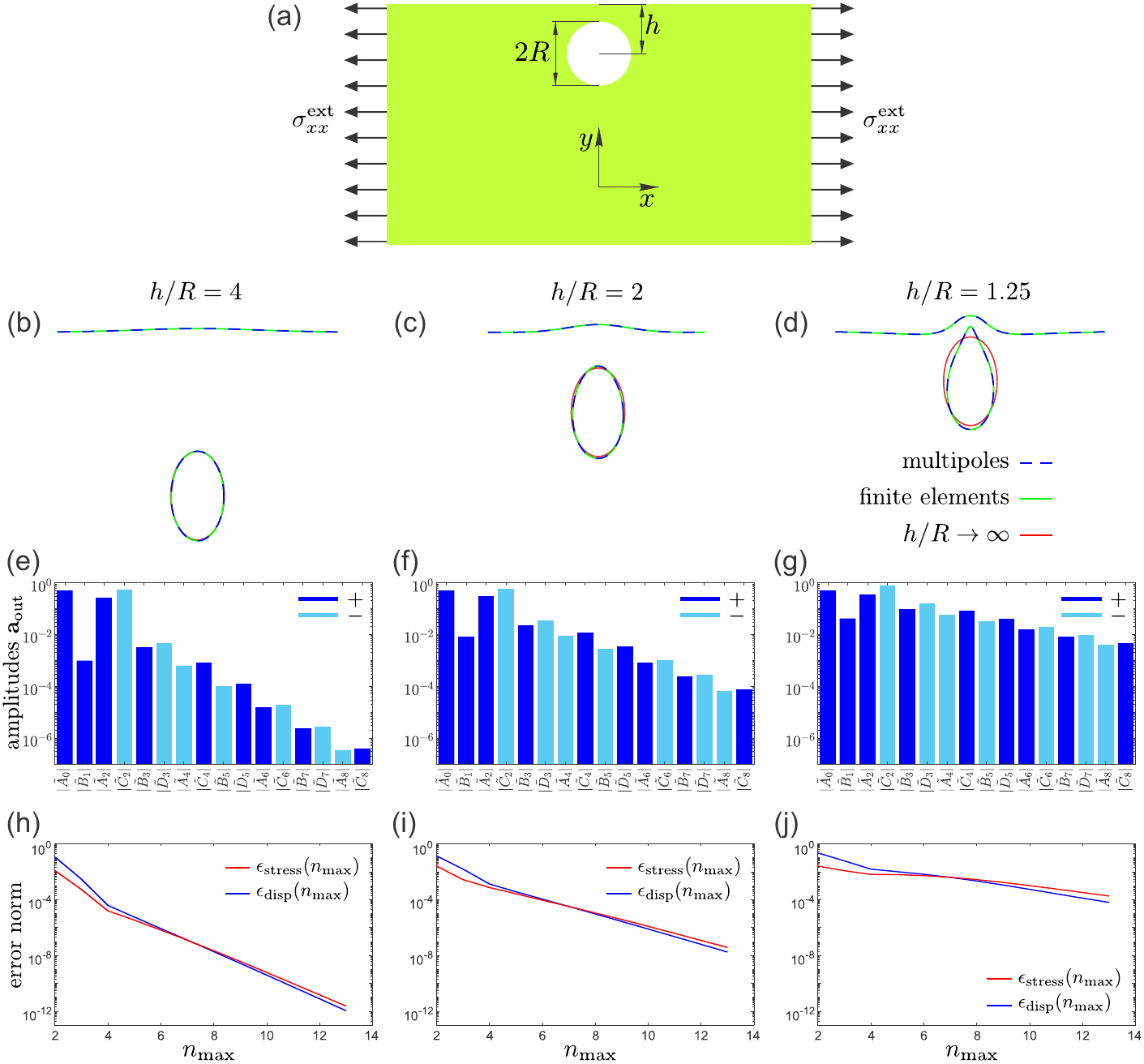}
  \caption{Uniaxial deformation of a semi-infinite elastic plate with a hole near a traction free edge. (a)~Schematic image describing the initial undeformed shape of the structure and the direction of applied load $\sigma_{xx}^\text{ext}=-0.11 E_0$, where $E_0$ is the Young's modulus of the elastic matrix, it's Poisson's ratio is $\nu_0=0.3$, and plane stress condition was used. The radius of the circular hole is $R$, which is at the distance $h$ from the traction-free edge.  
(b)-(d)~Contours of deformed holes and deformed boundaries for $h/R=4$, $2$ and $1.25$. Dashed blue lines show the contours obtained with elastic multipole method for $n_\text{max}=10$ and green solid lines show the contours obtained from finite element simulations. As a reference, we include red solid lines that correspond to shapes of holes that are far away from edges ($h/R \rightarrow \infty$). (e-g)~Absolute values of the amplitudes of induced multipoles $\mbf{a}_\text{out}$ at the center of the hole for $n_\text{max}=10$. Amplitudes are normalized, such that $\tilde{A}_n=A_n/\sigma_{xx}^\text{ext}$, $\tilde{B}_n=B_n/\sigma_{xx}^\text{ext}$, $\tilde{C}_n=C_n/\sigma_{xx}^\text{ext}$, $\tilde{D}_n=D_n/\sigma_{xx}^\text{ext}$. The dark and light blue colored bars correspond to the positive ($A_n, B_n, C_n, D_n>0$) and negative ($A_n, B_n, C_n, D_n<0$) amplitudes, respectively. Note that the amplitudes $B_{2m}=D_{2m}=A_{2m+1}=C_{2m+1}=0$ due to the symmetry of the problem. (h-j)~The normalized errors for displacements $\epsilon_\text{disp}(n_\text{max})$ (blue lines) and stresses $\epsilon_\text{stress}(n_\text{max})$ (red lines) obtained from Eq.~(\ref{eq:error}). The left column corresponds to results for $h/R=4$, the middle column to $h/R=2$ and the right column to $h/R=1.25$.
}
\label{Fig:hole_flat_edge}
\end{figure}

\subsubsection{Comparison with linear finite element simulations and experiments}

First, we tested the elastic multipole method for one circular hole of radius $R$ embedded in a semi-infinite plate $(y<0)$ subjected to uniaxial stress $\sigma_{xx}^{\text{ext}}$ (see Fig.~\ref{Fig:hole_flat_edge}). This way we investigated how a single hole interacts with its image near a traction-free edge ($\sigma^{\text{ext}}_{xy}(x,0)=\sigma^{\text{ext}}_{yy}(x,0)=0$). Three different values of the separation distance $h$ between the center of the hole and the traction-free edge were considered: $h=4R$, $h=2R$, and $h=1.25R$. The value of applied uniaxial stress was
$\sigma_{xx}^{\text{ext}} = -0.11 E_0$, where $E_0$ is the Young's modulus of  the elastic matrix, and we used the plane stress condition with the value of Kolosov's constant $\kappa_0=(3-\nu_0)/(1+\nu_0)$ and the Poisson's ratio $\nu_0=0.3$. Such large values of external loads were used only to exaggerate deformations. Note that in practical experiments these loads would cause nonlinear deformation.

In Fig.~\ref{Fig:hole_flat_edge} we show the contours of deformed holes and the amplitudes of induced multipoles $\mbf{a}_\text{out}$ for different values of the separation distance $h$ between the center of the hole and the traction-free edge, where we set $n_\text{max}=10$. Note that the amplitudes $\mbf{a}_\text{in}=\mbf{0}$ because the Young's modulus is zero for the hole. Results from the elastic multipole method are compared  with linear finite element simulations on a rectangular domain of size $800R \times 400R$ (see Appendix~\ref{app:FEM} for details). When the hole is far from the traction-free edge ($h \gg R$) it interacts very weakly with its image, which can be seen from the expansion of stresses and displacements in Eq.~(\ref{eq:BoundaryTractionsDisplacements}), where terms describing  interactions between the $i^\text{th}$ inclusions and its image contain powers of $R_i/a^*_{ii}\ll 1$, where $a^*_{ii}=2h$. This is the case for the separation distance $h=4R$, where we find that the contour of the  hole has elliptical shape (see Fig.~\ref{Fig:hole_flat_edge}b), which is characteristic for deformed holes embedded in an infinite elastic matrix~\cite{Eshelby,sarkar2019elastic}. When the hole is moved closer to the traction-free edge ($h=2R$, $h=1.25R$) it interacts more strongly with its image. As a consequence, the contour of the deformed hole becomes progressively more non-elliptical, approaching a teardrop shape, and the traction-free edge bulges out near the hole (see Fig.~\ref{Fig:hole_flat_edge}c,d). This is also reflected in the amplitudes $\mbf{a}_\text{out}$ of induced multipoles. They decrease exponentially with the degree of multipoles and  decrease more slowly when the hole is closer to the traction-free edge (Fig.~\ref{Fig:hole_flat_edge}e-g). 

To determine the proper number for the maximum degree $n_\text{max}$ of induced multipoles, we performed a convergence analysis for the spatial distributions of displacements $\mbf{u}^{(n_\text{max})}(x,y)$ and von Mises stress $\sigma^{(n_\text{max})}_\text{vM}(x,y)$, where $\sigma_\text{vM}=(\sigma_{xx}^2-\sigma_{xx}\sigma_{yy}+\sigma_{yy}^2+3\sigma_{xy}^2)^{1/2}$. Displacements and von Mises stresses were evaluated at $N_p$ discrete points $(x_i,y_j)=\left(iR/50,jR/50\right)$, where $i\in\{-500,-499,\ldots,500\}$ and $j\in\{-1000, -999, \ldots, 0\}$ and grid points that lie inside the hole were excluded. The normalized errors for displacements $\epsilon_\text{disp}(n_\text{max})$ and stresses $\epsilon_\text{stress}(n_\text{max})$ were obtained by calculating the relative changes of  spatial distributions of displacements and von Mises stresses when the maximum degree $n_\text{max}$ of induced multipoles is increased by one. The normalized errors are given by~\cite{SpecMethod}
\begin{subequations}
\begin{align}
\label{Eqn:errordisp}
    \epsilon_\text{disp}(n_\text{max}) &= \frac{1}{\sqrt{N_p}}\left[\sum_{i,j}\left|\frac{\mbf{u}^{(n_\text{max}+1)}(x_i,y_j)-\mbf{u}^{(n_\text{max})}(x_i,y_j)}{d\,\sigma_\text{vM}^\text{ext}/E_0}\right|^2 \right]^{1/2},\\
\label{Eqn:errorstress}
    \epsilon_\text{stress}(n_\text{max}) &= \frac{1}{\sqrt{N_p}}\left[\sum_{i,j}\left(\frac{\sigma^{(n_\text{max}+1)}_\text{vM}(x_i,y_j)-\sigma^{(n_\text{max})}_\text{vM}(x_i,y_j)}{\sigma_\text{vM}^\text{ext}}\right)^2 \right]^{1/2}.
\end{align}
\label{eq:error}%
\end{subequations}
Here, displacements and von Mises stresses are normalized by the characteristic scales $d \sigma_\text{vM}^\text{ext}/E_0$ and $\sigma_\text{vM}^\text{ext}$, respectively, where $d=2R$ is the diameter of the hole, $\sigma_\text{vM}^\text{ext}=|\sigma_{xx}^{\text{ext}}|$ is the value of  von Mises stress due to applied load, and $E_0$ is the Young's modulus of  surrounding elastic matrix. The normalized errors are plotted in Fig.~\ref{Fig:hole_flat_edge}h-j. As the maximum degree $n_\text{max}$ of induced multipoles is increased, the normalized errors for displacements $\epsilon_\text{disp}(n_\text{max})$ and stresses $\epsilon_\text{stress}(n_\text{max})$ decrease exponentially, because the
induced elastic multipoles form the basis for the biharmonic equation. This is akin to the spectral method, which is exponentially convergent when the functions and the shapes of boundaries are smooth~\cite{SpecMethod}.
Note that the normalized errors decrease more slowly when the hole is moved closer to the traction-free edge and the interaction of hole with its image becomes important (see Fig.~\ref{Fig:hole_flat_edge}h-j), which is reflected in larger magnitudes of amplitudes of higher-order multipoles  (Fig.~\ref{Fig:hole_flat_edge}e-g).

\begin{figure}[!t]
\centering
  \includegraphics[scale=1]{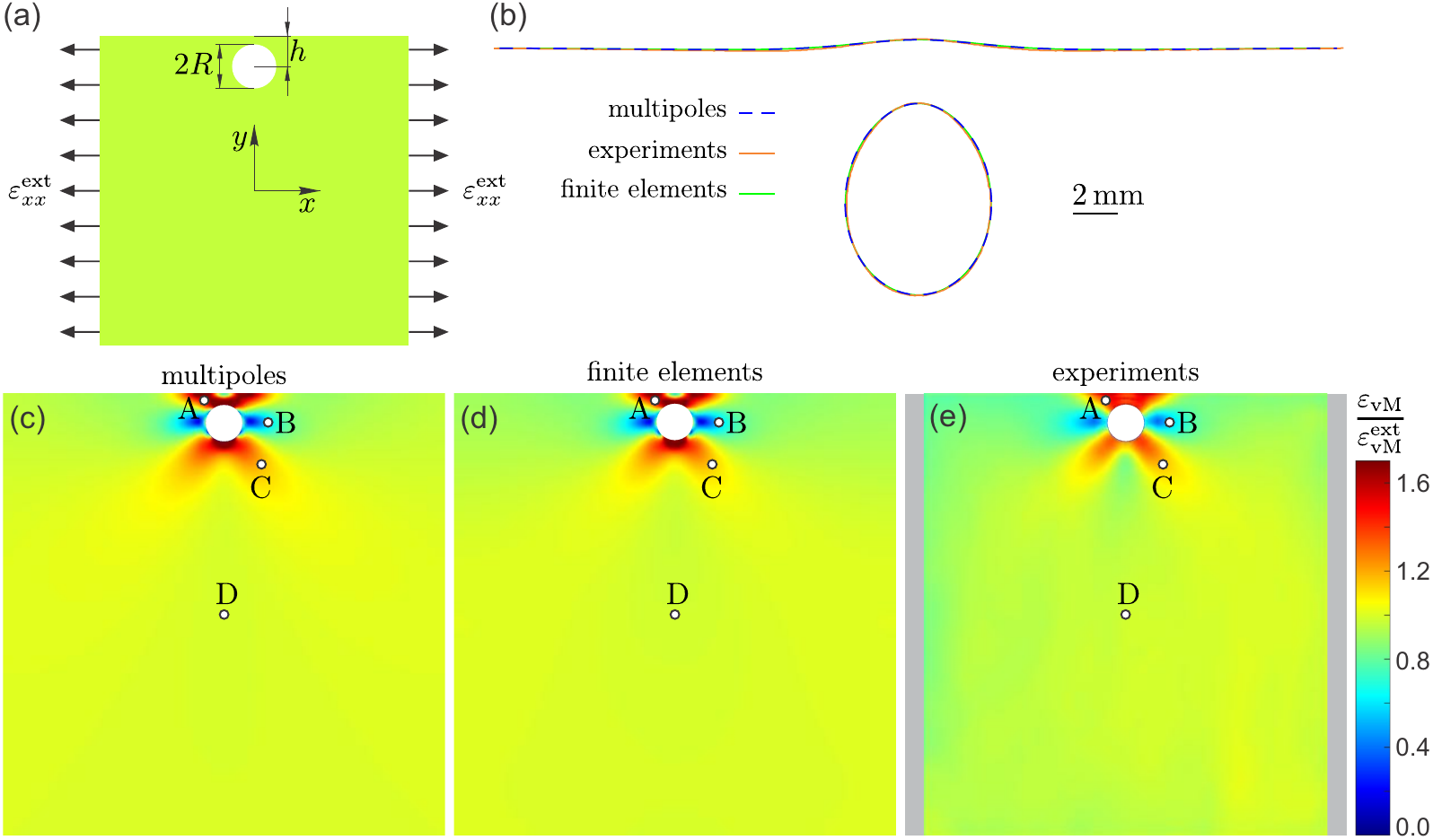}
  \caption{Uniaxial horizontal compression of an elastic structure with a hole near a traction-free edge. (a)~Schematic image describing the initial undeformed shape of the structure and applied strain $\epsilon_{xx}^\text{ext}=-0.05$.  The size of the experimental sample was $100 \text{ mm}\times 100\text{ mm}\times 25\text{ mm}$ with the hole of diameter $2R=8.11 \text{ mm}$, and the distance between the center of the hole and the traction-free edge was $h=6.25 \text{ mm}$.  The Young's modulus of the elastic matrix was $E_0=0.97\,\textrm{MPa}$ and  the Poisson's ratio was $\nu_0=0.49$. For linear finite element simulations, we used a 2D structure of size $100 \text{ mm}\times 100\text{ mm}$ under plane stress condition. The elastic multiple method describes the deformation of a semi-infinite 2D elastic matrix, where we used the load   $\sigma_{xx}^\text{ext}=\epsilon_{xx}^\text{ext} E_0$ and  plane stress condition. 
  (b)~Contours of deformed holes and traction-free edges obtained with elastic multipole method (dashed blue lines, $n_\text{max}=10$), experiments (solid orange lines) and finite element simulations (solid green lines). 
  (c-e)~Equivalent von Mises strain fields $\epsilon_\text{vM}$ were obtained with (c)~elastic multipole method ($n_\text{max}=10$), (d)~finite element simulations, and (e)~DIC (digital image correlation) analysis of experiments. Strain fields were normalized with the value of the equivalent von Mises strain $\epsilon_\text{vM}^\text{ext}=|\epsilon_{xx}^\text{ext}|$ imposed by the applied load. Note that the strain data was corrupted near the edges for this experimental sample due to oil stains on the speckle patterns near the boundary. For this reason, we omitted the affected border regions (grey) in the heat map~(e). Four marked points A-D were chosen for the quantitative comparison of strains $\epsilon_\text{vM}$. See Table~\ref{tab:free_edge} for details.
}
  \label{Fig:ExperimentalFreeEdge}
\end{figure}
\begin{table}[!b]
\centering
\caption{Quantitative comparison for the values of equivalent von Mises strains $\varepsilon_\text{vM}$ normalized with the value for the applied external load $\varepsilon_\text{vM}^\text{ext}=|\varepsilon_{xx}^{\text{ext}}|$ at points A-D (defined in Fig.~\ref{Fig:ExperimentalFreeEdge}) in compressed samples with one hole near a traction-free edge obtained with elastic multipole method (EMP), finite element simulations (FEM) and  DIC analysis of experiments (EXP). The relative percent errors between the EMP and FEM were calculated as $100\times(\varepsilon_\text{vM}^\text{(EMP)}-\varepsilon_\text{vM}^\text{(FEM)})/\varepsilon_\text{vM}^\text{(FEM)}$. The relative percent errors between the EMP and EXP were calculated as $100\times(\varepsilon_\text{vM}^\text{(EMP)}-\varepsilon_\text{vM}^\text{(EXP)})/\varepsilon_\text{vM}^\text{(EXP)}$.}
\label{tab:free_edge}
\def\arraystretch{1.1}
\begin{tabular}{|>{\centering}m{0.05\textwidth}|>{\centering}m{0.05\textwidth}>{\centering}m{0.05\textwidth}>{\centering}m{0.05\textwidth}|>{\centering}m{0.065\textwidth}>{\centering\arraybackslash}m{0.065\textwidth}|}
\hline
\multirow{2}{*}{points} & \multicolumn{3}{c|}{strain $\varepsilon_\text{vM}/\epsilon_\text{vM}^\text{ext}$} & \multicolumn{2}{c|}{error of EMP (\%)}\\
\cline{2-6}
 & EMP   & FEM   & DIC   & FEM & DIC \\
\hline
A 
& 1.5390 & 1.4868 & 1.3998 & 3.5 & 9.9 \\
B 
& 0.5782 & 0.5586 & 0.5506 & 3.5 & 5.0 \\
C 
& 1.1760 & 1.1406 & 1.1586 & 3.1 & 1.5 \\
D 
& 0.9932 & 0.9836 & 0.9792 & 1.0 & 1.4 \\
\hline            
\end{tabular}
\end{table}

Additionally, we compared the results of the elastic multipole method with the experiment (see Appendix~\ref{app:experiments} for details). The size of the elastomer sample in the experiment was $100 \text{ mm}\times 100\text{ mm}\times 25\text{ mm}$ with the hole of diameter $2R=8.11 \text{ mm}$, and the distance between the center of the hole and the traction-free edge was $h=6.25 \text{ mm}$ (see Fig.~\ref{Fig:ExperimentalFreeEdge}a). The contour of the deformed hole in the compressed experimental sample under external strain $\varepsilon_{xx}^{\text{ext}}=-0.05$ ($\sigma_{xy}^{\text{ext}}\approx0$ due to applied silicone oil to reduce friction) was compared with those obtained with elastic multipole method and linear finite element simulation (see Fig.~\ref{Fig:ExperimentalFreeEdge}b). For the elastic multipole method, we used external stress $\sigma_{yy}^{\text{ext}}=E_0 \varepsilon_{yy}^{\text{ext}}$ and  plane stress condition was assumed since the experimental sample was free to expand in the out-of-plane direction. Linear finite element simulation was performed for a $100 \text{ mm}\times100  \text{ mm}$ 2D structure with a circular hole located at the center under plane stress condition. In finite element simulation, the sample was compressed by  imposing a uniform displacement in  the $x$-direction on the left and right edges, while allowing the nodes on these edges to move freely in the $y$-direction. The midpoints of the left and right edge were constrained to prevent rigid body translations in the $y$-direction.

The contour of the deformed hole obtained with the experiment matched very well to those obtained with elastic multipole method ($n_\text{max}=10$) and finite element simulations (see Fig.~\ref{Fig:ExperimentalFreeEdge}b). We also compared the equivalent von Mises strain field defined as~\cite{Barber}
\begin{equation}
    \label{eq:vonMisesStrain}
  \varepsilon_{\text{vM}}=\frac{\sigma_\text{vM}}{E}=\frac{1}{1+\nu}\sqrt{ \varepsilon_{xx}^2-\varepsilon_{xx}\varepsilon_{yy}+\varepsilon_{yy}^2+3\varepsilon_{xy}^2+\frac{\nu}{(1-\nu)^2}(\varepsilon_{xx}+\varepsilon_{yy})^2}
\end{equation}
that were obtained with elastic multipole method ($n_\text{max}=10$), finite element simulations and  digital image correlation (DIC) analysis of experiment (see Appendix~\ref{app:experiments}). Strain fields show characteristic features of the induced quadrupoles (see Fig.~\ref{Fig:ExperimentalFreeEdge}c-e), because they have the largest amplitude and because the effect of higher-order multipoles decays faster away from holes and images. The effect of higher-order multipoles is  most pronounced in the region between the hole and the traction-free edge, where  contributions from both the hole and from its image are important. The strain fields agree very well between the elastic multipole method (Fig.~\ref{Fig:ExperimentalFreeEdge}c) and linear finite element simulations (Fig.~\ref{Fig:ExperimentalFreeEdge}d). The strain fields for experimental samples are qualitatively similar, but they deviate quantitatively near the hole, as can be seen from the heat map in Fig.~\ref{Fig:ExperimentalFreeEdge}e. The quantitative comparison of strains at four different points A-D (marked in Fig.~\ref{Fig:ExperimentalFreeEdge}) showed a relative error of $1$-$4\%$ between elastic multipole method and finite elements, and a relative error of $1$-$10\%$ between elastic multipole method and experiments (see Table~\ref{tab:free_edge}).
The discrepancy between elastic multipole method and finite element simulations is attributed to the finite size effects. For the elastic multipole method, we assumed a semi-infinite domain, while finite domains were modeled in finite element simulations to mimic experiments. The discrepancy between experiments and elastic multipole method is attributed to the confounding effects of nonlinear deformation due to moderately large compression ($\varepsilon_{xx}^{\text{ext}}=-0.05$), 3D deformation due to relatively thick samples, fabrication imperfections, nonzero friction between the sample and the mounting grips of the testing machine, the alignment of camera with the sample (2D DIC system was used), and the errors resulting from the choice of DIC parameters (see e.g. \cite{YuPan,Sutton}).

\subsection{Inclusions in an infinite elastic strip with prescribed tractions along the outer boundary of the strip}
\label{sec:strip}
In this section, we consider the deformation of an infinite elastic strip ($y_a<y<y_b$) with embedded circular holes and inclusions by prescribing uniform tractions along the outer boundary of an elastic strip. The deformation of the elastic matrix again induces elastic multipoles at the center of holes and inclusions. However, in this case, the induced multipoles generate an infinite number of image multipoles,  similar to an infinite number of images for charges between parallel conductive plates in electrostatics~\cite{Jackson}. The effect of all images can be captured with the integrals introduced by 
Howland~\cite{howland1929stress}, who considered deformation of an infinite elastic strip in response to a localized force. Here, however, we take a different approach, because we would like to explicitly show the contribution of each image.

\begin{figure}[!b]
  \centering
  \includegraphics[scale=1]{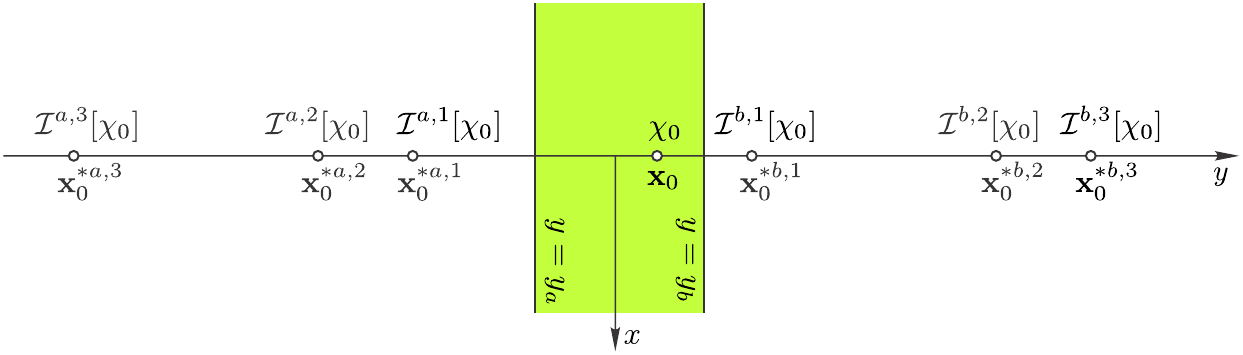}
  \caption{Images of an elastic multipole with the Airy stress function $\chi_0$  at position $\mbf{x_0}$ between the two parallel traction-free edges at $y=y_a$ and $y=y_b$. See text for the detailed description.}
  \label{Fig:StripImageIllus}
\end{figure}

To demonstrate how to enumerate all images let us first consider an elastic multipole with the Airy stress function $\chi_0$ that is
 at position $\mbf{x}_0=(x_0,y_0)$ between the two traction-free edges $a$ and $b$ at $y=y_a$ and $y=y_b$ (see Fig.~\ref{Fig:StripImageIllus}). To satisfy the boundary conditions at the edge $a$, we first place an image multipole $\mathcal{I}^{a}[\chi_0]\equiv \mathcal{I}^{a,1}[\chi_0]$ at position $\mbf{x}_0^{*a,1}=(x_0^{*a,1},y_0^{*a,1})=(x_0,2y_a-y_0)$ as  discussed in Section~\ref{sec:flat_edge_traction}. This first image ensures that the boundary conditions at the edge $a$ are satisfied, but the boundary conditions at the edge $b$ are still violated. To correct this, we need to place two more images with the Airy stress functions  $\mathcal{I}^{b}[\chi_0]\equiv \mathcal{I}^{b,1}[\chi_0]$ and $\mathcal{I}^{b}[\mathcal{I}^{a}[\chi_0]]\equiv\mathcal{I}^{b,2}[\chi_0]$ at positions $\mbf{x}_0^{*b,1}=(x_0^{*b,1},y_0^{*b,1})=(x_0,2y_b-y_0)$ and  $\mbf{x}_0^{*b,2}=(x_0^{*b,2},y_0^{*b,2})=(x_0,2y_b-y_0^{*a,1})$, respectively. These two additional images ensure that the boundary conditions at the edge $b$ are satisfied, but now the boundary conditions at the edge $a$ are violated. Thus we need to add two more images with the Airy stress functions  $\mathcal{I}^{a}[\mathcal{I}^{b}[\chi_0]]\equiv\mathcal{I}^{a,2}[\chi_0]$ and $\mathcal{I}^{a}[\mathcal{I}^{b}[\mathcal{I}^{a}[\chi_0]]]\equiv\mathcal{I}^{a,3}[\chi_0]$ at positions $\mbf{x}_0^{*a,2}=(x_0^{*a,2},y_0^{*a,2})=(x_0,2y_a-y_0^{*b,1})$ and  $\mbf{x}_0^{*a,3}=(x_0^{*a,3},y_0^{*a,3})=(x_0,2y_a-y_0^{*b,2})$, respectively. These two additional images  ensure that the boundary conditions at the edge $a$ are satisfied, but now the boundary conditions at the edge $b$ are violated. Thus, we need to keep adding more and more images, which are getting further and further away from the edges of the strip (see Fig.~\ref{Fig:StripImageIllus}). The infinite set of images can be constructed recursively with the Airy stress functions $\mathcal{I}^{a,k+1}[\chi_0]=\mathcal{I}^a[\mathcal{I}^{b,k}[\chi_0]]$ and $\mathcal{I}^{b,k+1}[\chi_0]=\mathcal{I}^b[\mathcal{I}^{a,k}[\chi_0]]$ that are  at positions $\mbf{x}_0^{*a,k+1}=(x_0^{*a,k+1},y_0^{*a,k+1})=(x_0,2y_a-y_0^{*b,k})$ and $\mbf{x}_0^{*b,k+1}=(x_0^{*b,k+1},y_0^{*b,k+1})=(x_0,2y_b-y_0^{*a,k})$, respectively. The initial conditions for the recursive relations are $\mathcal{I}^{a,0}[\chi_0]=\mathcal{I}^{b,0}[\chi_0]=\chi_0$ and $\mbf{x}_0^{*a,0}=\mbf{x}_0^{*b,0}=(x_0,y_0)$. Below we show how to use this approach to analyze the deformation of  elastic strips with circular inclusions and holes.

Consider a 2D infinite elastic strip with the Young's modulus $E_0$ and the Poisson's ratio $\nu_0$ filling the space $y_a<y<y_b$ subject to external  stress $\sigma_{xx}^\textrm{ext}$ with traction-free edges $a$ and $b$, where the boundary conditions are $\sigma_{yy}=\sigma_{xy}=0$ at $y=y_a$ and $y=y_b$. Embedded in the matrix are $N$ circular inclusions with radii $R_i$ centered at positions $\mbf{x}_i=(x_i,y_i)$ with Young's moduli $E_i$ and Poisson's ratios $\nu_i$, where $i\in\{1,\ldots, N\}$. Holes are described with the zero Young's modulus ($E_i=0$). 

The external load $\sigma_{xx}^\textrm{ext}$, represented with the Airy stress function $\chi_\textrm{ext}(x,y)=\frac{1}{2}\sigma_{xx}^\textrm{ext} y^2$, induces elastic multipoles at the centers of inclusions and holes. The Airy stress function $\chi_{\text{out}}\big(r_i,\varphi_i|\mbf{a}_{i,\text{out}}\big)$ outside the $i^{\textrm{th}}$ inclusion due to the induced multipoles can be expanded as shown in Eq.~(\ref{eq:AiryInducedOut}), where the origin of polar coordinates $(r_i,\varphi_i)$ is at the center $\mbf{x}_i$ of  $i^\text{th}$ inclusion and 
the set of  induced multipoles is $\mbf{a}_{i,\text{out}}=\{A_{i,0}, A_{i,1},\dots,B_{i,1}, B_{i,2},\dots, C_{i,2}, C_{i,3},\dots, D_{i,2}, D_{i,3},\dots\}$. The induced  multipoles $\mbf{a}_{i,\text{out}}$ at the center of the $i^\text{th}$ inclusion then further induce an infinite set of image multipoles at positions $\mbf{x}_i^{*a,k}$ and $\mbf{x}_i^{*b,k}$, where $k\in\{1,2,\ldots\}$, to satisfy the boundary conditions at the traction-free edges $a$ and $b$. The positions of these image multipoles can be generated recursively as 
$\mbf{x}_i^{*a,k+1}=(x_i^{*a,k+1},y_i^{*a,k+1})=(x_i,2y_a-y_i^{*b,k})$ and $\mbf{x}_i^{*b,k+1}=(x_i^{*b,k+1},y_i^{*b,k+1})=(x_i,2y_b-y_i^{*a,k})$ with the initial conditions $\mbf{x}_i^{*a,0}=\mbf{x}_i^{*b,0}=\mbf{x_i}=(x_i,y_i)$.
The Airy stress functions  $\mathcal{I}^{a,k}[\chi_{\text{out}}]$ and $\mathcal{I}^{b,k}[\chi_{\text{out}}]$ of image multipoles for the $i^\text{th}$ inclusion can be expanded as
\begin{subequations}
\begin{align}
  \mathcal{I}^{a,k}[\chi_{\text{out}}\big(r_i,\varphi_i|\mbf{a}_{i,\text{out}}\big)] = & \chi^*_{\text{out}}\big(r_i^{*a,k},\varphi_i^{*a,k}|\mbf{a}^{*a,k}_{i,\text{out}}\big),\\
  \mathcal{I}^{b,k}[\chi_{\text{out}}\big(r_i,\varphi_i|\mbf{a}_{i,\text{out}}\big)] = & \chi^*_{\text{out}}\big(r_i^{*b,k},\varphi_i^{*b,k}|\mbf{a}^{*b,k}_{i,\text{out}}\big),
\end{align}
\end{subequations}
\noindent where the function $\chi_\text{out}^*$ is defined in Eq.~(\ref{eq:AiryInducedOutImage2}), the origins of polar coordinates $(r_i^{*a,k},\varphi_i^{*a,k})$ and $(r_i^{*b,k},\varphi_i^{*b,k})$ are at $\mbf{x}_i^{*a,k}$ and $\mbf{x}_i^{*b,k}$, respectively, and the sets of amplitudes of image multipoles are $\mbf{a}^{*a,k}_{i,\text{out}}=\{A_{i,0}^{*a,k}, A_{i,1}^{*a,k},\dots,B_{i,0}^{*a,k}, B_{i,1}^{*a,k},\dots, C_{i,2}^{*a,k}, C_{i,3}^{*a,k},\dots, D_{i,2}^{*a,k}, D_{i,3}^{*a,k},\dots\}$ and $\mbf{a}^{*b,k}_{i,\text{out}}=\{A_{i,0}^{*b,k}, A_{i,1}^{*b,k},\dots,B_{i,0}^{*b,k}, B_{i,1}^{*b,k},\dots, C_{i,2}^{*b,k}, C_{i,3}^{*b,k},\dots, D_{i,2}^{*b,k}, D_{i,3}^{*b,k},\dots\}$. The amplitudes of image multipoles $\mbf{a}^{*a,k}_{i,\text{out}}$ and $\mbf{a}^{*b,k}_{i,\text{out}}$ are related to the amplitudes $\mbf{a}_{i,\text{out}}$ of  induced multipoles at the center of the  $i^{\text{th}}$ inclusion and they can be generated recursively.  The initial conditions for recursive relations are $\mbf{a}^{*a,0}_{i,\text{out}}=\mbf{a}^{*b,0}_{i,\text{out}}=\mbf{a}_{i,\text{out}}$, and the amplitudes of image multipoles  $\mbf{a}^{*a,k+1}_{i,\text{out}}$ and $\mbf{a}^{*b,k+1}_{i,\text{out}}$ are related to the amplitudes of image multipoles $\mbf{a}^{*b,k}_{i,\text{out}}$ and $\mbf{a}^{*a,k}_{i,\text{out}}$ as
\begin{subequations}
\begin{align}
\begin{split}
    A_{i,n}^{*a,k+1} &= \begin{cases}
    -A_{i,0}^{*b,k}, & n=0, \vspace{2mm} \\
    \begin{aligned}
    &-(n+1)A_{i,n}^{*b,k}-(n+2)C_{i,n+2}^{*b,k} -2(n-1)\alpha_{i}^{*b,k}B_{i,n-1}^{*b,k}\\
    &-2(2n+1)\alpha_{i}^{*b,k}D_{i,n+1}^{*b,k} +4(n-1)(\alpha_{i}^{*b,k})^2C_{i,n}^{*b,k},
    \end{aligned} &  n\geq 1,
    \end{cases}\\
    B_{i,n}^{*a,k+1} &= \begin{cases}
    0,  & n=0,\\
    +2B_{i,1}^{*b,k} +3D_{i,3}^{*b,k} +2\alpha_{i}^{*b,k}A_{i,0}^{*b,k} -6\alpha_{i}^{*b,k}C_{i,2}^{*b,k}, & n = 1, \vspace{2mm} \\
    \begin{aligned}
    &+(n+1)B_{i,n}^{*b,k} +(n+2)D_{i,n+2}^{*b,k} -2(n-1)\alpha_{i}^{*b,k}A_{i,n-1}^{*b,k}\\ &-2(2n+1)\alpha_{i}^{*b,k}C_{i,n+1}^{*b,k} -4(n-1)(\alpha_{i}^{*b,k})^2D_{i,n}^{*b,k},
    \end{aligned} & n\geq 2,
    \end{cases}\\
    C_{i,n}^{*a,k+1} &= \begin{cases}
    +C_{i,2}^{*b,k} - A_{i,0}^{*b,k}, & n = 2,\\
    +(n-1)C_{i,n}^{*b,k}+(n-2)A_{i,n-2}^{*b,k}+2(n-2) \alpha_{i}^{*b,k} D_{i,n-1}^{*b,k}, {\hskip .5mm} & n\geq 3,
    \end{cases}\\
    D_{i,n}^{*a,k+1} &=\begin{cases} -(n-1)D_{i,n}^{*b,k}-(n-2)B_{i,n-2}^{*b,k}+2(n-2) \alpha_{i}^{*b,k}C_{i,n-1}^{*b,k},  {\hskip 1.5mm} & n\geq 2, \end{cases}
\end{split}
\end{align}
\begin{align}
\begin{split}
    A_{i,n}^{*b,k+1} &= \begin{cases}
    -A_{i,0}^{*a,k}, & n=0, \\
    \begin{aligned}
    &-(n+1)A_{i,n}^{*a,k}-(n+2)C_{i,n+2}^{*a,k} -2(n-1)\alpha_{i}^{*a,k}B_{i,n-1}^{*a,k}\\
    &-2(2n+1)\alpha_{i}^{*a,k}D_{i,n+1}^{*a,k} +4(n-1)(\alpha_{i}^{*a,k})^2C_{i,n}^{*a,k},
    \end{aligned} &  n\geq 1,
    \end{cases}\\
    B_{i,n}^{*b,k+1} &= \begin{cases}
    0,  & n=0\\
    +2B_{i,1}^{*a,k} +3D_{i,3}^{*a,k} +2\alpha_{i}^{*a,k}A_{i,0}^{*a,k} -6\alpha_{i}^{*a,k}C_{i,2}^{*a,k}, & n = 1,\\
    \begin{aligned}
    &+(n+1)B_{i,n}^{*a,k} +(n+2)D_{i,n+2}^{*a,k} -2(n-1)\alpha_{i}^{*a,k}A_{i,n-1}^{*a,k}\\ &-2(2n+1)\alpha_{i}^{*a,k}C_{i,n+1}^{*a,k} -4(n-1)(\alpha_{i}^{*a,k})^2D_{i,n}^{*a,k},
    \end{aligned} & n\geq 2,
    \end{cases}\\
    C_{i,n}^{*b,k+1} &= \begin{cases}
    +C_{i,2}^{*a,k} - A_{i,0}^{*a,k}, & n = 2\\
    +(n-1)C_{i,n}^{*a,k}+(n-2)A_{i,n-2}^{*a,k}+2(n-2) \alpha_{i}^{*a,k} D_{i,n-1}^{*a,k}, {\hskip .5mm} & n\geq 3,
    \end{cases}\\
    D_{i,n}^{*b,k+1} &=\begin{cases} -(n-1)D_{i,n}^{*a,k}-(n-2)B_{i,n-2}^{*a,k}+2(n-2) \alpha_{i}^{*a,k}C_{i,n-1}^{*a,k},  {\hskip 1.5mm} & n\geq 2, \end{cases}
\end{split}
\end{align}
\end{subequations}
where $\alpha_{i}^{*a,k}=(y_b-y_i^{*a,k})/R_i$ and 
$\alpha_{i}^{*b,k}=(y_a-y_i^{*b,k})/R_i$.

The total Airy stress function outside all inclusions can then be written as 
\begin{equation}
\label{eq:AiryOutStrip}
\begin{split}
\chi^{\text{tot}}_{\text{out}}\big(x,y|\mbf{a}_{\text{out}}\big)=&\chi_{\text{ext}}(x,y)+\sum_{i=1}^N \chi_{\text{out}}\big(r_i(x,y),\varphi_i(x,y)|\mbf{a}_{i,\text{out}}\big)\\
&+ \sum_{i=1}^N \sum_{k=1}^\infty \Big( \mathcal{I}^{a,k}\left[\chi_{\text{out}}\big(r_i(x,y),\varphi_i(x,y)|\mbf{a}_{i,\text{out}}\big)\right]
+  \mathcal{I}^{b,k}\left[\chi_{\text{out}}\big(r_i(x,y),\varphi_i(x,y)|\mbf{a}_{i,\text{out}}\big)\right] \Big),
\end{split}
\end{equation}
where the first term is due to external stress, the first summation describes contributions due to induced multipoles at the centers of inclusions and the last two summations describe contributions due to images of induced multipoles. The set of amplitudes of induced multipoles for all inclusions is defined as $\mbf{a}_{\text{out}}=\{\mbf{a}_{1,\text{out}}, \ldots, \mbf{a}_{N,\text{out}}\}$. 

We also expand the induced Airy stress function $\chi_{\text{in}}\big(r_i,\varphi_i|\mbf{a}_{i,\text{in}}\big)$ inside the $i^\text{th}$ inclusion, as shown in Eq.~(\ref{eq:AiryInducedIn}), where the set of amplitudes of induced multipoles is  $\mbf{a}_{i,\text{in}}=\{a_{i,2}, a_{i,3},\dots,b_{i,2}, b_{i,3},\dots, c_{i,0}, c_{i,1},\dots, d_{i,1}, d_{i,2},\dots\}$. We again define the total Airy stress function $\chi_{\text{in}}^{\text{tot}}\big(x,y|\mbf{a}_{i,\text{in}}\big)$ inside the $i^\text{th}$ inclusion, which also includes the effect of the external load ($\chi_{\text{ext}}$) as shown in Eq.~(\ref{eq:AiryIn}). The rest of the steps are the same as the ones described in Section~\ref{sec:flat_edge_traction}. The amplitudes of induced multipoles $\mbf{a}_{i,\text{out}}$ and $\mbf{a}_{i,\text{in}}$ are obtained by satisfying the boundary conditions that tractions and displacements are continuous across the circumference of each inclusion in Eq.~(\ref{eq:BC}), which can be 
converted to the matrix equation similar to that in Eq.~(\ref{eq:matrix}). To numerically solve the matrix equation we truncate the degrees of multipoles at $n_\text{max}$ as discussed in Section~\ref{sec:flat_edge_traction}. Furthermore, we truncate the summation for the number of images at $k_\text{max}$ in Eq.~(\ref{eq:AiryOutStrip}).

\begin{figure}[!t]
  \centering
  \includegraphics[scale=1]{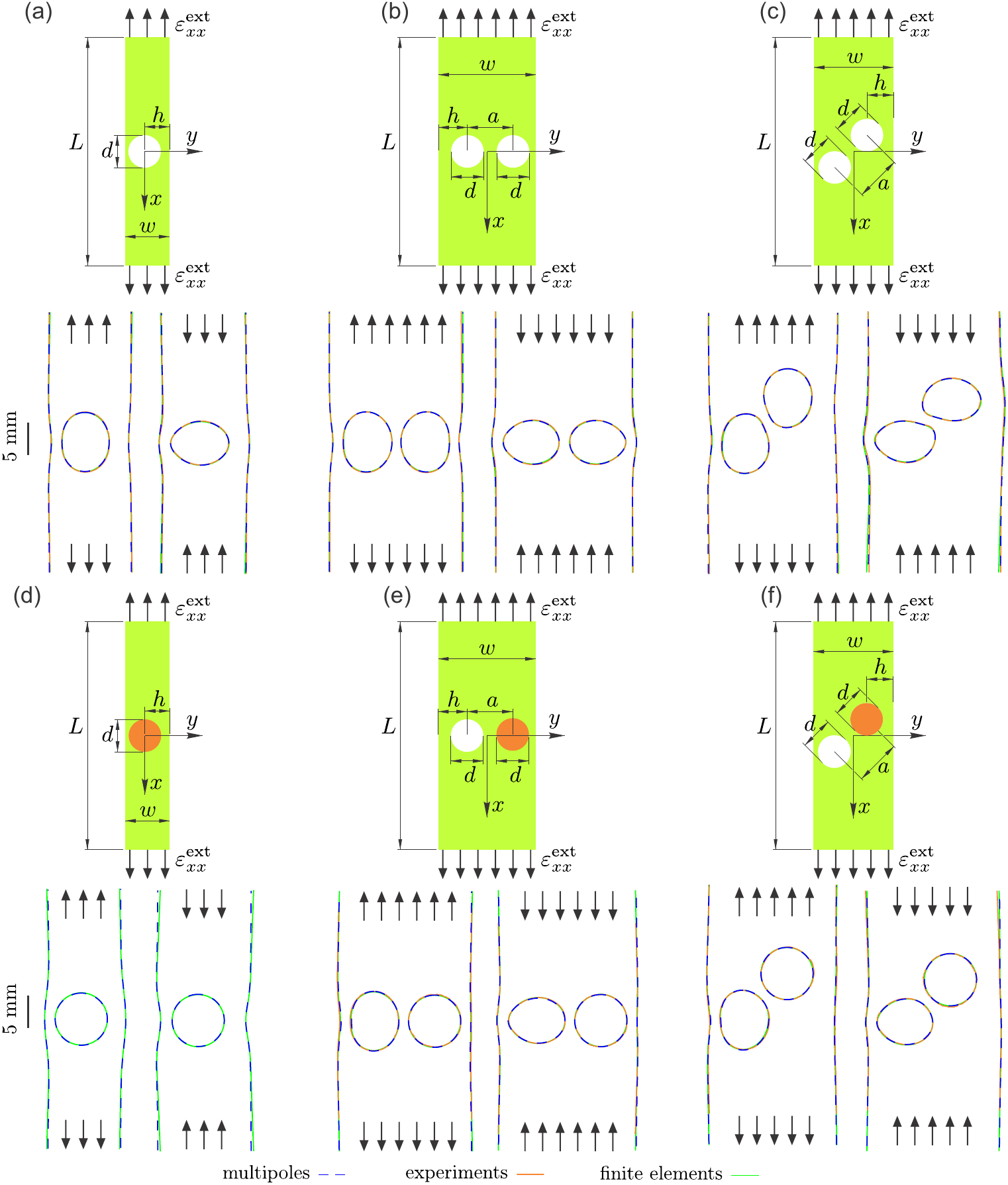}
   \caption{Deformation of elastic strips with holes and inclusions. (a-f) Schematic images describing the initial undeformed shapes of  structures with holes and the direction of applied strain $\varepsilon_{xx}^\text{ext}$.  The Young's modulus of the elastic matrix is $E_0=0.97\,\textrm{MPa}$ and  Poisson's ratio is $\nu_0=0.49$. Holes are represented with white circles and inclusions with orange circles ($E_\text{inc}=2.9\,\textrm{GPa}$, $\nu_\text{inc}=0.37$). (a-f) For each structure, we show contours of deformed holes, inclusions and traction-free edges under tensile (left) and compressive (right) loads.  Note that contours are only shown in the vicinity of deformed holes and inclusions and they do not span the whole length of  samples. Dashed blue lines show the contours obtained with elastic multipole method for $n_\text{max}=10$ and $k_\text{max}=6$. Orange and green solid lines show the contours obtained with experiments and linear finite element simulations, respectively. For all samples the length was $L=100$~mm and the thickness was $25$~mm. The diameters of all inclusions and holes were $d=8.00$~mm. Other geometrical properties were: (a,d)~$h=7.06$~mm, $w=12.75$~mm, (b,e)~$h=5.67$~mm, $a=9.49$~mm, $w=20.72$~mm, and (c,f)~$h=7.56$~mm, $a=9.57$~mm, $w=19.97$~mm. Tensile and compressive loads were: (a)~$\varepsilon_{xx}^\text{ext}=+0.027$ and  $\varepsilon_{xx}^\text{ext}=-0.033$, (b)~$\varepsilon_{xx}^\text{ext}=+0.027$ and  $\varepsilon_{xx}^\text{ext}=-0.029$, (c)~$\varepsilon_{xx}^\text{ext}=+0.020$ and  $\varepsilon_{xx}^\text{ext}=-0.030$, (d)~$\varepsilon_{xx}^\text{ext}=+0.250$ and  $\varepsilon_{xx}^\text{ext}=-0.250$, (e)~$\varepsilon_{xx}^\text{ext}=+0.034$ and  $\varepsilon_{xx}^\text{ext}=-0.033$,
   (f)~$\varepsilon_{xx}^\text{ext}=+0.038$ and  $\varepsilon_{xx}^\text{ext}=-0.033$.
}
  \label{fig:strip_holes}
\end{figure}

The elastic multipole method described above was compared with linear finite element simulations and experiments. Three different configurations of elastic strips with length $L=100$~mm and thickness $25$~mm were considered: $w=12.75$~mm wide strips with one hole, $w=19.97$~mm wide strips with two holes placed orthogonal to the long axis of the strip and $w=20.72$~mm wide strips with two holes at $45^\circ$ angle relative to the long axis of the strip (see  Fig.~\ref{fig:strip_holes}a-c). The diameter of all holes was $d=8.00$~mm. Experiments were carried out using displacement controlled compressive and tensile loading (strain $\varepsilon_{xx}^{\text{ext}}$), where both ends of the strip were glued to aluminum plates, while the outer edges of the strip were free to move (see Appendix~\ref{app:experiments}). For the elastic multipole method, we set the external load to $\sigma_{xx}^{\text{ext}}=E_0 \varepsilon_{xx}^{\text{ext}}$, $\sigma_{yy}^{\text{ext}}=\sigma_{xy}^{\text{ext}}=0$ and assumed plane stress condition. For 2D linear finite element simulations we used plane stress condition with the same external loads as for elastic multipole method (see Appendix~\ref{app:FEM}). Note that the elastic multipole method considers infinitely long strips ($L\rightarrow \infty$), while the length $L$ is finite in experiments and simulations. In experiments and simulations, we observed a slight bending of strips under both compression and tension because holes were placed asymmetrically. However, this bending is not captured by the elastic multipole method. Bending could in principle be captured by adding additional terms to the Airy stress functions, such as $y^3$ and $xy^3$~\cite{Barber}, which were omitted in our expansion for the induced multipoles. Note that the coefficients of these additional terms cannot be obtained from the continuity of tractions and displacements across the circumference of inclusions in Eq.~(\ref{eq:BC}), but additional equations would be needed, such as the boundary conditions at the two ends of the strip. This is beyond the scope of this paper and hence we expect to observe some errors for the elastic multipole method due to the neglected bending.
 
In Fig.~\ref{fig:strip_holes}a-c, we show contours of deformed holes obtained with elastic multipole method ($n_\text{max}=10$, $k_\text{max}=6$), linear finite elements, and experiments. The results for all 3 different approaches agree very well  in the vicinity of holes where the effect of bending is small. In Fig.~\ref{fig:strip_holes}a, we observe  more pronounced deformations of the hole and the outer edges of the strip in the regions where they approach each other due to the strong interaction between the hole and its images. This is similar to the deformation of the hole near the single traction-free edge in Figs.~\ref{Fig:hole_flat_edge} and ~\ref{Fig:ExperimentalFreeEdge}, but here such deformation is observed on both sides (Fig.~\ref{fig:strip_holes}a). Note that  deformations are more pronounced near the left edge because the hole is closer to the left edge and thus the interactions between the hole and its images become more dominant. In Figs.~\ref{fig:strip_holes}b-c, we observe interactions between holes as well as interactions between holes and their images, which is reflected in more pronounced deformations in the regions, where holes are close to each other, and in the regions, where holes are close to the outer edges of the strip. 

In Fig.~\ref{fig:strip_holes}d-f we repeated the analysis, where for each sample one of the holes was filled with a PMMA rod that was glued to the elastic matrix with a cyanoacrylate adhesive. Compared to the base elastic matrix with Young's modulus $E_0=0.97$~MPa, the PMMA inclusion is very rigid ($E_{\text{inc}}=2.9$~GPa). This is reflected in the contours of deformed samples, where inclusions remain circular, while holes and traction-free edges are deformed (see Fig.~\ref{fig:strip_holes}d-f). We observe interactions between holes and inclusions as well as interactions between them and  their images, which is reflected in more pronounced deformations in the regions, where holes are close to inclusions, and in the regions, where holes and inclusions are close to the outer edges of the strip. Note that for a narrow elastic strip with a single inclusion (see Fig.~\ref{fig:strip_holes}d) we had to apply a very large compressive/tensile strain of $\varepsilon_{xx}^{\text{ext}}=\pm25\%$ to observe a notable deformation of the outer edges of the strip in the vicinity of inclusion. At such large compressive/tensile loads there was a failure of the glue between the PMMA rod and the elastic matrix as well as between the elastic matrix and the outer aluminum plates at the two ends of the strip. Thus we were not able to perform experiments in this case. Note that the strip edges near the rigid inclusion (Fig.~\ref{fig:strip_holes}d) bulge in the opposite direction than the strip edges near holes (e.g. Fig.~\ref{fig:strip_holes}a), which is related to the opposite sign of amplitudes of induced quadrupoles as discussed in the companion paper~\cite{sarkar2019elastic}.

 \begin{figure}[!b]
  \centering
  \includegraphics[scale=1]{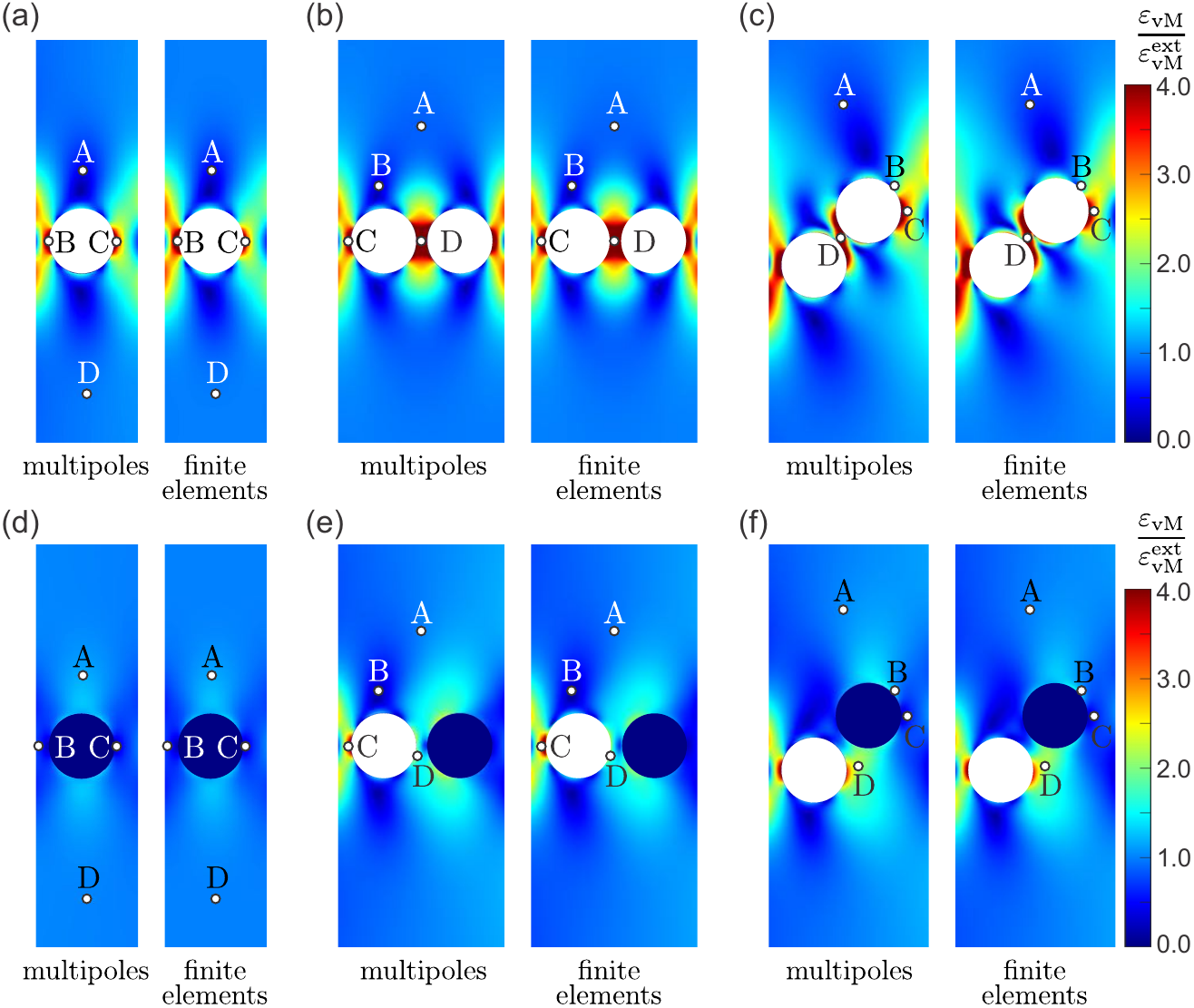}
   \caption{Equivalent von Mises strain fields $\varepsilon_\text{vM}$ were obtained with elastic multipole method ($n_\text{max}=10$, $k_\text{max}=6$) and linear finite element simulations. The values of the strain fields $\varepsilon_\text{vM}$ were normalized with the value of the equivalent von Mises strain $\varepsilon_\text{vM}^\text{ext}=|\varepsilon_{xx}^\text{ext}|$ due to the external load.
   The geometrical properties and the magnitudes of applied loads for the samples in (a-f) were the same as in Fig.~\ref{fig:strip_holes}. For each of the samples in (a-f), four marked points A-D were chosen for the quantitative comparison of strains $\varepsilon_\text{vM}$. See Table~\ref{tab:strip} for details.
}
  \label{fig:strip_holes_strain}
\end{figure}

 \begin{table}[!t]
\centering
\caption{Quantitative comparison for the values of equivalent von Mises strains $\varepsilon_\text{vM}$ normalized with the value for the applied external load $\varepsilon_\text{vM}^\text{ext}=|\varepsilon_{xx}^\text{ext}|$ at points A-D for 6 different samples defined in Fig.~\ref{fig:strip_holes_strain} obtained with elastic multipole method (EMP) and finite element simulations (FEM). The relative percent errors  between the EMP and FEM were calculated as $\epsilon=100\times|\varepsilon_\text{vM}^\text{(EMP)}-\varepsilon_\text{vM}^\text{(FEM)}|/\varepsilon_\text{vM}^\text{(FEM)}$.}
\small
\def\arraystretch{1.15}
\label{tab:strip}
\begin{tabular}{|>{\centering}m{0.05\textwidth}|*{2}{>{\centering}m{0.042\textwidth}}>{\centering}m{0.035\textwidth}|*{2}{>{\centering}m{0.042\textwidth}}>{\centering}m{0.035\textwidth}|*{2}{>{\centering}m{0.042\textwidth}}>{\centering}m{0.035\textwidth}|*{2}{>{\centering}m{0.042\textwidth}}>{\centering}m{0.035\textwidth}|*{2}{>{\centering}m{0.042\textwidth}}>{\centering}m{0.035\textwidth}|*{2}{>{\centering}m{0.042\textwidth}}>{\centering\arraybackslash}m{0.035\textwidth}|}
\hline
\multirow{3}{*}{points}& \multicolumn{3}{c|}{sample (a)}& \multicolumn{3}{c|}{sample (b)}& \multicolumn{3}{c|}{sample (c)}&
\multicolumn{3}{c|}{sample (d)}&
\multicolumn{3}{c|}{sample (e)}&
\multicolumn{3}{c|}{sample (f)}
\\
\cline{2-19}
& \multicolumn{3}{c|}{strain $\epsilon_\text{vM}/\epsilon_\text{vM}^\text{ext}$} & \multicolumn{3}{c|}{strain $\epsilon_\text{vM}/\epsilon_\text{vM}^\text{ext}$}& \multicolumn{3}{c|}{strain $\epsilon_\text{vM}/\epsilon_\text{vM}^\text{ext}$}& \multicolumn{3}{c|}{strain $\epsilon_\text{vM}/\epsilon_\text{vM}^\text{ext}$}& \multicolumn{3}{c|}{strain $\epsilon_\text{vM}/\epsilon_\text{vM}^\text{ext}$}& \multicolumn{3}{c|}{strain $\epsilon_\text{vM}/\epsilon_\text{vM}^\text{ext}$}  \\
\cline{2-19}
  &  EMP   & FEM   & $\epsilon(\%)$  &  EMP   & FEM   & $\epsilon(\%)$ &  EMP   & FEM   & $\epsilon(\%)$ &  EMP   & FEM   & $\epsilon(\%)$ &  EMP   & FEM   & $\epsilon(\%)$ &  EMP   & FEM   & $\epsilon(\%)$\\
\hline
A & 0.439 & 0.454 & 3.3 & 0.875 & 0.883 & 0.9 & 0.675 & 0.692 & 2.4 & 1.168 & 1.154 & 1.2 & 1.002 & 0.952 & 5.3 & 1.045 & 1.005 & 4.0\\
B & 5.740 & 6.077 & 5.5 & 0.331 & 0.330 & 0.2 & 1.301 & 1.245 & 4.5 & 0.643 & 0.612 & 5.0 & 0.231 & 0.223 & 3.6 & 1.190 & 1.135 & 4.9\\
C & 3.783 & 3.746 & 1.0 & 4.561 & 4.562 & 0.0 & 3.311 & 3.203 & 3.4 & 0.273 & 0.283 & 3.6 & 3.443 & 3.497 & 1.5 & 0.496 & 0.524 & 5.3\\
D & 0.971 & 0.984 & 1.3 & 5.760 & 5.803 & 0.7 & 4.188 & 4.308 & 2.8 & 1.014 & 1.004 & 0.9 & 1.955 & 2.020 & 3.2 & 2.166 & 2.141 & 1.2\\
\hline            
\end{tabular}
\end{table}
 
Furthermore, in Fig.~\ref{fig:strip_holes_strain} we compared the equivalent von Mises strain obtained with elastic multipole method and linear finite element simulations for all 6 samples described above (see Fig.~\ref{fig:strip_holes}). The strain fields obtained with the two methods agree very well. Note that we were unable to obtain strain fields in experiments because of the delamination of the spray-painted speckle pattern  near the edges of the strip, which is required for  DIC analysis. We observe large strain concentrations in regions where holes are close to the traction-free edges, inclusions and other holes. In contrast, we observe reduced strains in regions where inclusions are close to free edges. The quantitative comparison of strains at four different point A-D (marked in Fig.~\ref{fig:strip_holes_strain}) showed a relative error of $0$-$6\%$ (see Table~\ref{tab:strip}). The relative error was the largest in the regions that were affected by the bending of strips in finite element simulations, which was neglected in the elastic multipole method.

\begin{figure}[!t]
\centering
\includegraphics[scale=1]{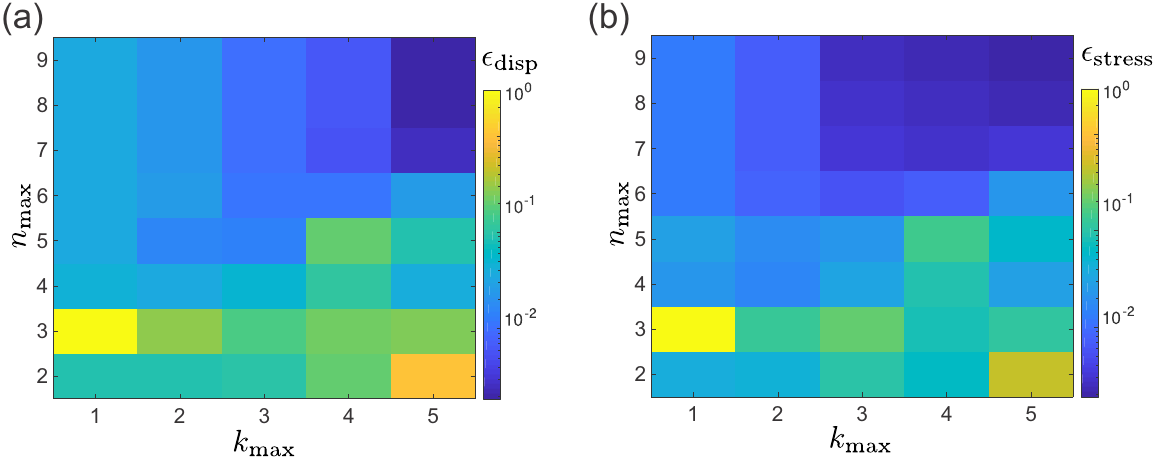}
  \caption{The normalized errors for (a)~displacements  $\epsilon_\text{disp}(n_\text{max},k_\text{max})$ and (b)~stresses $\epsilon_\text{stress}(n_\text{max},k_\text{max})$ as defined in Eq.~(\ref{eq:error_strip}) for the deformation of an elastic strip with a single hole (see Fig.~\ref{fig:strip_holes}a).}
\label{Fig:error_strip}
\end{figure}

Finally, we also comment on the convergence analysis for  spatial distributions of displacements $\mbf{u}^{(n_\text{max},k_\text{max})}(x,y)$ and von Mises stresses $\sigma^{(n_\text{max},k_\text{max})}_\text{vM}(x,y)$ as a function of the numbers $n_\text{max}$ for the maximum degree of induced multipoles and $k_\text{max}$ for the maximum number of images. Displacements and von Mises stresses were evaluated at $N_p$ discrete points $(x_i,y_j)=\left(iR/50,jw/200\right)$, where $i\in\{-250,-249,\ldots,250\}$ and $j\in\{-100, -99, \ldots, 100\}$ and grid points that lie inside holes and inclusions were excluded. Here, $R$ is the radius of holes and $w$ is the width of the strip. The normalized errors for displacements $\epsilon_\text{disp}(n_\text{max},k_\text{max})$ and stresses $\epsilon_\text{stress}(n_\text{max},k_\text{max})$ were obtained by calculating the relative changes of  spatial distributions of displacements and von Mises stresses when the maximum degree $n_\text{max}$ of induced multipoles and the maximum number $k_\text{max}$ of images are both increased by one. The normalized errors are given by~\cite{SpecMethod}
\begin{subequations}
\begin{align}
\label{Eqn:errordisp_strip}
    \epsilon_\text{disp}(n_\text{max},k_\text{max}) &= \frac{1}{\sqrt{N_p}}\left[\sum_{i,j}\left|\frac{\mbf{u}^{(n_\text{max}+1,k_\text{max}+1)}(x_i,y_j)-\mbf{u}^{(n_\text{max},k_\text{max})}(x_i,y_j)}{d\,\sigma_\text{vM}^\text{ext}/E_0}\right|^2 \right]^{1/2},\\
\label{Eqn:errorstress_strip}
    \epsilon_\text{stress}(n_\text{max},k_\text{max}) &= \frac{1}{\sqrt{N_p}}\left[\sum_{i,j}\left(\frac{\sigma^{(n_\text{max}+1,k_\text{max}+1)}_\text{vM}(x_i,y_j)-\sigma^{(n_\text{max},k_\text{max})}_\text{vM}(x_i,y_j)}{\sigma_\text{vM}^\text{ext}}\right)^2 \right]^{1/2}.
\end{align}
\label{eq:error_strip}%
\end{subequations}
Here, displacements and von Mises stresses are normalized with the characteristic scales $d \sigma_\text{vM}^\text{ext}/E_0$ and $\sigma_\text{vM}^\text{ext}$, respectively, where $d$ is the diameter of holes, $\sigma_\text{vM}^\text{ext}=|\sigma_{xx}^{\text{ext}}|$ is the value of  von Mises stress due to external load, and $E_0$ is the Young's modulus of the surrounding elastic matrix.
The normalized errors are plotted in Fig.~\ref{Fig:error_strip} for the elastic strip with one hole (see Fig.~\ref{fig:strip_holes}a).
As the maximum degree $n_\text{max}$ of induced multipoles and the maximum number $k_\text{max}$ of images are increased, the normalized errors for displacements $\epsilon_\text{disp}(n_\text{max},k_\text{max})$ and stresses $\epsilon_\text{stress}(n_\text{max},k_\text{max})$ decrease exponentially. As  discussed in Section~\ref{sec:flat_edge_traction} and in Ref.~\cite{sarkar2019elastic} a higher number $n_\text{max}$ is needed when either holes and inclusions are close to each other or when holes and inclusions are close to the boundary. For sufficiently large $n_\text{max}$, the errors are quite low already for $k_\text{max}=1$ (see Fig.~\ref{Fig:error_strip}), because the most dominant interactions occur with the nearest image. The errors can then be further reduced by increasing the number of images $k_\text{max}$.

\subsection{Inclusions in a semi-infinite elastic matrix with prescribed displacements along the outer boundary of the matrix}
\label{sec:flat_edge_displacements}

In this section, we consider the deformation of a semi-infinite elastic matrix ($y<y_b$) with embedded circular holes and inclusions by prescribing uniform displacements along the outer boundary of the elastic matrix. We follow the same procedure as  described in Section~\ref{sec:flat_edge_traction} for prescribed uniform tractions along the outer boundary of the elastic matrix. Deformation of the elastic matrix induces elastic multipoles at the centers of holes and inclusions, which further induce image multipoles in order to satisfy boundary conditions at the outer edge. The first step is thus to find the appropriate Airy stress functions for image multipoles. As described in Section~\ref{sec:images}, it is sufficient to find the Airy stress function for the image of disclination, which can then be used to derive expressions for the Airy stress functions that correspond to images for all other elastic multipoles.

Thus we first consider a semi-infinite elastic medium with Young’s modulus $E_0$ and Poisson's ratio $\nu_0$ filling  half-space $y < y_b$ with a disclination with charge $s$  at position $\mathbf{x}_0=(x_0, y_0) = (x_0,y_b - h)$, which is at the distance $h$ from the fixed edge at $y=y_b$. The Airy stress function $\chi(\mbf{x})$ for this case can be found by solving the equation $\Delta\Delta\chi=E_0 s \delta(\mbf{x}-\mbf{x_0})$ with the boundary conditions $u_x(x,y_b)=u_y(x,y_b)=0$. This boundary condition can be achieved by bonding the elastic matrix to an infinitely rigid material ($E\rightarrow \infty$) on the half-plane ($y>y_b$) at $y=y_b$. Such problem was previously considered by Adeerogba in Ref.~\cite{Adeerogba}, who demonstrated that for the eigenstresses in the elastic matrix described with the Airy stress function $\chi_0(x,y)$, boundary conditions at the fixed edge can be satisfied with the addition of the image Airy stress function 
\begin{equation}
\label{eq:AdeerogbaOp}
\mathcal{I}\big[\chi_0(x,y)\big]=\frac{1}{\kappa_0}\left(1-2(y-y_b)\frac{\partial}{\partial y}+(y-y_b)^2\Delta\right)\chi_0\big(x,2y_b-y\big)+\frac{1}{4}\left(\frac{1}{\kappa_0}-\kappa_0\right)\int dy \int dy\,\Delta \chi_0\big(x,2y_b-y\big),
\end{equation}
where the value of Kolosov's constant  $\kappa_0$ is $(3-\nu_0)/(1+\nu_0)$ for  plane stress and $3-4\nu_0$ for  plane strain condition~\cite{Barber}. The Airy stress function for   disclinations with charge $s$  at position $\mbf{x}_0$ is  $\chi_m(\mbf{x}-\mbf{x}_0|s)= \frac{E_0 s}{8\pi}|\mbf{x}-\mbf{x}_0|^2 \big(\ln|\mbf{x}-\mbf{x}_0| -1/2\big)$. Using Eq.~(\ref{eq:AdeerogbaOp}), we obtain the Airy stress function for its image  at position $\mbf{x}_0^*=(x_0^*,y_0^*)=(x_0,y_b+h)$ as
\begin{equation}
\begin{split}
    \mathcal{I}\big[\chi_m(\mbf{x}-\mbf{x}_0|s)\big] = \frac{E_0s}{8\pi\kappa_0}\Big[&{r^*}^2\ln r^* +\frac{(5 + 2 \kappa_0 + \kappa_0^2)}{8}{r^*}^2+\frac{(-9 - \kappa_0 - 7 \kappa_0^2 + \kappa_0^3)}{16}{r^*}^2\cos(2\varphi^*)\\
    &-\frac{(1-\kappa_0^2)}{2}{r^*}^2\ln {r^*}\cos(2\varphi^*)+\frac{(1-\kappa_0^2)}{2}{r^*}^2\varphi^*\sin(2\varphi^*)+4h r^* \ln r^* \sin \varphi^* + 4 h^2 \ln r^* \Big],
\end{split}
\end{equation}
where $(r_i^* = \sqrt{(x-x_0^*)^2+(y-y_0^*)^2},\varphi_i^* = \arctan[(y-y_0^*)/(x-x_0^*)])$ are  polar coordinates centered at the image of the disclination. Note that the terms ${r^*}^2\ln {r^*}\cos(2\varphi^*)$ and ${r^*}^2\varphi^*\sin(2\varphi^*)$ are absent in the Michell solution for the biharmonic equation~\cite{Michell}, but their difference ${r^*}^2\cos(2\varphi^*)\ln {r^*}-{r^*}^2\varphi^*\sin(2\varphi^*)$ is harmonic and  thus satisfies the biharmonic equation. The images for all other multipoles can be obtained by using the operator in Eq.~(\ref{eq:AdeerogbaOp}) or by taking appropriate derivatives of the Airy stress function for the image of disclination as demonstrated in Section~\ref{sec:images}.

Now, let us consider a 2D semi-infinite elastic matrix with the Young's modulus $E_0$ and the Poisson's ratio $\nu_0$ filling the half-space $y<y_b$ subjected to external strain ($\varepsilon_{xx}^\textrm{ext}$, $\varepsilon_{yy}^\textrm{ext}$). Embedded in the matrix are $N$ circular inclusions with radii $R_i$ centered at positions $\mbf{x}_i=(x_i,y_i)=(x_i,y_b-h_i)$ with Young's moduli $E_i$ and Poisson's ratios $\nu_i$, where $i\in\{1,\ldots, N\}$ and $h_i$ is the distance between the center of the  $i^\text{th}$ inclusion and the displacement-controlled edge at $y=y_b$. Holes are described with the zero Young's modulus ($E_i=0$).

The external strain ($\varepsilon_{xx}^\textrm{ext}$, $\varepsilon_{yy}^\textrm{ext}$) can be described with the equivalent external stresses $\sigma_{xx}^\textrm{ext}=\frac{E_0 }{(1-\nu_0^2)}\left[\varepsilon_{xx}^\textrm{ext}+\nu_0\varepsilon_{yy}^\textrm{ext}\right]$,  $\sigma_{yy}^\textrm{ext}=\frac{E_0}{(1-\nu_0^2) }\left[\varepsilon_{yy}^\textrm{ext}+\nu_0\varepsilon_{xx}^\textrm{ext}\right]$ for  plane stress condition and with stresses $\sigma_{xx}^\textrm{ext}=\frac{E_0 }{(1+\nu_0)(1-2\nu_0)}\left[(1-\nu_0)\varepsilon_{xx}^\textrm{ext}+\nu_0\varepsilon_{yy}^\textrm{ext}\right]$,  $\sigma_{yy}^\textrm{ext}=\frac{E_0 }{(1+\nu_0)(1-2\nu_0)}\left[(1-\nu_0)\varepsilon_{yy}^\textrm{ext}+\nu_0\varepsilon_{xx}^\textrm{ext}\right]$ for  plane strain condition. External load represented with the Airy stress function $\chi_\textrm{ext}(x,y)=\frac{1}{2}\sigma_{xx}^\textrm{ext} y^2 + \frac{1}{2}\sigma_{yy}^\textrm{ext} x^2$ then induces elastic multipoles at the centers of inclusions and holes. The Airy stress function $\chi_{\text{out}}\big(r_i,\varphi_i|\mbf{a}_{i,\text{out}}\big)$ outside the $i^{\textrm{th}}$ inclusion due to the induced multipoles can be expanded as shown in Eq.~(\ref{eq:AiryInducedOut}), where the origin of polar coordinates $(r_i,\varphi_i)$ is at the center $\mbf{x}_i$ of the $i^\text{th}$ inclusion and 
the set of amplitudes of induced multipoles is $\mbf{a}_{i,\text{out}}=\{A_{i,0}, A_{i,1},\dots,B_{i,1}, B_{i,2},\dots, C_{i,2}, C_{i,3},\dots, D_{i,2}, D_{i,3},\dots\}$. The induced elastic multipoles at the center of the $i^\text{th}$ inclusion then further induce image multipoles at position $\mbf{x}_i^*=(x_i^*,y_i^*)=(x_i,y_b+h_i)$ to satisfy  boundary conditions at the edge ($u_{x}(x,y_b)=x \varepsilon_{xx}^\textrm{ext}$,   $u_{y}(x,y_b)=y_b \varepsilon_{yy}^\textrm{ext}$). The Airy stress function $\mathcal{I}\left[\chi_{\text{out}}\big(r_i,\varphi_i|\mbf{a}_{i,\text{out}}\big)\right]$ of image multipoles due to the $i^\text{th}$ inclusion can be expanded as shown in Eq.~(\ref{eq:AiryInducedOutImage2}), where the set of amplitudes of image multipoles is $\mbf{a}^*_{i,\text{out}}=\{A^*_{i,0}, A^*_{i,1},\dots,B^*_{i,0}, B^*_{i,1},\dots, C^*_{i,2}, C^*_{i,3},\dots, D^*_{i,2}, D^*_{i,3},\dots\}$.
 The amplitudes $\mbf{a}^*_{i,\text{out}}$ of image multipoles can be obtained with the help of the image operator in Eq.~(\ref{eq:AdeerogbaOp}) and they are related to the amplitudes of induced multipoles $\mbf{a}_{i,\text{out}}$ for the $i^\text{th}$ inclusion as
\begin{subequations}
\small
\begin{align}
    A_{i,n}^{*} &= \begin{cases}
    +\frac{1}{\kappa_0}A_{i,0}+\left(-\frac{1}{\kappa_0}+\kappa_0\right)C_{i,2}, & n=0, \\
    +\frac{(n+1)}{\kappa_0}A_{i,n}+\frac{1}{n}\left(\frac{(n+1)^2}{\kappa_0}-\kappa_0\right)C_{i,n+2} +\frac{2(n-1)}{\kappa_0}\frac{h_i}{R_i}B_{i,n-1}+\frac{2(2n+1)}{\kappa_0}\frac{h_i}{R_i}D_{i,n+1} -\frac{4(n-1)}{\kappa_0}\frac{h_i^2}{R_i^2}C_{i,n}, &  n\geq 1,
    \end{cases}\\
    B_{i,n}^{*} &= \begin{cases}
    \left(-\frac{1}{\kappa_0}+\kappa_0\right)D_{i,2}, & n=0,\\
    -\frac{2}{\kappa_0}B_{i,1} -\left(\frac{4}{\kappa_0}-\kappa_0\right)D_{i,3} -\frac{2}{\kappa_0}\frac{h_i}{R_i}A_{i,0} +\frac{6}{\kappa_0}\frac{h_i}{R_i}C_{i,2}, & n = 1,\\
    -\frac{(n+1)}{\kappa_0}B_{i,n} -\frac{1}{n}\left(\frac{(n+1)^2}{\kappa_0}-\kappa_0\right)D_{i,n+2} +\frac{2(n-1)}{\kappa_0}\frac{h_i}{R_i}A_{i,n-1} +\frac{2(2n+1)}{\kappa_0}\frac{h_i}{R_i}C_{i,n+1} +\frac{4(n-1)}{\kappa_0}\frac{h_i^2}{R_i^2}D_{i,n}, & n\geq 2,
    \end{cases}\\
    C_{i,n}^{*} &= \begin{cases}
    -\frac{1}{\kappa_0} C_{i,2} +\frac{1}{\kappa_0} A_{i,0}, & n = 2,\\
    -\frac{(n-1)}{\kappa_0}C_{i,n}-\frac{(n-2)}{\kappa_0}A_{i,n-2}-\frac{2(n-2)}{\kappa_0} \frac{h_i}{R_i} D_{i,n-1}, {\hskip 67mm} & n\geq 3,
    \end{cases}\\
    D_{i,n}^{*} &=\begin{cases} +\frac{(n-1)}{\kappa_0}D_{i,n}+\frac{(n-2)}{\kappa_0}B_{i,n-2}-\frac{2(n-2)}{\kappa_0} \frac{h_i}{R_i}C_{i,n-1},  {\hskip 67.5mm} & n\geq 2. \end{cases}
\end{align}
\end{subequations}
The total Airy stress function $\chi^{\text{tot}}_{\text{out}}\big(x,y|\mbf{a}_{\text{out}}\big)$ outside all inclusions is then the sum of contributions due to external load, $\chi_{\text{ext}}(x,y)$, due to induced multipoles at the center of the $i^\text{th}$ inclusion, $\chi_{\text{out}}\big(r_i,\varphi_i|\mbf{a}_{i,\text{out}}\big)$, and due to their images, $\mathcal{I}\left[\chi_{\text{out}}\big(r_i,\varphi_i|\mbf{a}_{i,\text{out}}\big)\right]$, as shown in Eq.~(\ref{eq:AiryOut}). We also expand the induced Airy stress function inside the $i^\text{th}$ inclusion $\chi_{\text{in}}\big(r_i,\varphi_i|\mbf{a}_{i,\text{in}}\big)$, as shown in Eq.~(\ref{eq:AiryInducedIn}), and define the total Airy stress function $\chi^\text{tot}_{\text{in}}\big(r_i,\varphi_i|\mbf{a}_{i,\text{in}}\big)$ inside the $i^\text{th}$ inclusion, which also includes the effect of external load, $\chi_{\text{ext}}(x,y)$ as shown in Eq.~(\ref{eq:AiryIn}). The rest of the steps are the same as those described in Section~\ref{sec:flat_edge_traction}. The amplitudes of induced multipoles $\mbf{a}_{i,\text{out}}$ and $\mbf{a}_{i,\text{in}}$ are obtained by satisfying the boundary conditions that tractions and displacements are continuous across the circumference of each inclusion in Eq.~(\ref{eq:BC}), which can be converted to the matrix equation similar to that in Eq.~(\ref{eq:matrix}). 

\begin{figure}[!t]
  \centering
  \includegraphics[scale=1]{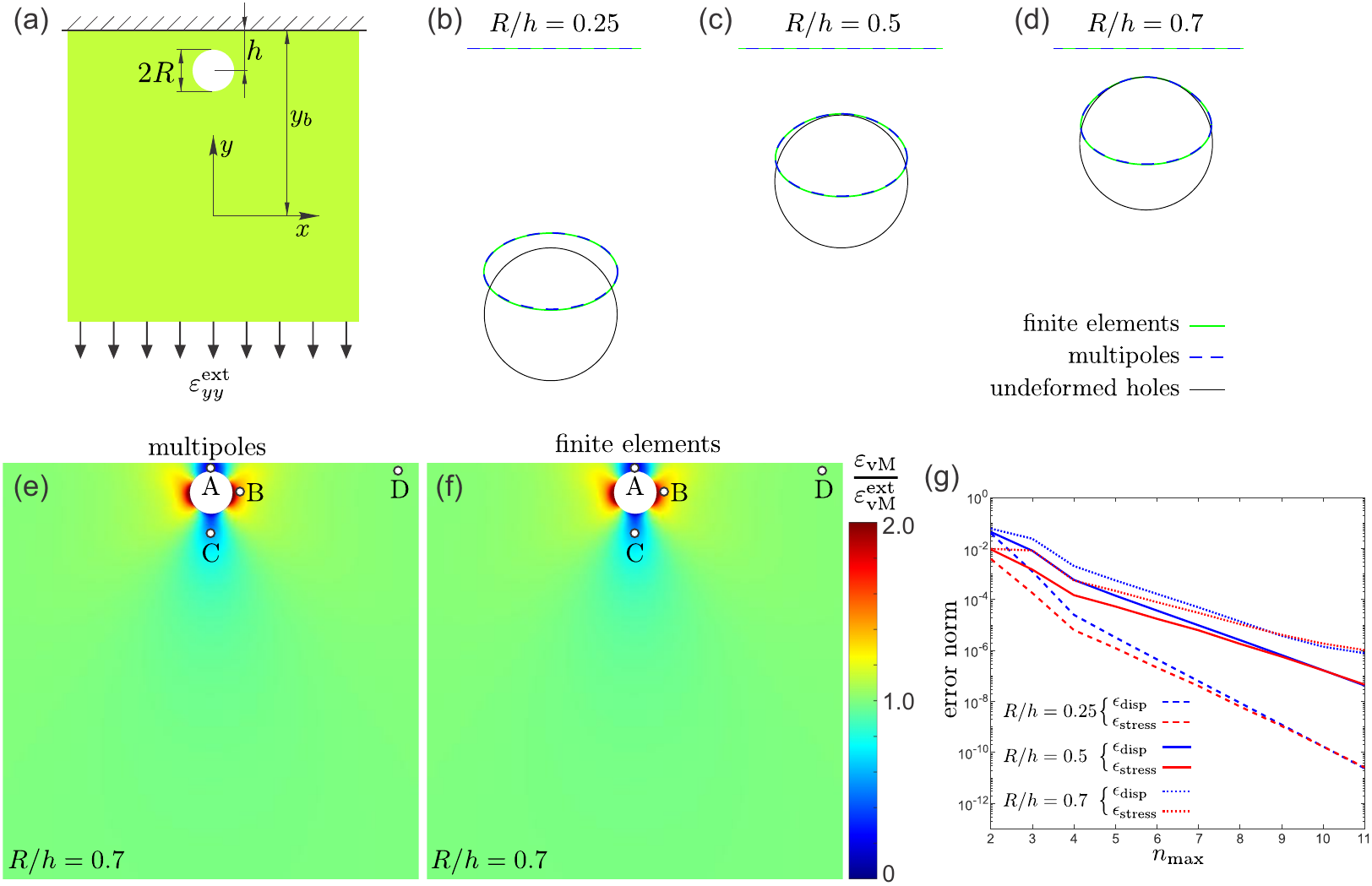}
  \caption{Deformation of a circular hole near a straight rigid edge under external strain. (a)~Schematic image describing the initial undeformed shape of the structure and the direction of applied strain $\varepsilon_{yy}^{\text{ext}}=-0.15$ ($\varepsilon_{xx}^{\text{ext}}=0$). The Poisson's ratio for the elastic matrix was $\nu_0=0.30$. (b-d)~Contours of deformed holes for $R/h=0.25$, $0.5$, and $0.7$, where $R$ is the hole radius and $h$ is the distance between the center of the hole and the rigid edge. Dashed blue lines show the contours obtained with elastic multipole method for $n_\text{max}=10$.  Green solid lines show the contours obtained with linear finite element simulations. As a reference, we include black solid lines that correspond to the undeformed configurations of  holes at their original positions. (e-f)~Equivalent von Mises strain fields $\varepsilon_\text{vM}$ for the case with $R/h=0.7$ were obtained with (e)~elastic multipole method ($n_\text{max}=10$) and  (f)~linear finite element simulations. Strain fields were normalized  with the value of the equivalent von Mises strain $\varepsilon_\text{vM}^\text{ext}=|\varepsilon_{yy}^{\text{ext}}|$ imposed by external loads. Four marked points A-D were chosen for the quantitative comparison of strains $\varepsilon_\text{vM}$. See Table~\ref{tab:rigid_edge} for details. (g)~The normalized errors for displacements $\epsilon_\text{disp}(n_\text{max})$ (blue lines) and stresses $\epsilon_\text{stress}(n_\text{max})$ (red lines) are defined in Eq.~(\ref{eq:error}).
  }
\label{Fig:flat_rigid_edge}
\end{figure}

The elastic multipole method described above was tested for one circular hole of radius $R$ embedded in a semi-infinite plate ($y<0$)  subjected to uniaxial strain $\varepsilon_{yy}^{\text{ext}} = -0.15$ under plane stress condition (see Fig.~\ref{Fig:flat_rigid_edge}a).  Three different values of the separation distance $h$ between the center of the hole and the rigid edge were considered: $h=4R$, $h=2R$, and $h=R/0.7$. In Fig.~\ref{Fig:flat_rigid_edge}b-d, we show the contours of deformed holes for different values of the separation distance $h$ of the hole from the rigid edge, where we set $n_\text{max}=10$. When the hole is far away from the edge ($h=4R$) its deformed shape is elliptical because the interaction of hole with its image is very weak (see Fig.~\ref{Fig:flat_rigid_edge}b). However, when the hole is moved closer to the rigid edge ($h=2R$, $h=R/0.7$), its deformed shape deviates from the ellipse, because the hole interacts more strongly with its image (see Fig.~\ref{Fig:flat_rigid_edge}c,d). In particular, for $h=R/0.7$, the portion of  the contour of the deformed hole facing the rigid edge overlaps with that of the undeformed hole, whereas the portion of the circumference of the deformed hole facing away from the rigid edge is still elliptical. This occurs because displacements  near the rigid edge are very small.

The normalized errors obtained from the convergence analysis for displacements $\epsilon_\text{disp}(n_\text{max})$ and stresses $\epsilon_\text{stress}(n_\text{max})$ are evaluated according to  Eq.~(\ref{eq:error}) and they are plotted in Fig.~\ref{Fig:flat_rigid_edge}g. As the maximum degree $n_\text{max}$ of induced multipoles is increased, the normalized errors decrease exponentially and they decrease more slowly when the hole is brought close to the rigid edge and their interaction with its image becomes important (see Fig.~\ref{Fig:flat_rigid_edge}d).

Results from the elastic multipole method are compared with linear finite element simulations on a rectangular domain of size $800R \times 400R$ with prescribed displacements along the boundary (see Appendix~\ref{app:FEM} for details).  The contours of deformed holes in simulations matched very well with those obtained with the elastic multipole method with $n_\text{max}=10$ (Fig.~\ref{Fig:flat_rigid_edge}b-d). We  also compared the equivalent von Mises strain fields $\varepsilon_\text{vM}$ (see Eq.~(\ref{eq:vonMisesStrain})) obtained with elastic multipole method (Fig.~\ref{Fig:flat_rigid_edge}e) and linear finite elements (Fig.~\ref{Fig:flat_rigid_edge}f). The agreement between the two strain fields is very good and the quantitative comparison of strains at four different points A-D (marked in Fig.~\ref{Fig:flat_rigid_edge}e,f) showed a relative error of only $\sim0.1\%$ (see Table~\ref{tab:rigid_edge}), which is much smaller than errors for the example with one hole near the traction-free edge in Section~\ref{sec:flat_edge_traction}. This is because the sample of size $\approx 25 R \times 25 R$ was used for linear finite element simulations to match the size of the experimental sample for the traction-free edge case, but here we used a much larger  sample of size $800 R \times 400 R$, which better approximates semi-infinite elastic matrix that is analyzed with the elastic multipole method. Moreover, the hole was further away from the rigid edge than for the traction-free case in Section~\ref{sec:flat_edge_traction}, which results in lower amplitudes of induced higher-order multipoles.

\begin{table}[!t]
\centering
\caption{Quantitative comparison for the values of equivalent von Mises strains $\varepsilon_\text{vM}$ normalized with the value for the applied external load $\varepsilon_\text{vM}^\text{ext}=|\varepsilon_{yy}^{\text{ext}}|$ at points A-D (defined in Fig.~\ref{Fig:flat_rigid_edge}) in compressed samples with one hole near a rigid edge obtained with elastic multipole method (EMP) and finite element simulations (FEM). The relative percent errors between  EMP and  FEM were calculated as $100\times(\varepsilon_\text{vM}^\text{(EMP)}-\varepsilon_\text{vM}^\text{(FEM)})/\varepsilon_\text{vM}^\text{(FEM)}$.}
\def\arraystretch{1.1}
\label{tab:rigid_edge}
\begin{tabular}{|>{\centering}m{0.05\textwidth}|>{\centering}m{0.06\textwidth}>{\centering}m{0.06\textwidth}>{\centering\arraybackslash}m{0.034\textwidth}|}
\hline
\multirow{2}{*}{points} & \multicolumn{3}{c|}{strain $\varepsilon_\text{vM}/\varepsilon_\text{vM}^\text{ext}$}\\
\cline{2-4}
 & EMP   & FEM   & $\epsilon(\%)$\\
\hline
A & 0.22208 & 0.22232 & 0.11\\
B & 1.48299 & 1.48307 & 0.02\\
C & 0.55360 & 0.55369 & 0.01\\
D & 1.01509 & 1.01508 & 0.00\\
\hline            
\end{tabular}
\end{table}

Note that we  tried to perform experiments for this case as well by gluing the edge to aluminum plates. Upon compression of the sample, we observed a  pronounced out-of-plane deformation in the vicinity of the hole, which is not captured by the elastic multipole method and by  2D linear finite element simulations. Hence, we skipped the comparison with the experiment in this case.

\subsection{Inclusions in deformed elastic disks}
\label{sec:disk}

In this section we consider the deformation of an elastic disk with embedded circular holes and inclusions by prescribing three different types of boundary conditions at the outer edge of the disk: hydrostatic stress ($\sigma_{rr}\ne0$, $\sigma_{r\varphi}=0$), no-slip condition ($u_{r}\ne0$, $u_{\varphi}=0$), and slip condition ($u_{r}\ne 0$, $\sigma_{r\varphi}=0$). As  described in  previous sections, the deformation of the elastic matrix induces elastic multipoles at the centers of holes and inclusions, which further induce image multipoles to satisfy boundary conditions at the outer edge of the disk.  

\begin{figure}[!t]
\centering
\includegraphics[scale=1]{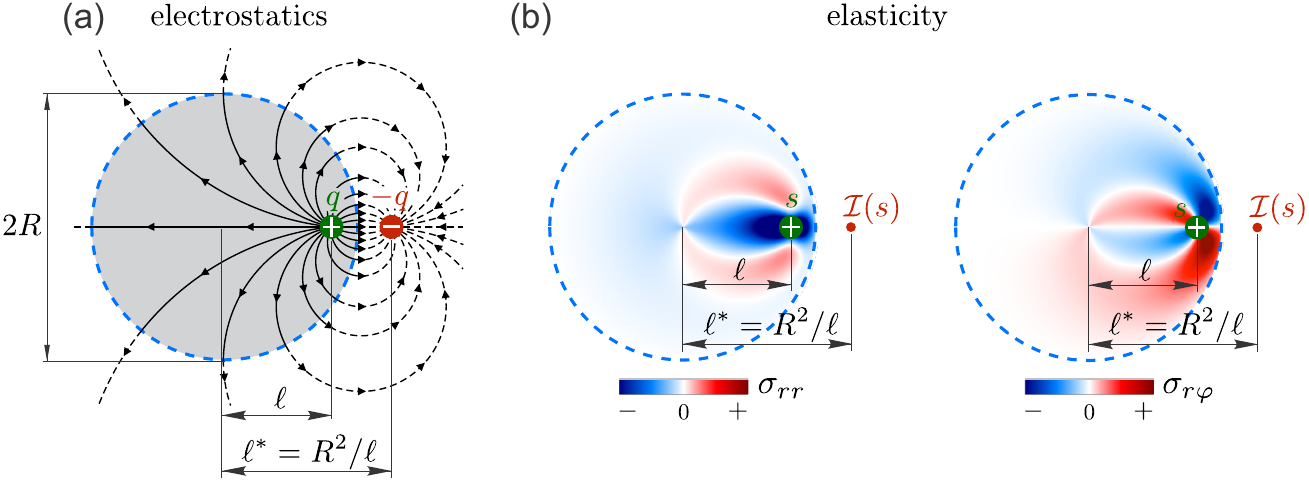}
  \caption{Image charges in electrostatics and elasticity for circular geometry. (a)~A point electric charge $q$ (green) at distance $\ell$ from the center of a circular conductive shell with radius $R$ (dashed blue line) induces an opposite image charge $\mathcal{I}(q)=-q$ (red) at the inverse point at distance $\ell^*=R^2/\ell$ to satisfy the boundary condition that the electric field lines (black lines) are orthogonal to the conductive plate. (b)~A disclination with charge $s$ (green) embedded in an elastic disk near a traction-free edge induces an image charge $\mathcal{I}(s)$ (red) at the inverse point to satisfy the boundary conditions at the edge, where tractions must vanish ($\sigma_{rr}=\sigma_{r\varphi}=0$).}
\label{Fig:ImageIllus_circle}
\end{figure}

Image charges in circular geometries have been considered in electrostatics, where it is known that a charge $q$ inside a conductive circular shell of radius $R$ induces an image charge $-q$ at the inverse point (see Fig.~\ref{Fig:ImageIllus_circle}a)~\cite{Jackson,Smythe}. Both the charge and the image charge are located along the same radial line and their distances to the center of the conductive circular shell are $\ell$ and $\ell^*=R^2/\ell$, respectively. Similarly, a disclination with charge $s$ embedded inside an elastic disk induces image disclination at the inverse point to satisfy boundary conditions at the edge of the elastic disk (see Fig.~\ref{Fig:ImageIllus_circle}b). Using the method described below, we can show that for a disclination with the Airy stress function $\chi_m(r_0,\varphi_0|s) = \frac{E_0s}{8\pi} r_0^2(\ln r_0-1/2)$ embedded inside an elastic disk with Young's modulus $E_0$ and traction-free boundary ($\sigma_{rr}=\sigma_{r\varphi}=0$), the induced image charge is described with the Airy stress function $\mathcal{I}[\chi_m(r_0,\varphi_0|s)] = \frac{E_0s}{8\pi} \left[-{r_0^*}^2(\ln r_0^*-1/2)-2 (\ell^* - \ell ) r_0^* \ln r_0^* \cos (\varphi^*-\theta) - (\ell^* - \ell)^2 \ln r_0^* - (1 - \ell^2/R^2 + 2 \ln (\ell/R)){r_0^*}^2/2\right]$. Here, the origins of polar coordinates $(r,\varphi)$, $(r_0,\varphi_0)$, and $(r_0^*,\varphi_0^*)$ are at the center of the elastic disk, at the position of disclination $(r=\ell,\varphi=\theta)$, and at the position of its image $(r=\ell^*,\varphi=\theta)$, respectively. The expressions for the Airy stress functions for images of all other multipoles can then be obtained similarly as discussed in  Section~\ref{sec:images} or by using the compact expression for the image operator $\mathcal{I}$ that was derived by Ogbonna in Ref.~\cite{Ogbonna}. While compact analytical expressions for the images of disclination are possible to obtain for traction-free and no-slip boundary conditions, this is not possible for the slip boundary condition. Below we describe how images multipoles  can be systematically obtained for all boundary conditions by following the standard procedure from electrostatics~\cite{Smythe}.

Now, let us consider a 2D elastic disk of radius $R$ with the Young's modulus $E_0$ and the Poisson's ratio $\nu_0$. Embedded in the disk are $N$ circular inclusions of radii $R_i$ with  Young's moduli $E_i$, Poisson's ratios $\nu_i$, where $i\in\{1,\ldots, N\}$. They are centered at positions $\mbf{x}_i=(r=\ell_i,\varphi=\theta_i)$, where $\ell_i$ is the distance from the center of  the $i^\text{th}$ inclusion to the center of the disk. Holes are described with the zero Young's modulus ($E_i=0$). We consider 3 different loading conditions: hydrostatic stress ($\sigma_{rr}(r=R,\varphi)=\sigma_{rr}^\text{ext}$, $\sigma_{r\varphi}(r=R,\varphi)=0$), no-slip condition ($u_{r}(r=R,\varphi)=u_r^\text{ext}$, $u_{\varphi}(r=R,\varphi)=0$), and slip condition ($u_{r}(r=R,\varphi)=u_r^\text{ext}$, $\sigma_{r\varphi}(r=R,\varphi)=0$). In the absence of inclusions, all 3 loading conditions are equivalent and they produce isotropic deformation of disk described with the Airy stress function $\chi_\text{ext}(r,\varphi)=\frac{1}{2} \sigma_{rr}^\text{ext} r^2$, where the radial displacement $u_r^\text{ext}$ at the edge of the disk is related to the radial stress $\sigma_{rr}^\text{ext}$ as $u_r^\text{ext}=(\kappa_0-1)\sigma_{rr}^\text{ext} R/(4 \mu_0)$. Here $\mu_0=E_0/[2(1+\nu_0)]$ is the shear modulus and the value of Kolosov's constant $\kappa_0$ for  plane stress is $\kappa_0=(3-\nu_0)/(1+\nu_0)$ and for  plane strain is  $\kappa_0=3-4\nu_0$.

External stress induces elastic multipoles at the centers of inclusions.  Thus the Airy stress function outside the $i^{\text{th}}$ inclusion due to the induced multipoles can be expanded as
\begin{equation}
  \label{eq:AiryInducedOutDisk}
  \begin{split}
  \chi_{\text{out}}\big(r_i,\varphi_i|\mbf{a}_{i,\text{out}}\big) = &A_{i,0} R_i^2 \ln \left(\frac{r_i}{R_i}\right) + \sum_{n=1}^\infty R_i^2 \left[ A_{i,n} \left(\frac{r_i}{R_i}\right)^{-n}\cos(n\varphi_i) +B_{i,n}
  \left(\frac{r_i}{R_i}\right)^{-n}\sin(n \varphi_i)\right]\\
  &\quad \quad \quad \quad \quad\quad\quad  +\sum_{n=2}^{\infty} R_i^2 \left[ C_{i,n} \left(\frac{r_i}{R_i}\right)^{-n+2}\cos (n
  \varphi_i) + D_{i,n}
  \left(\frac{r_i}{R_i}\right)^{-n+2}\sin (n \varphi_i)\right],
  \end{split}
\end{equation}
where the origin of polar coordinates $(r_i,\varphi_i)$ is at the center $\mbf{x}_i$ of the $i^\text{th}$ inclusion with radius $R_i$ and the set of amplitudes of induced multipoles is $\mbf{a}_{i,\text{out}}=\{A_{i,0}, A_{i,1},\dots,B_{i,1}, B_{i,2},\dots, C_{i,2}, C_{i,3},\dots, D_{i,2}, D_{i,3},\dots\}$. To satisfy boundary conditions at the outer edge of  disk these multipoles further induces image multipoles $\mathcal{I}\left[\chi_{\text{out}}\big(r_i,\varphi_i|\mbf{a}_{i,\text{out}}\big)\right]$. Since the boundary conditions have to be satisfied at the outer edge of the elastic disk, it is more convenient to expand all Airy stress functions with respect to the polar coordinates $(r,\varphi)$ centered at the origin of the disk. Polar coordinates $(r_i,\varphi_i)$ centered at the $i^\text{th}$ inclusion can be expressed in terms of polar coordinates centered at the disk as $r_i(r,\varphi) =\big(r^2+\ell_{i}^2-2 r \ell_{i} \cos(\varphi - \theta_{i})\big)^{1/2}$ and
$\varphi_i(r,\varphi) =\pi +\theta_{i} - \arctan\left[\big(r \sin(\varphi  - \theta_{i})\big)/\big(\ell-r \cos(\varphi  - \theta_{i})\big)\right]$. Since we are interested in the region where $r\rightarrow R$ we expand the Airy stress functions $\chi_{\text{out}}\big(r_i(r,\varphi),\varphi_i(r,\varphi)|\mbf{a}_{i,\text{out}}\big)$ in terms of powers of $\ell_i/r<1$ as
\begin{align}
  \chi_{\text{out}}\big(r_i(r,\varphi),\varphi_i(r,\varphi)|\mbf{a}_{i,\text{out}}\big)=&\chi_{\text{out}}\big(r,\varphi|\mbf{a}^d_{i,\text{out}}\big),\\
  \begin{split}
  \chi_{\text{out}}\big(r_i(r,\varphi),\varphi_i(r,\varphi)|\mbf{a}_{i,\text{out}}\big)=& A^d_{i,0} R_i^2 \ln \left(\frac{r}{R_i}\right) + \sum_{n=1}^\infty R_i^2 \left[ A^d_{i,n} \left(\frac{r}{R_i}\right)^{-n}\cos(n\varphi) +B^d_{i,n}
  \left(\frac{r}{R_i}\right)^{-n}\sin(n \varphi)\right]\\
  &\quad \quad \quad \quad \quad\quad\quad  +\sum_{n=2}^{\infty} R_i^2 \left[ C^d_{i,n} \left(\frac{r}{R_i}\right)^{-n+2}\cos (n
  \varphi) + D^d_{i,n}
  \left(\frac{r}{R_i}\right)^{-n+2}\sin (n \varphi)\right],
  \end{split}\nonumber
\end{align}
where the amplitudes $\mbf{a}^d_{i,\text{out}}=\{A^d_{i,0}, A^d_{i,1},\dots,B^d_{i,1}, B^d_{i,2},\dots, C^d_{i,2}, C^d_{i,3},\dots, D^d_{i,2}, C^d_{i,3},\dots\}$ relative to the center of  disk are related to the amplitudes $\mbf{a}_{i,\text{out}}$ relative to the center of the $i^\text{th}$ inclusion as
\begin{subequations}
\label{eq:disk_amplitudes_disk_inclusion}
\begin{align}
    A_{i,n}^{d} & = \begin{cases}
    A_{i,0}, & n = 0, \vspace{4mm}\\
    \begin{aligned}
    &+\sum_{m=1}^{n} {n-1 \choose m-1}\left(\frac{\ell_i}{R_i}\right)^{n-m}\Big[A_{i,m}\cos \big((n-m)\theta_i\big)-B_{i,m}\sin \big((n-m)\theta_i\big)\Big]\\
    &+\sum_{m=2}^{n+1} {n-1 \choose m-2}\left(\frac{\ell_i}{R_i}\right)^{n-m+2}\Big[-C_{i,m}\cos \big((n-m)\theta_i\big)+D_{i,m}\sin \big((n-m)\theta_i\big)\Big]\\
    &-\frac{1}{n}\left(\frac{\ell_i}{R_i}\right)^nA_{i,0}\cos (n \theta_i),
    \end{aligned} & n \geq 1,
    \end{cases}\\
    B_{i,n}^{d} & = \begin{cases}
    \begin{aligned}
    &+\sum_{m=1}^{n} {n-1 \choose m-1}\left(\frac{\ell_i}{R_i}\right)^{n-m}\Big[A_{i,m}\sin \big((n-m)\theta_i\big)+B_{i,m}\cos \big((n-m)\theta_i\big)\Big]\\
    &+\sum_{m=2}^{n+1} {n-1 \choose m-2}\left(\frac{\ell_i}{R_i}\right)^{n-m+2}\Big[-C_{i,m}\sin \big((n-m)\theta_i\big)-D_{i,m}\cos \big((n-m)\theta_i\big)\Big]\\
    &-\frac{1}{n}\left(\frac{\ell_i}{R_i}\right)^n A_{i,0}\sin (n\theta_i),
    \end{aligned} & n \geq 1
    \end{cases},\\
    C_{i,n}^{d} & = \begin{cases}
    \begin{aligned}
    \sum_{m=2}^{n} {n-2 \choose m-2}\left(\frac{\ell_i}{R_i}\right)^{n-m}\Big[C_{i,m}\cos \big((n-m)\theta_i\big)-D_{i,m}\sin \big((n-m)\theta_i\big)\Big],
    \end{aligned} {\hskip 12mm} & n \geq 2
    \end{cases},\\
    D_{i,n} & = \begin{cases}
    \begin{aligned}
    \sum_{m=2}^{n} {n-2 \choose m-2}\left(\frac{\ell_i}{R_i}\right)^{n-m}\Big[C_{i,m}\sin \big((n-m)\theta_i\big)+D_{i,m}\cos \big((n-m)\theta_i\big)\Big],
    \end{aligned} {\hskip 12mm} & n \geq 2
    \end{cases}.
\end{align}
\end{subequations}
Image multipoles $\mathcal{I}[\chi_{\text{out}}\big(r_i(r,\varphi),\varphi_i(r,\varphi)|\mbf{a}_{i,\text{out}}\big)]$ for the $i^\text{th}$ inclusion are at positions $\mbf{x}^*_i=(r=\ell_i^*,\varphi=\theta_i)$, where $\ell_i^*=R^2/\ell_i > R$. Thus their Airy stress function is expanded in terms of powers $r/\ell_i^*$ as
  \begin{align}
  \mathcal{I}[\chi_{\text{out}}\big(r_i(r,\varphi),\varphi_i(r,\varphi)|\mbf{a}_{i,\text{out}}\big)]=&\chi_{\text{in}}\big(r,\varphi|\mbf{a}^{*d}_{i,\text{out}}\big),\nonumber \\
  \begin{split}
  \mathcal{I}[\chi_{\text{out}}\big(r_i(r,\varphi),\varphi_i(r,\varphi)|\mbf{a}_{i,\text{out}}\big)]=& c^{*d}_{i,0} r^2 +\sum_{n=2}^{\infty} R_i^2 \left[a^{*d}_{i,n} \left(\frac{r}{R_i}\right)^n\cos(n\varphi)+ b^{*d}_{i,n}
  \left(\frac{r}{R_i}\right)^n\sin(n\varphi)\right]\\
  &\quad \quad\    +\sum_{n=1}^{\infty} R_i^2 \left[c^{*d}_{i,n} \left(\frac{r}{R_i}\right)^{n+2}\cos
  (n\varphi_i) + d^{*d}_{i,n}
  \left(\frac{r}{R_i}\right)^{n+2}\sin (n\varphi) \right],
  \end{split}
  \end{align}
where the amplitudes of image multipoles relative to the disk center are $\mbf{a}^{*d}_{i,\text{out}}=\{a^{*d}_{i,2}, a^{*d}_{i,3},\dots, b^{*d}_{i,2}, b^{*d}_{i,3},\dots, c^{*d}_{i,0}, c^{*d}_{i,1},\dots, d^{*d}_{i,1}, d^{*d}_{i,2},\dots\}$ and factors of $\ell_i^*/R_i$ are absorbed in the amplitudes $\mbf{a}^{*d}_{i,\text{out}}$. The amplitudes of image multipoles $\mbf{a}^{*d}_{i,\text{out}}$ are related to the amplitudes of induced multipoles $\mbf{a}^{d}_{i,\text{out}}$, such that the boundary conditions for tractions and displacements at the edge of the disk are satisfied. To evaluate tractions and displacements at the edge we define the total Airy stress function outside all inclusions as
\begin{align}
\chi^{\text{tot}}_{\text{out}}\big(r,\varphi|\mbf{a}_{\text{out}}\big)=&\externalStress{\chi_{\text{ext}}(r,\varphi)}+\sum_{i=1}^N \inclusionI{\chi_{\text{out}}\big(r_i(r,\varphi),\varphi_i(r,\varphi)|\mbf{a}_{i,\text{out}}\big)}+ \sum_{i=1}^N \imageInclusion{\mathcal{I}\left[\chi_{\text{out}}\big(r_i(r,\varphi),\varphi_i(r,\varphi)|\mbf{a}_{i,\text{out}}\big)\right]},\nonumber\\
\begin{split}
    \chi^{\text{tot}}_{\text{out}}\big(r,\varphi|\mbf{a}_{\text{out}}\big) = & \externalStress{\chi_{\text{ext}}(r,\varphi)} +
    \sum_{i=1}^N R_i^2 \left[\inclusionI{A_{i,0}^{d} \ln \left(\frac{r}{R_i}\right)} + \imageInclusion{c_{i,0}^{*d}\left(\frac{r}{R_i}\right)^2} \right]\\
    &+\sum_{i=1}^N\sum_{n=1}^{\infty}R_i^2\left[\inclusionI{A_{i,n}^{d}\left(\frac{r}{R_i}\right)^{-n}+C_{i,n}^{d}\left(\frac{r}{R_i}\right)^{-n+2}}+\imageInclusion{a_{i,n}^{*d}\left(\frac{r}{R_i}\right)^{n}+c_{i,n}^{*d}\left(\frac{r}{R_i}\right)^{n+2}}\right]\cos (n\varphi)\\
    &+\sum_{i=1}^N\sum_{n=1}^{\infty}R_i^2\left[\inclusionI{B_{i,n}^{d}\left(\frac{r}{R_i}\right)^{-n}+D_{i,n}^{d}\left(\frac{r}{R_i}\right)^{-n+2}}+\imageInclusion{b_{i,n}^{*d}\left(\frac{r}{R_i}\right)^{n}+d_{i,n}^{*d}\left(\frac{r}{R_i}\right)^{n+2}}\right]\sin (n\varphi),
\end{split}
\end{align}
where we set $a_{i,1}^{*d}=b_{i,1}^{*d}=C_{i,1}^d=D_{i,1}^d=0$. With the help of Table~\ref{table:MichellStressDisplacement}, we then evaluate tractions and displacements as
\begin{subequations}
\small
\begin{align}
\label{eq:disk_tractionRR}
\sigma_{\text{tot},rr}^\text{out}(r=R,\varphi|\mbf{a}_{\text{out}}) = & \externalStress{\sigma_{rr}^\text{ext}} + \sum_{i=1}^N \left[\inclusionI{A_{i,0}^{d} \left(\frac{R}{R_i}\right)^{-2}} + \imageInclusion{ 2c_{i,0}^{*d}}\right]\nonumber\\
    &\hspace{0cm}+\sum_{i=1}^N\sum_{n=1}^{\infty}\Bigg[\inclusionI{-n(n+1)A_{i,n}^{d}\left(\frac{R}{R_i}\right)^{-n-2}-(n-1)(n+2)C_{i,n}^{d}\left(\frac{R}{R_i}\right)^{-n}}\nonumber\\
    &\hspace{2cm}\imageInclusion{-n(n-1)a_{i,n}^{*d}\left(\frac{R}{R_i}\right)^{n-2}-(n+1)(n-2)c_{i,n}^{*d}\left(\frac{R}{R_i}\right)^{n}}\Bigg]\cos (n\varphi)\nonumber\\
    &\hspace{0cm}+\sum_{i=1}^N\sum_{n=1}^{\infty}\Bigg[\inclusionI{-n(n+1)B_{i,n}^{d}\left(\frac{R}{R_i}\right)^{-n-2}-(n-1)(n+2)D_{i,n}^{d}\left(\frac{R}{R_i}\right)^{-n}}\nonumber\\
    &\hspace{2cm}\imageInclusion{-n(n-1)b_{i,n}^{*d}\left(\frac{R}{R_i}\right)^{n-2}-(n+1)(n-2)d_{i,n}^{*d}\left(\frac{R}{R_i}\right)^{n}}\Bigg]\sin (n\varphi),\\
\label{eq:disk_tractionRF}
\sigma_{\text{tot},r\varphi}^\text{out}(r=R,\varphi|\mbf{a}_{\text{out}}) = &+\sum_{i=1}^N\sum_{n=1}^{\infty}\Bigg[\inclusionI{-n(n+1)A_{i,n}^{d}\left(\frac{R}{R_i}\right)^{-n-2}-n(n-1)C_{i,n}^{d}\left(\frac{R}{R_i}\right)^{-n}}\nonumber\\
    &\hspace{2cm}\imageInclusion{+n(n-1)a_{i,n}^{*d}\left(\frac{R}{R_i}\right)^{n-2}+n(n+1)c_{i,n}^{*d}\left(\frac{R}{R_i}\right)^{n}}\Bigg]\sin (n\varphi)\nonumber\\
    &+\sum_{i=1}^N\sum_{n=1}^{\infty}\Bigg[\inclusionI{n(n+1)B_{i,n}^{d}\left(\frac{R}{R_i}\right)^{-n-2}+n(n-1)D_{i,n}^{d}\left(\frac{R}{R_i}\right)^{-n}}\nonumber\\
    &\hspace{2cm}\imageInclusion{-n(n-1)b_{i,n}^{*d}\left(\frac{R}{R_i}\right)^{n-2}-n(n+1)d_{i,n}^{*d}\left(\frac{R}{R_i}\right)^{n}}\Bigg]\cos (n\varphi),\\
\label{eq:disk_displacementR}
\frac{2\mu_0}{R}u_{\text{tot},r}^\text{out}(r=R,\varphi|\mbf{a}_{\text{out}})= &\frac{2\mu_0}{R}\left(\externalStress{u_r^\text{ext}}+u_x \cos \varphi + u_y \sin \varphi\right) +\sum_{i=1}^N\left[\inclusionI{-A_{i,0}^{d} \left(\frac{R}{R_i}\right)^{-2}} + \imageInclusion{ (\kappa_0-1)c_{i,0}^{*d}}\right]\nonumber\\
    &+\sum_{i=1}^N\sum_{n=1}^{\infty}\Bigg[\inclusionI{n A_{i,n}^{d}\left(\frac{R}{R_i}\right)^{-n-2}+(\kappa_0+n-1)C_{i,n}^{d}\left(\frac{R}{R_i}\right)^{-n}}\nonumber\\
    &\hspace{2cm}\imageInclusion{-na_{i,n}^{*d}\left(\frac{R}{R_i}\right)^{n-2}+(\kappa_0-n-1)c_{i,n}^{*d}\left(\frac{R}{R_i}\right)^{n}}\Bigg]\cos (n\varphi)\nonumber\\
    &+\sum_{i=1}^N\sum_{n=1}^{\infty}\Bigg[\inclusionI{nB_{i,n}^{d}\left(\frac{R}{R_i}\right)^{-n-2}+(\kappa_0+n-1)D_{i,n}^{d}\left(\frac{R}{R_i}\right)^{-n}}\nonumber\\
    &\hspace{2cm}\imageInclusion{-nb_{i,n}^{*d}\left(\frac{R}{R_i}\right)^{n-2}+(\kappa_0-n-1)d_{i,n}^{*d}\left(\frac{R}{R_i}\right)^{n}}\Bigg]\sin (n\varphi)\\
\label{eq:disk_displacementF}
\frac{2\mu_0}{R}u_{\text{tot},\varphi}^\text{out}(r=R,\varphi|\mbf{a}_{\text{out}}) = &\frac{2\mu_0}{R}\left(-u_x \sin\varphi +u_y\cos\varphi\right)\nonumber\\
    &+\sum_{i=1}^N\sum_{n=1}^{\infty}\Bigg[\inclusionI{n A_{i,n}^{d}\left(\frac{R}{R_i}\right)^{-n-2}-(\kappa_0-n+1)C_{i,n}^{d}\left(\frac{R}{R_i}\right)^{-n}}\nonumber\\
    &\hspace{2cm}\imageInclusion{+na_{i,n}^{*d}\left(\frac{R}{R_i}\right)^{n-2}+(\kappa_0+n+1)c_{i,n}^{*d}\left(\frac{R}{R_i}\right)^{n}}\Bigg]\sin (n\varphi)\nonumber\\
    &+\sum_{i=1}^N\sum_{n=1}^{\infty}\Bigg[\inclusionI{-nB_{i,n}^{d}\left(\frac{R}{R_i}\right)^{-n-2}+(\kappa_0-n+1)D_{i,n}^{d}\left(\frac{R}{R_i}\right)^{-n}}\nonumber\\
    &\hspace{2cm}\imageInclusion{-nb_{i,n}^{*d}\left(\frac{R}{R_i}\right)^{n-2}-(\kappa_0+n+1)d_{i,n}^{*d}\left(\frac{R}{R_i}\right)^{n}}\Bigg]\cos (n\varphi),
\end{align}
\end{subequations}
where  colors correspond to the  Airy stress functions \externalStress{$\chi_\textrm{ext}(r,\varphi)$}, \inclusionI{$\chi_{\text{out}}\big(r_i,\varphi_i|\mbf{a}_{i,\text{out}}\big)$}, and \imageInclusion{$\mathcal{I}\left[\chi_{\text{out}}\big(r_i,\varphi_i|\mbf{a}_{i,\text{out}}\big)\right]$}. We also included displacements  $u_x$ and $u_y$ due to a rigid body translation and introduced the shear modulus $\mu_i=E_i/[2(1+\nu_i)]$ and the Kolosov's constant $\kappa_i$ for the $i^\text{th}$ inclusion, where the value of Kolosov's constants is $\kappa_i=(3-\nu_i)/(1+\nu_i)$ for  plane stress and $\kappa_i=3-4\nu_i$ for  plane strain conditions~\cite{Barber}. Similarly, we define the shear modulus $\mu_0=E_0/[2(1+\nu_0)]$ and the Kolosov's constant $\kappa_0$ for the elastic matrix of the disk.

For the hydrostatic stress boundary condition, we require that $\sigma_{\text{tot},rr}^\text{out}(r=R,\varphi|\mbf{a}_{\text{out}})=\sigma_{rr}^\text{ext}$ in Eq.~(\ref{eq:disk_tractionRR}) and that $\sigma_{\text{tot},r\varphi}^\text{out}(r=R,\varphi|\mbf{a}_{\text{out}})=0$ in Eq.~(\ref{eq:disk_tractionRF}). From this system of equations we obtain 
\begin{subequations}
\begin{align}
    a_{i,n}^{*d} &= \begin{cases}\begin{aligned}
    -(n+1)\left(\frac{R_i}{R}\right)^{2n}A_{i,n}^{d}-n\left(\frac{R_i}{R}\right)^{2n-2}C_{i,n}^{d}\end{aligned}, & n \geq 2,
    \end{cases}\\
    b_{i,n}^{*d} &= \begin{cases}\begin{aligned}
    -(n+1)\left(\frac{R_i}{R}\right)^{2n}B_{i,n}^{d}-n\left(\frac{R_i}{R}\right)^{2n-2}D_{i,n}^{d}\end{aligned}, & n \geq 2,
    \end{cases}\\
    c_{i,n}^{*d} &= \begin{cases}\begin{aligned}
    -\frac{1}{2}\left(\frac{R_i}{R}\right)^{2}A_{i,0}^{d}\end{aligned}, & n = 0,\\
    \begin{aligned}n\left(\frac{R_i}{R}\right)^{2n+2}A_{i,n}^{d}+(n-1)\left(\frac{R_i}{R}\right)^{2n}C_{i,n}^{d}\end{aligned}, {\hskip 2.5mm} & n \geq 1,
    \end{cases}\\
    d_{i,n}^{*d} &= \begin{cases}\begin{aligned}
    n\left(\frac{R_i}{R}\right)^{2n+2}B_{i,n}^{d}+(n-1)\left(\frac{R_i}{R}\right)^{2n}D_{i,n}^{d}\end{aligned}, {\hskip 2.5mm} & n \geq 1.
    \end{cases}
\end{align}
\end{subequations}
For the no-slip boundary condition, we require that $u_{\text{tot},r}^\text{out}(r=R,\varphi|\mbf{a}_{\text{out}})=u_{r}^\text{ext}$ in Eq.~(\ref{eq:disk_displacementR}) and that $u_{\text{tot},\varphi}^\text{out}(r=R,\varphi|\mbf{a}_{\text{out}})=0$ in Eq.~(\ref{eq:disk_displacementF}). From this system of equations we obtain 
\begin{subequations}
\begin{align}
    a_{i,n}^{*d} &= \begin{cases}\begin{aligned}
    \frac{1}{\kappa_0}(n+1)\left(\frac{R_i}{R}\right)^{2n}A_{i,n}^{d}+\left(\frac{n}{\kappa_0}+\frac{1}{n}\left(\kappa_0-\frac{1}{\kappa_0}\right)\right)\left(\frac{R_i}{R}\right)^{2n-2}C_{i,n}^{d}\end{aligned}, & n \geq 2,
    \end{cases}\\
    b_{i,n}^{*d} &= \begin{cases}\begin{aligned}
    \frac{1}{\kappa_0}(n+1)\left(\frac{R_i}{R}\right)^{2n}B_{i,n}^{d}+\left(\frac{n}{\kappa_0}+\frac{1}{n}\left(\kappa_0-\frac{1}{\kappa_0}\right)\right)\left(\frac{R_i}{R}\right)^{2n-2}D_{i,n}^{d}\end{aligned}, & n \geq 2,
    \end{cases}\\
    c_{i,n}^{*d} &= \begin{cases}
    \begin{aligned}\frac{1}{(\kappa_0-1)}A_{i,0}^{d}\left(\frac{R_i}{R}\right)^{2}\end{aligned}, & n = 0,\\
    \begin{aligned}-\frac{1}{\kappa_0}n\left(\frac{R_i}{R}\right)^{2n+2}A_{i,n}^{d}-\frac{1}{\kappa_0}(n-1)\left(\frac{R_i}{R}\right)^{2n}C_{i,n}^{d}\end{aligned}, {\hskip 25.5mm}  & n \geq 1,
    \end{cases}\\
    d_{i,n}^{*d} &= \begin{cases}\begin{aligned}
     -\frac{1}{\kappa_0}n\left(\frac{R_i}{R}\right)^{2n+2}B_{i,n}^{d}-\frac{1}{\kappa_0}(n-1)\left(\frac{R_i}{R}\right)^{2n}D_{i,n}^{d}\end{aligned}, {\hskip 25.5mm}  & n \geq 1,
    \end{cases}\\
    u_x&=\sum_{i=1}^N 
    -\frac{2}{\kappa_0}\frac{R_i}{2\mu_0}\left(\frac{R_i}{R}\right)^2 A_{i,1}^d,\\
    u_y&=\sum_{i=1}^N 
    -\frac{2}{\kappa_0}\frac{R_i}{2\mu_0}\left(\frac{R_i}{R}\right)^2 B_{i,1}^d,
\end{align}
\end{subequations}
For the slip boundary conditions, we require that $u_{\text{tot},r}^\text{out}(r=R,\varphi|\mbf{a}_{\text{out}})=u_{r}^\text{ext}$ in Eq.~(\ref{eq:disk_displacementR}) and that $\sigma_{\text{tot},r\varphi}^\text{out}(r=R,\varphi|\mbf{a}_{\text{out}})=0$ in Eq.~(\ref{eq:disk_tractionRF}). From this system of equations we obtain 
\begin{subequations}
\begin{align}
    a_{i,n}^{*d} &= \begin{cases}\begin{aligned}
    \frac{(n+1)(\kappa_0-1)}{(\kappa_0(n-1)+n+1)}\left(\frac{R_i}{R}\right)^{2n}A_{i,n}^{d}+\frac{2n\kappa_0}{(\kappa_0(n-1)+n+1)}\left(\frac{R_i}{R}\right)^{2n-2}C_{i,n}^{d}\end{aligned}, & n \geq 2,
    \end{cases}\\
    b_{i,n}^{*d} &= \begin{cases}\begin{aligned}
    \frac{(n+1)(\kappa_0-1)}{(\kappa_0(n-1)+n+1)}\left(\frac{R_i}{R}\right)^{2n}B_{i,n}^{d}+\frac{2n\kappa_0}{(\kappa_0(n-1)+n+1)}\left(\frac{R_i}{R}\right)^{2n-2}D_{i,n}^{d}\end{aligned}, & n \geq 2,
    \end{cases}\\
    c_{i,n}^{*d} &= \begin{cases}\begin{aligned}
    \frac{1}{(\kappa_0-1)}A_{i,0}^{d}\left(\frac{R_i}{R}\right)^{2}\end{aligned}, & n = 0,\\
    \begin{aligned}\frac{2n}{(\kappa_0(n-1)+n+1)}\left(\frac{R_i}{R}\right)^{2n+2}A_{i,n}^{d}+\frac{(n-1)(1-\kappa_0)}{(\kappa_0(n-1)+n+1)}\left(\frac{R_i}{R}\right)^{2n}C_{i,n}^{d}\end{aligned}, & n \geq 1,
    \end{cases}\\
    d_{i,n}^{*d} &= \begin{cases}\begin{aligned}
     \frac{2n}{(\kappa_0(n-1)+n+1)}\left(\frac{R_i}{R}\right)^{2n+2}B_{i,n}^{d}+\frac{(n-1)(1-\kappa_0)}{(\kappa_0(n-1)+n+1)}\left(\frac{R_i}{R}\right)^{2n}D_{i,n}^{d}\end{aligned}, & n \geq 1,
    \end{cases}\\
    u_x&=\sum_{i=1}^N (1-\kappa_0)\frac{R_i}{2\mu_0}\left(\frac{R_i}{R}\right)^2 A_{i,1}^d,\\
    u_y&=\sum_{i=1}^N (1-\kappa_0)\frac{R_i}{2\mu_0}\left(\frac{R_i}{R}\right)^2 B_{i,1}^d.
\end{align}
\end{subequations}

Above we showed how the amplitudes for image multipoles $\mbf{a}^{*d}_{i,\text{out}}$ relative to the center of the disk are related to the amplitudes of induced multipoles $\mbf{a}^{d}_{i,\text{out}}$, which can be further expressed in terms of the amplitudes of induced multipoles $\mbf{a}_{i,\text{out}}$ relative to the center of the $i^\text{th}$ inclusion by using Eq.~(\ref{eq:disk_amplitudes_disk_inclusion}). In principle we could also expand the Airy stress function for the image multipoles $\mathcal{I}\left[\chi_{\text{out}}\big(r_i,\varphi_i|\mbf{a}_{i,\text{out}}\big)\right]=\chi^*_{\text{out}}\big(r_i^*,\varphi_i^*|\mbf{a}^*_{i,\text{out}}\big)$ in terms of polar coordinates $(r_i^*,\varphi_i^*)$ relative to the center $\mbf{x}_i^*$ of the image of the $i^\text{th}$ inclusion as shown in Eq.~(\ref{eq:AiryInducedOutImage2}) and we could also relate the amplitudes $\mbf{a}^{*d}_{i,\text{out}}$ to the amplitudes $\mbf{a}^*_{i,\text{out}}$. However, we omit this step, because we never actually need to evaluate the Airy stress functions $\mathcal{I}\left[\chi_{\text{out}}\big(r_j,\varphi_j|\mbf{a}_{j,\text{out}}\big)\right]$ with respect to $\mbf{x}_i^*$, but we need to expand them around the center $\mbf{x}_i$ of the  $i^\text{th}$ inclusion to evaluate boundary conditions at the edge of that inclusion as was discussed in details in Section~\ref{sec:flat_edge_traction}. The resulting expression is
\begin{align}
  \label{eq:AiryImageCircleGeneralTaylor}
  \mathcal{I}\left[\chi_{\text{out}}\big(r_j(r_i,\varphi_i),\varphi_j(r_i,\varphi_i)|\mbf{a}_{j,\text{out}}\big)\right]&=\chi_{\text{in}}\big(r_i,\varphi_i|\mbf{a}^{*i}_{j,\text{out}}\big), \\
  \begin{split}
  \mathcal{I}\left[\chi_{\text{out}}\big(r_j(r_i,\varphi_i),\varphi_j(r_i,\varphi_i)|\mbf{a}_{j,\text{out}}\big)\right]&=    \phantom{+c_{j,0}^{*i} r_i^2 +}\sum_{n=2}^{\infty} R_j^2 \left[a_{j,n}^{*i} \left(\frac{r_i}{R_j}\right)^n\cos(n\varphi_i)+ b_{j,n}^{*i} \left(\frac{r_i}{R_j}\right)^n
  \sin(n\varphi_i)\right]\nonumber \\
  &\phantom{=}+c_{j,0}^{*i} r_i^2 +\sum_{n=1}^{\infty} R_
  j^2 \left[c_{j,n}^{*i} \left(\frac{r_i}{R_j}\right)^{n+2}\cos
  (n\varphi_i) + d_{j,n}^{*i}
  \left(\frac{r_i}{R_j}\right)^{n+2}\sin (n\varphi_i) \right],
  \end{split}
\end{align}
where the amplitudes $\mbf{a}^{*i}_{j,\text{out}}$ of image multipoles relative to the center $\mbf{x}_i=(r=\ell_i,\varphi=\theta_i)$ of fifth $i^\text{th}$ inclusion are related to the amplitudes $\mbf{a}^{*d}_{j,\text{out}}$ relative to the center of disk as
\begin{subequations}
\begin{align}
    a_{j,n}^{*i} =& 
        +\sum_{m=n}^{\infty}{m \choose n}\left(\frac{\ell_i}{R_j}\right)^{m-n}\Big[a_{j,m}^{*d}\cos\big((n-m)\theta_i\big)-b_{j,m}^{*d}\sin\big((n-m)\theta_i\big)\Big]\nonumber\\
        &+ \sum_{m=n-1}^{\infty}{m+1 \choose n}\left(\frac{\ell_i}{R_j}\right)^{m-n+2}\Big[c_{j,m}^{*d}\cos\big((n-m)\theta_i\big)-d_{j,m}^{*d}\sin\big((n-m)\theta_i\big)\Big],\\
    b_{j,n}^{*i} =& 
        +\sum_{m=n}^{\infty}{m \choose n}\left(\frac{\ell_i}{R_j}\right)^{m-n}\Big[a_{j,m}^{*d}\sin\big((n-m)\theta_i\big)+b_{j,m}^{*d}\cos\big((n-m)\theta_i\big)\Big] \nonumber \\
        &+ \sum_{m=n-1}^{\infty}{m+1 \choose n}\left(\frac{\ell_i}{R_j}\right)^{m-n+2}\Big[c_{j,m}^{*d}\sin\big((n-m)\theta_i\big)+d_{j,m}^{*d}\cos\big((n-m)\theta_i\big)\Big],\\
    c_{j,n}^{*i} =& 
        \sum_{m=n}^{\infty}{m+1 \choose n+1}\left(\frac{\ell_i}{R_j}\right)^{m-n}\Big[c_{j,m}^{*d}\cos\big((n-m)\theta_i\big)-d_{j,m}^{*d}\sin\big((n-m)\theta_i\big)\Big],\\
    d_{j,n}^{*i} =& 
        \sum_{m=n}^{\infty}{m+1 \choose n+1}\left(\frac{\ell_i}{R_j}\right)^{m-n}\Big[c_{j,m}^{*d}\sin\big((n-m)\theta_i\big)+d_{j,m}^{*d}\cos\big((n-m)\theta_i\big)\Big].
\end{align}
\end{subequations}

The rest of the steps are the same as  in Section~\ref{sec:flat_edge_traction}.
We expand the induced Airy stress function inside  $i^\text{th}$ inclusion $\chi_{\text{in}}\big(r_i,\varphi_i|\mbf{a}_{i,\text{in}}\big)$ as shown in Eq.~(\ref{eq:AiryInducedIn}), where the set of amplitudes of induced multipoles is represented as $\mbf{a}_{i,\text{in}}=\{a_{i,2}, a_{i,3},\dots,b_{i,2}, b_{i,3},\dots, c_{i,0}, c_{i,1},\dots, d_{i,1}, d_{i,2},\dots\}$. We again define the total Airy stress function $\chi_{\text{in}}^{\text{tot}}\big(x,y|\mbf{a}_{i,\text{in}}\big)$ inside the $i^\text{th}$ inclusion, which also includes the effect of the external load ($\chi_{\text{ext}}$) as shown in Eq.~(\ref{eq:AiryIn}). The amplitudes of induced multipoles $\mbf{a}_{i,\text{out}}$ and $\mbf{a}_{i,\text{in}}$ are obtained by satisfying the boundary conditions that tractions and displacements are continuous across the circumference of each inclusion in Eq.~(\ref{eq:BC}), which can be 
converted to a matrix equation similar to that in Eq.~(\ref{eq:matrix}). The only difference in this procedure is that we use the above Eq.~(\ref{eq:AiryImageCircleGeneralTaylor}) instead of Eq.~(\ref{eq:ImageChiJOutI}) when  expanding the Airy stress functions  $\mathcal{I}\left[\chi_{\text{out}}\big(r_j,\varphi_j|\mbf{a}_{j,\text{out}}\big)\right]$ for the image multipoles around the center $\mbf{x}_i$ of the $i^\text{th}$ inclusion.
To solve the matrix equation numerically, we truncate the degrees of  multipoles at $n_\text{max}$ as discussed in Section~\ref{sec:flat_edge_traction}.

\begin{figure}[!t]
\centering
\includegraphics[scale=1]{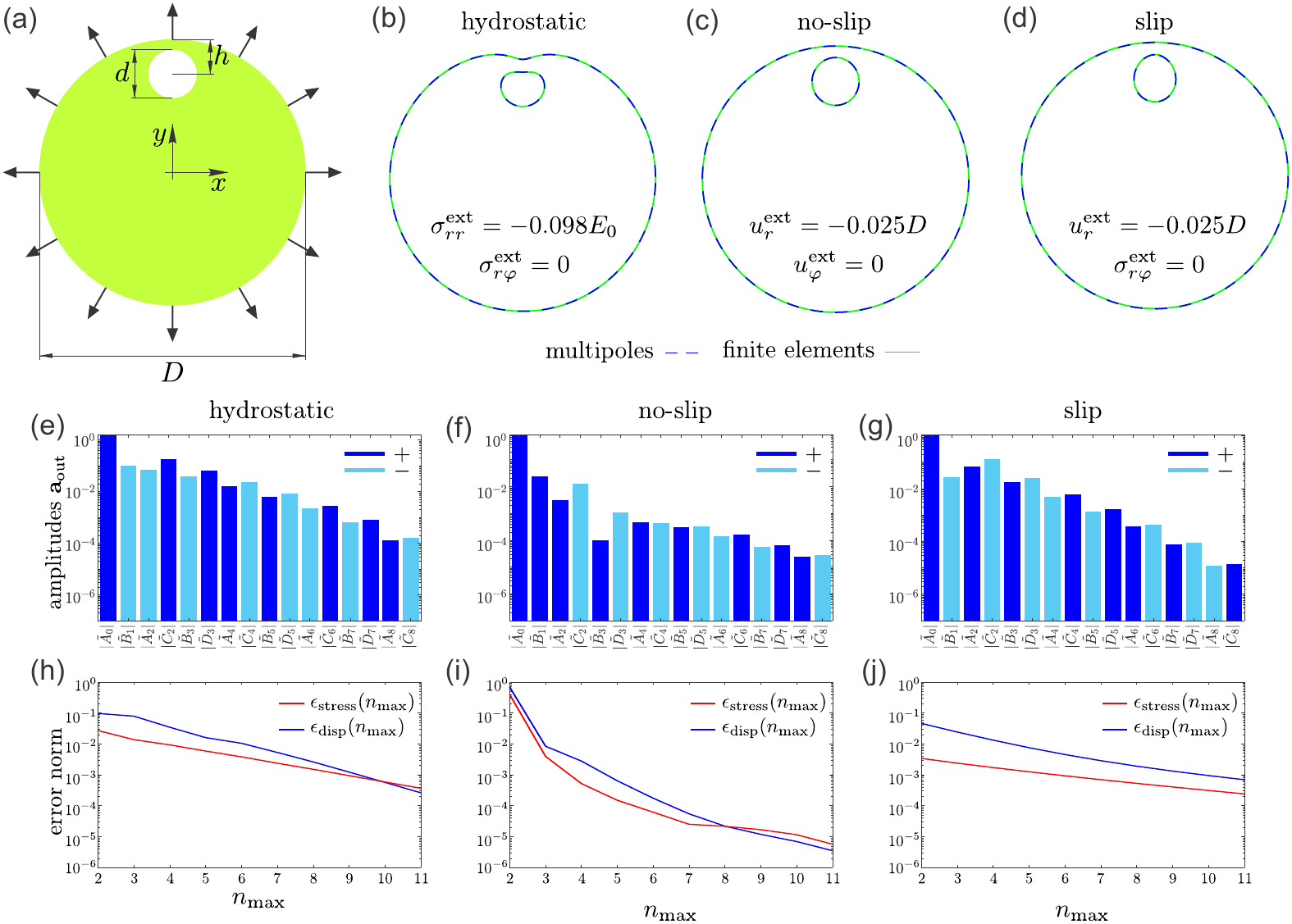}
  \caption{Deformation of a circular hole near the edge of a deformed elastic disk. (a)~Schematic image describing the initial undeformed shape of the structure and the direction of applied deformation. The Poisson's ratio for the elastic matrix was $\nu_0=0.49$ and the Young's modulus was $E_0$. Hole is represented with the white circle. Geometrical parameters are $d/D=0.20$ and $h/D=0.14$.
  (b-d)~Contours of deformed holes and disk boundaries for (b)~hydrostatic stress ($\sigma_{rr}^\text{ext}=-0.098E_0$, $\sigma_{r\varphi}^\text{ext}=0$), (c)~no-slip ($u_{r}^\text{ext}=-0.025D$, $u_{\varphi}^\text{ext}=0$), and (d)~slip ($u_{r}^\text{ext}=-0.025D$, $\sigma_{r\varphi}^\text{ext}=0$) boundary conditions. For all 3 cases, plane stress condition was used with the Kolosov's constant $\kappa_0=(3-\nu_0)/(1+\nu_0)$.
  Dashed blue lines show the contours obtained with elastic multipole method for $n_\text{max}=10$. Green solid lines show the contours obtained from linear finite element simulations.
 (e-g)~Absolute values of the amplitudes of induced multipoles $\mbf{a}_\text{out}$ at the center of hole obtained with $n_\text{max}=10$ for  (e)~hydrostatic stress, (f)~no-slip, and (g)~slip boundary conditions. The amplitudes are normalized, such that $\tilde{A}_n=A_n/\sigma_{rr}^\text{ext}$, $\tilde{B}_n=B_n/\sigma_{rr}^\text{ext}$, $\tilde{C}_n=C_n/\sigma_{rr}^\text{ext}$, $\tilde{D}_n=D_n/\sigma_{rr}^\text{ext}$. For the no-slip and slip boundary conditions we use the relation $\sigma_{rr}^\text{ext}=8 \mu_0 u_r^\text{ext}/[D(\kappa_0-1)]$, where $\mu_0=E_0/[2(1+\nu_0)]$ is the shear modulus of elastic material.
 The dark and light blue colored bars correspond to the positive ($A_n, B_n, C_n, D_n>0$) and negative ($A_n, B_n, C_n, D_n<0$) amplitudes, respectively. Note that the amplitudes $B_{2m}=D_{2m}=A_{2m+1}=C_{2m+1}=0$ due to the symmetry of the problem. 
  (h-j)~The normalized errors for displacements $\epsilon_\text{disp}(n_\text{max})$ (blue lines) and stresses $\epsilon_\text{stress}(n_\text{max})$ (red lines) obtained from Eq.~(\ref{eq:disk_error}) for the (h)~hydrostatic stress, (i)~no-slip, and (j)~slip boundary conditions.}
  \label{fig:disk_all_boundary_conditions}
\end{figure}

The elastic multipole method described above was tested for a compressed elastic disk of diameter $D$ with a single circular hole with diameter $d=0.20D$ for all 3 different boundary conditions discussed above under  plane stress condition (see Fig.~\ref{fig:disk_all_boundary_conditions}). The distance between the center of the hole and the edge of  disk was $h=0.14D$. In Fig.~\ref{fig:disk_all_boundary_conditions}b-d, we show contours of deformed holes and disk boundaries for  (b)~hydrostatic stress, (c)~no-slip, and (d)~slip boundary conditions obtained with the elastic multipole method ($n_\text{max}=10$), which match  very well with the contours obtained with linear finite element simulations (see Appendix~\ref{app:FEM} for details). For easier comparison between different boundary conditions, we chose the value of the hydrostatic stress $\sigma_{rr}^\text{ext}=8 \mu_0 u_r^\text{ext}/[D(\kappa_0-1)]$ such that the pristine elastic disk without holes would have the same displacement of the edge $u_r^\text{ext}$ as was used for the no-slip and slip boundary conditions, where $\mu_0=E_0/[2(1+\nu_0)]$ is the shear modulus and $\kappa_0=(3-\nu_0)/(1+\nu_0)$ is the Kolosov's constant for  plane stress condition. 
For the hydrostatic stress boundary condition in Fig.~\ref{fig:disk_all_boundary_conditions}b, we observe a pronounced deformation in the region where the hole is close to the outer edge of  the disk, which is similar to the case of one hole near the traction-free edge in Fig.~\ref{Fig:hole_flat_edge}. For the no-slip boundary condition in Fig.~\ref{fig:disk_all_boundary_conditions}c we observe that the hole remains circular in the region where the hole is close to the outer edge of the disk, which is similar to the case for one hole near the straight rigid edge in Fig.~\ref{Fig:flat_rigid_edge}. For the slip boundary condition in Fig.~\ref{fig:disk_all_boundary_conditions}d we observe that the deformation of the hole is more pronounced in the region where the hole is close to the outer edge of the disk, but this deformation is not as striking as for the hydrostatic stress boundary condition in Fig.~\ref{fig:disk_all_boundary_conditions}b.

These observations are also reflected in the amplitudes $\mbf{a}_\text{out}$ of induced multipoles at the center of the hole (Fig.~\ref{fig:disk_all_boundary_conditions}e-g). For the hydrostatic stress boundary condition, which results in the most  pronounced deformation of the hole, the amplitudes of induced multipoles decrease very slowly with the degree of multipoles (see Fig.~\ref{fig:disk_all_boundary_conditions}e). In contrast, the no-slip  boundary condition results in the least pronounced deformation of the hole and the amplitudes of induced multipoles decrease much more rapidly with the degree of multipoles (see Fig.~\ref{fig:disk_all_boundary_conditions}f). The results for the slip boundary condition are somewhere in between (see Fig.~\ref{fig:disk_all_boundary_conditions}g).

The convergence analysis for the spatial distributions of displacements $\mbf{u}^{(n_\text{max})}(r,\varphi)$ and von Mises stress $\sigma^{(n_\text{max})}_\text{vM}(r,\varphi)$ in the elastic multipole method was used to asses how many multipoles are needed. Displacements and von Mises stresses were evaluated at $N_p$ discrete points $(r_i,\varphi_j)=\left(iR/500,2\pi j/500\right)$, where $i\in\{1,2,\ldots,500\}$ and $j\in\{0, 1, \ldots, 499\}$ and grid points that lie inside the hole were excluded. The normalized errors for displacements $\epsilon_\text{disp}(n_\text{max})$ and stresses $\epsilon_\text{stress}(n_\text{max})$ were obtained by calculating the relative changes of  spatial distributions of displacements and von Mises stresses when the maximum degree $n_\text{max}$ of induced multipoles is increased by one. The normalized errors are given by~\cite{SpecMethod}
\begin{subequations}
\begin{align}
\label{Eqn:disk_errordisp}
    \epsilon_\text{disp}(n_\text{max}) &= \frac{1}{\sqrt{N_p}}\left[\sum_{i,j}\left|\frac{\mbf{u}^{(n_\text{max}+1)}(r_i,y\varphi_j)-\mbf{u}{(n_\text{max})}(r_i,\varphi_j)}{d\,\sigma_\text{vM}^\text{ext}/E_0}\right|^2 \right]^{1/2},\\
\label{Eqn:disk_errorstress}
    \epsilon_\text{stress}(n_\text{max}) &= \frac{1}{\sqrt{N_p}}\left[\sum_{i,j}\left(\frac{\sigma^{(n_\text{max}+1)}_\text{vM}(r_i,y\varphi_j)-\sigma^{(n_\text{max})}_\text{vM}(r_i,y\varphi_j)}{\sigma_\text{vM}^\text{ext}}\right)^2 \right]^{1/2}.
\end{align}
\label{eq:disk_error}%
\end{subequations}
Here, displacements and von Mises stresses are normalized with the characteristic scales $d \sigma_\text{vM}^\text{ext}/E_0$ and $\sigma_\text{vM}^\text{ext}$, respectively, where $d=2R$ is the diameter of the hole, $\sigma_\text{vM}^\text{ext}$ is the value of  von Mises stress due to external load, and $E_0$ is the Young's modulus of  surrounding elastic matrix. The normalized errors for displacements $\epsilon_\text{disp}(n_\text{max})$ and stresses $\epsilon_\text{stress}(n_\text{max})$ for all 3 boundary conditions are shown in Fig.~\ref{fig:disk_all_boundary_conditions}h-j. The normalized errors decrease exponentially with the maximum degree of multipoles $n_\text{max}$, which mimics the exponential decay of amplitude of induced multipoles in Fig.~\ref{fig:disk_all_boundary_conditions}e-g. The normalized errors for the  hydrostatic stress boundary condition were the largest (Fig.~\ref{fig:disk_all_boundary_conditions}h), which is reflecting the observation that the amplitudes of multipoles decrease very slowly with the degree of multipoles (see Fig.~\ref{fig:disk_all_boundary_conditions}e). In contrast, the normalized errors for the no-slip boundary condition reduced very quickly with the maximum degree of multipoles $n_\text{max}$ (see Fig.~\ref{fig:disk_all_boundary_conditions}i), which is again mimicking the distribution of  amplitudes of induced multipoles (see Fig.~\ref{fig:disk_all_boundary_conditions}f). The results for the slip boundary condition are somewhere in between (see Fig.~\ref{fig:disk_all_boundary_conditions}j).

\begin{figure}[!t]
\centering
\includegraphics[scale=1]{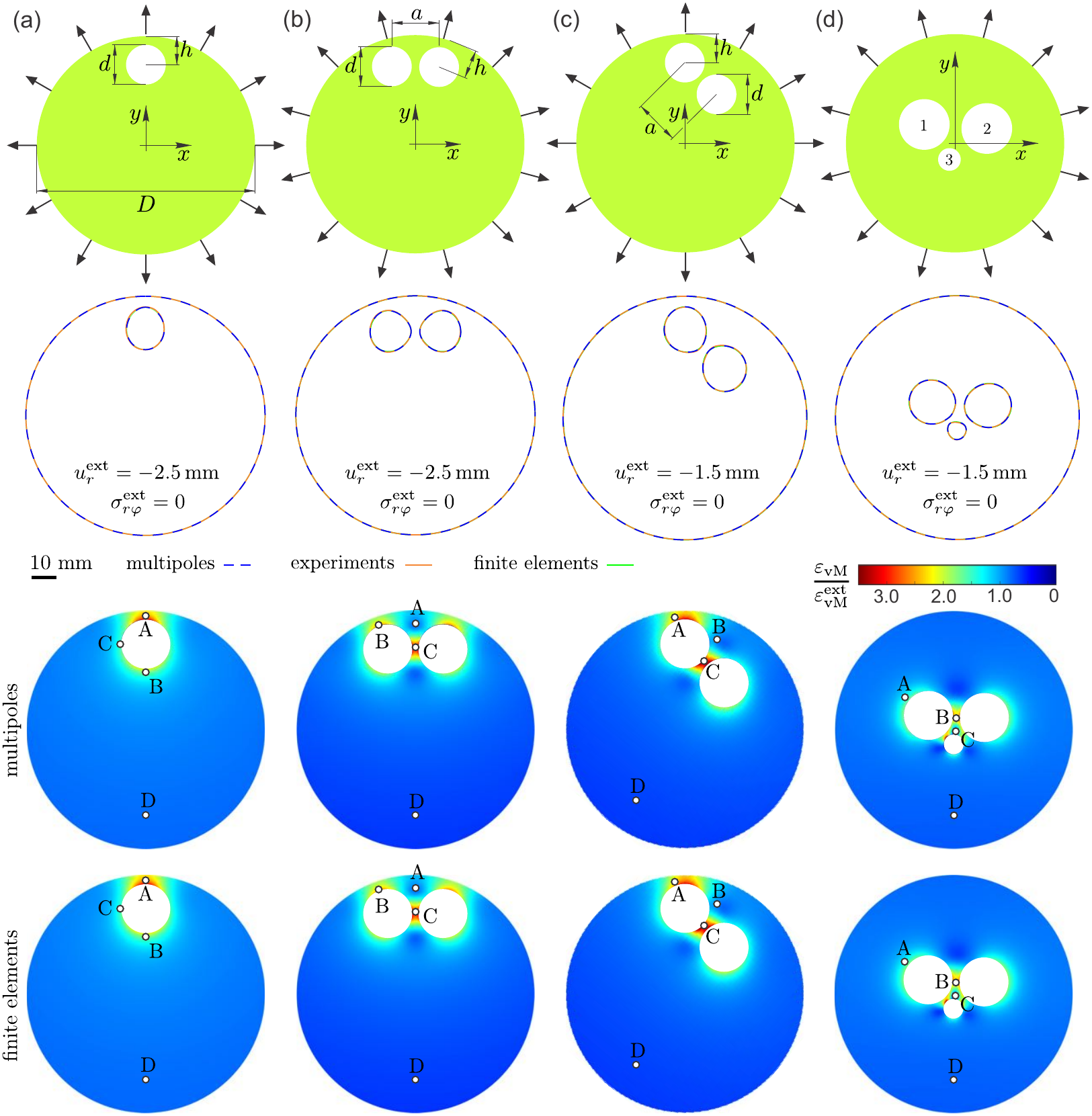}
  \caption{Deformation of elastic disks with holes. (a-d)~Schematic images describing the initial undeformed shapes of structures with holes and the direction of applied deformation $u_r^\text{ext}$ for the slip boundary condition.  Holes are represented with white circles.  For all elastic disks, the diameter was $D=100$~mm and the thickness was $25$~mm. They were made of rubber with the Young's modulus $E_0=0.97\,\textrm{MPa}$ and  Poisson's ratio is $\nu_0=0.49$.   For each structure we show the contours of deformed holes and disk boundaries and we report the value of applied compression $u_r^\text{ext}$. Plane stress condition was used for all cases with the Kolosov's constant $\kappa_0=(3-\nu_0)/(1+\nu_0)$.  Dashed blue lines show the contours obtained with elastic multipole method for $n_\text{max}=10$. Orange and green solid lines show the contours obtained with experiments and linear finite element simulations, respectively. For each of the 4 structures, we show equivalent von Mises strain fields $\varepsilon_\text{vM}$ obtained with elastic multipole method ($n_\text{max}=10$) and finite element simulations. Strain fields were normalized with the value of the equivalent von Mises strain $\varepsilon_\text{vM}^\text{ext}=4u_r^\text{ext}/[D(1+\nu_0)(\kappa_0-1)]$ imposed by external loads. Four marked points A-D were chosen for the quantitative comparison of strains $\varepsilon_\text{vM}$. See Table~\ref{tab:disk} for details. 
 The diameters of holes in (a-c) were $d=20.0$~mm. In (a-c) the distance $h$ from the centers of holes to the edge of the elastic disk were (a)~$h=14.00$~mm, (b)~$h=13.95$~mm, and (c)~$h=14.00$~mm. In (b-c) the separation distance $a$ between holes was (b)~$a=23.07$~mm and (c)~$a=23.10$~mm. In panel (d) the diameters of holes  were $d_1=d_2=20.0$~mm and $d_3=8.0$~mm and they were centered at $(x_1,y_1)=(-10.64~\text{mm},6.11~\text{mm})$, $(x_2,y_2)=(12.76~\text{mm},5.17~\text{mm})$ and $(x_3,y_3)=(-0.11~\text{mm},-5.97~\text{mm})$.
  }
  \label{fig:circle_exp_deformation}
\end{figure}

The elastic multipole method was also tested against experiments. We developed a simple experimental system for  radial compressive loading of elastic disks with holes. The elastic disks were placed in rigid rings with smaller diameters  made from polyethylene plastic. Elastic discs with diameter $D=100$~mm and thickness $25$~mm were pushed sequentially through the rigid rings of decreasing diameters ranging from $99$~mm to $94$~mm in steps of $1$~mm. Silicone oil was applied to reduce  friction between the rubber sample and the walls of rigid rings. This experimental setup is thus approximating the slip boundary condition discussed above. The contours of deformed holes were then obtained by scanning the surface of compressed samples.

In Fig.~\ref{fig:circle_exp_deformation} we show initial and deformed configurations of disks with 4 different arrangements of holes. In experiments, disks were pushed through all 6 rigid rings. For each ring, the surface of disks was scanned and the contours of deformed holes were extracted with the Image Processing Toolbox in MATLAB 2018b. At higher compression, we observed a noticeable out-of-plane deformation in the vicinity of holes, which is not captured with the elastic multipole method. Therefore we used only those experimental results, where the out-of-plane deformation was minimal.
The deformed contours of holes obtained with experiments were compared to the contours obtained with the elastic multiple method ($n_\text{max}=10$) and  linear finite element simulations  for the slip boundary condition and under  plane stress condition. An excellent match was observed between the contours obtained with different methods (second row in Fig.~\ref{fig:circle_exp_deformation}). As in previous examples, we observed more pronounced deformations of holes in the regions, where holes were close to each other, and in regions, where holes were close to the edge of the disk.
We have also compared the equivalent von Mises strain fields $\varepsilon_\text{vM}$ (see Eq.~(\ref{eq:vonMisesStrain})) obtained with the elastic multipole method (third row in Fig.~\ref{fig:circle_exp_deformation}) and finite elements (fourth row in Fig.~\ref{fig:circle_exp_deformation}). In strain fields, we observe strain concentration in regions where holes are close to each other and in regions where holes are close to the disk boundary, which is similar to the observations for holes near traction-free edges in Sections~\ref{sec:flat_edge_traction} and \ref{sec:strip}.
The agreement between the strain fields obtained with two different methods is very good and the quantitative comparison of strains at four different points A-D (marked in Fig.~\ref{fig:circle_exp_deformation}) showed a relative error of $\sim1\%$ (see Table~\ref{tab:disk}). Note that the relative errors were so small because we used exactly the same geometry of elastic disks for both finite elements and elastic multipole method.

 \begin{table}[!t]
\centering
\caption{Quantitative comparison for the values of equivalent von Mises strains $\varepsilon_\text{vM}$ normalized with the value for the applied external load $\varepsilon_\text{vM}^\text{ext}=4|u_r^\text{ext}|/[D(1+\nu_0)(\kappa_0-1)]$ at points A-D for 4 different disk samples with holes defined in Fig.~\ref{fig:circle_exp_deformation} obtained with the elastic multipole method (EMP) and linear finite element simulations (FEM). The relative percent errors  between  EMP and FEM were calculated as $\epsilon=100\times|\varepsilon_\text{vM}^\text{(EMP)}-\varepsilon_\text{vM}^\text{(FEM)}|/\varepsilon_\text{vM}^\text{(FEM)}$.}
\label{tab:disk}
\def\arraystretch{1.1}
\begin{tabular}{|>{\centering}m{0.052\textwidth}|*{2}{>{\centering}m{0.052\textwidth}}>{\centering}m{0.036\textwidth}|*{2}{>{\centering}m{0.052\textwidth}}>{\centering}m{0.036\textwidth}|*{2}{>{\centering}m{0.052\textwidth}}>{\centering}m{0.036\textwidth}|*{2}{>{\centering}m{0.052\textwidth}}>{\centering\arraybackslash}m{0.036\textwidth}|}
\hline
\multirow{3}{*}{points}& \multicolumn{3}{c|}{sample (a)}& \multicolumn{3}{c|}{sample (b)}& \multicolumn{3}{c|}{sample (c)}&
\multicolumn{3}{c|}{sample (d)}\\
\cline{2-13}
& \multicolumn{3}{c|}{strain $\epsilon_\text{vM}/\epsilon_\text{vM}^\text{ext}$} & \multicolumn{3}{c|}{strain $\epsilon_\text{vM}/\epsilon_\text{vM}^\text{ext}$}& \multicolumn{3}{c|}{strain $\epsilon_\text{vM}/\epsilon_\text{vM}^\text{ext}$}&
\multicolumn{3}{c|}{strain $\epsilon_\text{vM}/\epsilon_\text{vM}^\text{ext}$}\\
\cline{2-13}
  &  EMP   & FEM   & $\epsilon(\%)$  &  EMP   & FEM   & $\epsilon(\%)$ &  EMP   & FEM   & $\epsilon(\%)$ &  EMP   & FEM   & $\epsilon(\%)$ \\
\hline
A & 2.5205 & 2.5396 & 0.75 & 0.8505 & 0.8475 & 0.36 & 2.2558 & 2.2703 & 0.64 & 1.3196 & 1.3227 & 0.23 \\
B & 1.5034 & 1.4946 & 0.59 & 2.3422 & 2.3406 & 0.07 & 0.9111 & 0.9047 & 0.71  & 2.3110 & 2.2805 & 1.34\\
C & 1.7716 & 1.7723 & 0.04 & 2.8214 & 2.8076 & 0.49 & 3.3440 & 3.3368 & 0.19 & 0.7336 & 0.7248 & 1.21\\
D & 0.8077 & 0.8075 & 0.03 & 0.6439 & 0.6441 & 0.03 & 0.6586 & 0.6584 & 0.03 & 0.7592 & 0.7593 & 0.01\\
\hline            
\end{tabular}
\end{table}

\section{Conclusion}
\label{sec:conclusion}

In this paper, we demonstrated how  image charges and induction, which are common concepts in electrostatics, can be effectively used also for the analysis of linear deformation of bounded 2D elastic structures with circular holes and inclusions for both plane stress and plane strain conditions. The multipole expansion of induced fields described in this work is a so-called \textit{far-field} method and hence it is extremely efficient when holes and inclusions are far apart from each other and when they are far  from the boundaries. In this case, very accurate results can be obtained by considering only induced quadrupoles, since the effect of higher-order multipoles decays more rapidly at large distances. When holes and inclusions are closer to each other or when they are closer to the boundaries, their interactions with each other via induced higher-order multipoles and their interactions with induced images become important as well. The accuracy of the results increases exponentially with the maximum degree of elastic multipoles, which is also the case in electrostatics, and this is characteristic for spectral methods~\cite{SpecMethod}. The results of the elastic multipole method matched very well with both  linear finite element simulations and experiments.

The elastic multipole method presented here could also be extended to other geometries of structures with holes and inclusions. The analysis for elastic disks in Section~\ref{sec:disk} could be directly extended to the analysis for linear deformation of an elastic annulus with circular holes and inclusion by constructing an infinite set of image multipoles similarly as was done for the elastic strip in Section~\ref{sec:strip}. For structures with other geometries, the first step is to identify the Airy stress function for the image of disclination, which is directly related to the Green's function for biharmonic equation (see Section~\ref{sec:images}). The Green's function for a rectangular geometry can be obtained with a Fourier series~\cite{Barber}. The Green's function for an infinite elastic wedge with traction-free boundary conditions was previously obtained by Gregory in Ref.~\cite{Gregory}. Interestingly, for a sufficiently small wedge angle the Airy stress functions exhibit an infinite number of oscillations near the wedge tip~\cite{moffatt1964viscous,Osher,Seif,Gregory}. Once the Airy stress functions for the images of disclinations are found, then the Airy stress functions for images of all other multipoles can be obtained by following the procedure outlined in Section~\ref{sec:images}. These results could then be used to analyze deformations of  structures with holes and inclusions, where external load induces elastic multipoles at the center of  holes and inclusions, which further induce image multipoles to satisfy boundary conditions at the outer boundaries of these structures. The amplitudes of induced multipoles can then be obtained from the continuity of tractions and displacements at the boundary of each hole as described in Section~\ref{sec:flat_edge_traction}.

While the elastic multipole method presented here focused  on linear deformation, these concepts can also be extended to the nonlinear regime. Recently, Bar-Sinai \textit{et al.} used induced elastic quadrupoles to describe the nonlinear deformation of compressed structures with periodic arrays of holes to estimate the initial linear deformation, the critical buckling load, as well as the buckling mode~\cite{Moshe6}. The accuracy of their method could be  improved by expanding the induced fields to higher order multipoles and by including image multipoles. 
Thus the elastic multipole method has the potential to significantly advance our understanding of deformation patterns in structures with holes and inclusions.

\section*{Acknowledgements}
This work was supported by NSF through the Career Award DMR-1752100 and by the Slovenian Research Agency through the grant P2-0263. We would like to acknowledge useful discussions with Michael Moshe (Hebrew University) and thank Jonas Trojer (University of Ljubljana) for the help with experiments.

\appendix

\section{Linear finite element simulations}
\label{app:FEM}
Linear analyses in finite element simulations were performed with the commercial software Ansys\textsuperscript{\textregistered} Mechanical Release 17.2. Geometric models of 2D plates with holes and inclusions were discretized with 2D eight-node, quadratic elements of type PLANE183 set to the plane stress option. The material for plates and inclusions was modeled as a linear isotropic elastic material. To ensure high accuracy, we used a fine mesh with 360 quadratic elements evenly spaced around the circumference of each hole and each inclusion, and elements of similar sizes were also used inside the inclusions. To keep the number of elements at a manageable level, the size of elements was allowed to increase at a rate of approximately $2\%$ per element, when moving away from  inclusions.

For the simulation of a single hole with radius $R$ embedded in a compressed plate near a traction-free edge in Fig.~\ref{Fig:hole_flat_edge}, we used a sufficiently large rectangular domain of size $L \times w$ with length $L=800R$ and width $w=400R$ to minimize the effect of boundaries far away from the hole. 
Along the left and right boundaries, we prescribed tractions $\sigma_{xx}(\pm L/2,y)=\sigma_{xx}^\text{ext}$ and $\sigma_{xy}(\pm L/2,y)=0$, and along the top and bottom boundaries, we prescribed tractions $\sigma_{xy}(x,0)=\sigma_{yy}(x,0)=\sigma_{xy}(x,-w)=\sigma_{yy}(x,-w)=0$. To prevent rigid body motions of the structure, the midpoints of the left and right edges of the domain were constrained to move only in the $x$-direction ($u_y(\pm L/2,0)=0$), while the midpoint of the bottom edge of the domain was fixed in the $x$-direction ($u_x(0,-w)=0$).

For comparison with the experiment of the sample with one hole near a traction-free edge in Fig.~\ref{Fig:ExperimentalFreeEdge} we simulated a plate of length $L=100$~mm and width $w=100$~mm. To mimic experimental conditions, the plate was compressed by prescribing a uniform horizontal displacement on the left and right surfaces ($u_x(\pm L/2,y)=\pm L/2\varepsilon_{xx}^\text{ext}$), while allowing the nodes on these surfaces to move freely in the $y$-direction ($\sigma_{xy}(\pm L/2,y)=0$). The top and bottom boundaries were traction-free ($\sigma_{xy}(x,0)=\sigma_{yy}(x,0)=\sigma_{xy}(x,-w)=\sigma_{yy}(x,-w)=0$). The midpoints of the left and right edges were again constrained ($u_y(\pm L/2,-w/2)=0$) to prevent rigid body translation in the $y$-direction.

For the simulation of elastic strips of length $L=100$~mm and width $w$ in Figs.~\ref{fig:strip_holes} and \ref{fig:strip_holes_strain} we used a uniform external pressure load at the two ends of the strip ($\sigma_{xx}(\pm L/2,y)=\sigma_{xx}^\text{ext}$, $\sigma_{xy}(\pm L/2,y)=0$), while the other boundaries were traction-free ($\sigma_{xy}(x,\pm w/2)=\sigma_{yy}(x,\pm w/2)=0$).  To prevent rigid body motions of the strip we restricted 3 degrees of freedom ($u_x(0,-w/2)=0$, $u_y(0,-w/2)=0$, $u_x(0,w/2)=0$).

For the simulation of a single hole with radius $R$ located near a straight rigid edge in Fig.~\ref{Fig:flat_rigid_edge}, we used a domain of length $L=800R$ and width $w=400R$. Points on the top boundary of the domain were fixed ($u_x(x,0)=u_y(x,0)=0$) and the sample was compressed by applying uniform displacement in the $y$-direction on the bottom boundary of the plate ($u_y(x,-w)=w \varepsilon_{yy}^\text{ext}$, $\sigma_{xy}(x,-w)=0$). Points on the left and right boundaries were constrained to move only in the $y$-direction ($u_x(\pm L/2,y)=0$ and $\sigma_{xy}(\pm L/2,y)=0$).

For the simulations of elastic disks with radius $R$ in Figs.~\ref{fig:disk_all_boundary_conditions} and \ref{fig:circle_exp_deformation}  we considered three different boundary conditions: hydrostatic ($\sigma_{\text{tot},rr}^\text{out}(r=R,\varphi)=\sigma_{rr}^\text{ext}$, $\sigma_{\text{tot},r\varphi}^\text{out}(r=R,\varphi)=0$), no-slip ($u_{\text{tot},r}^\text{out}(r=R,\varphi)=u_{r}^\text{ext}$, $u_{\text{tot},\varphi}^\text{out}(r=R,\varphi)=0$), and slip ($u_{\text{tot},r}^\text{out}(r=R,\varphi)=u_{r}^\text{ext}$, $\sigma_{\text{tot},r\varphi}^\text{out}(r=R,\varphi)=0$).
To prevent the rigid-body motions of elastic discs with holes, we fixed 3 degrees of freedom for the hydrostatic load ($u_x(x=0,y=0)=u_y(x=0,y=0)=u_x(x=0,y=R)=0$) and 1 degree of freedom for the slip boundary condition ($u_x(x=0,y=R)=0$). No additional constraints were needed for the no-slip boundary condition.

\section{Experimental Methods}
\label{app:experiments}
Experimental samples were prepared by casting Elite Double 32 (Zhermack) elastomers with a measured Young's modulus $E_0=0.97$~MPa and assumed Poisson's ratio $\nu=0.49$~\cite{babaee20133d}.
Molds were fabricated from 5 mm thick acrylic plates with laser cut circular holes, which were then filled with acrylic cylinders in the assembled molds to create cylindrical holes in elastomer samples. Approximately 30~min after casting, molds were disassembled and solid samples were placed in a convection oven at 40\;$^\circ$C for 12~hours for further curing. The cylindrical inclusions made from acrylic (Young's modulus $E=2.9$~GPa, Poisson's ratio $\nu=0.37$~\cite{acrylic})
were inserted into the holes in elastomer samples and glued by a cyanoacrylate adhesive where required.

\begin{figure}[!t]
\centering
\includegraphics[scale=1]{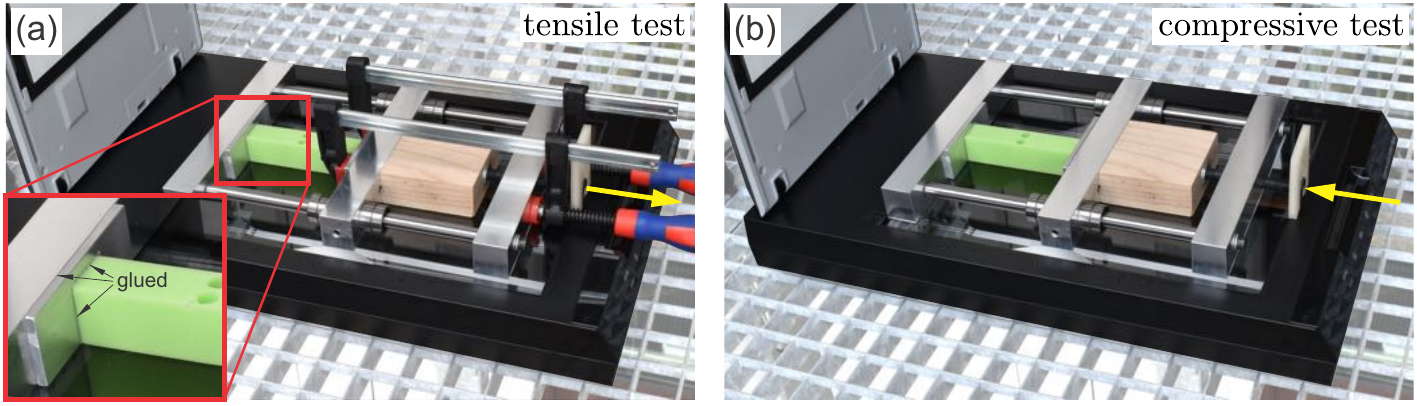}
  \caption{Experimental systems for displacement controlled (a) tensile and (b) compressive tests for thin strips with holes and inclusions. The mechanism sits on a scanning device, which is used to extract the contours of deformed holes and inclusions. The two ends of the strip were glued to the aluminum plates (see the zoomed-in photo in (a)) to prevent bending due to the asymmetric position of holes/inclusion for both tensile and compressive loading. F-clamps were used to apply the tensile load to the elastic strip in (a).  A 3D printed plastic wrench was used to precisely control the screw turns to achieve the desired compression in (b).}
\label{Fig:exp_both_edges}
\end{figure}

We designed a custom testing system for extracting the contours of deformed holes and inclusions in compressed experimental samples. The system comprises a custom-made loading mechanism and a flatbed photo scanner (see Fig.~\ref{Fig:exp_both_edges}).
For the experiment in Section~\ref{sec:flat_edge_traction}, the sample was placed between aluminum plates and silicone oil was applied to reduce friction. Compressive displacement loading was applied incrementally in 1/3 mm steps via $120^{\circ}$ turns  of the M10x1 screw (metric thread with 10 mm diameter and 1 mm pitch) in the mechanism controlled by a 3D printed plastic wrench (see Fig.~\ref{Fig:exp_both_edges}b). For the experiments with elastic strips in Section~\ref{sec:strip}, samples were glued to aluminum plates to prevent bending due to the asymmetric position of holes/inclusions (see Fig.~\ref{Fig:exp_both_edges}a). Compressive loading was applied as described above. For the tensile displacement loading, we again turned the M10x1 screw in increments of 1/3 mm. The outward movement of the screw created a gap between the screw and the wooden block, which was then closed by applying compression with F-style clamps (see Fig.~\ref{Fig:exp_both_edges}a).

The loading mechanism was placed on an Epson V550 photo scanner to scan the surface of deformed samples and silicone oil was applied between the sample and the glass surface of the scanner to reduce friction between them. Scanned images were post-processed with the Corel PHOTO-PAINT X8 and Image Processing Toolbox in MATLAB 2018b. First, the dust particles and air-bubbles trapped in a thin silicone oil film were digitally removed from scanned images. Scanned grayscale images were then converted to black and white binary images from which the contours were extracted with MATLAB. Note that when samples were mounted in the loading mechanism they could be slightly compressed even before we start turning the screw to apply additional loads. To identify the value of this offset, we fitted contours of holes/inclusions/edges for each sample in experiments to those obtained with finite elements at a specific wrench position (usually, at the third increment, i.e. $\sim 1.0$~mm displacement). This offset was then used when we compared experiments to the multipole method and simulations at a different value of the applied load.

We designed another testing system for capturing the displacement and strain fields in compressed samples via  digital image correlation (DIC) technique (see Fig.~9b in Ref.~\cite{sarkar2019elastic}). Black and white speckle patterns were sprayed onto the surface of  samples with slow-drying acrylic paint to prevent the pattern from hardening too quickly, which could lead to delamination under applied compressive loads. Using a Zwick Z050 universal material testing machine, we applied  a  compressive displacement in 0.2~mm increments, where  again a silicone oil was applied between the steel plates and the elastomer samples to prevent sticking and to reduce friction. A Nikon D5600 photo camera was used at each step to take a snapshot of the compressed sample. These photos were then used to calculate the displacements and strains fields via Ncorr, an open-source 2D DIC MATLAB based software.~\cite{DIC}

\bibliography{refer.bib}
\bibliographystyle{ieeetr}
\end{document}